\pdfoutput=1
\documentclass[12pt,a4paper]{article}

\usepackage{ifthen} 
\newboolean{pdflatex}
\setboolean{pdflatex}{true} 

\newboolean{articletitles}
\setboolean{articletitles}{true} 

\newboolean{uprightparticles}
\setboolean{uprightparticles}{false} 

\def\paperauthors{LHCb collaboration}
\def\paperasciititle{Search for time-dependent CP violation in D0 -> K+ K- and D0 -> pi+ pi- decays} 
\def\papertitle{Search for time-dependent\\ $C\!P$ violation in $D^0 \to K^+ K^-$\\ and $D^0 \to \pi^+ \pi^-$ decays} 
\def\paperkeywords{{flavour physics}, {charm physics}, {CP violation}, {oscillation}}
\def\papercopyright{\the\year\ CERN for the benefit of the LHCb collaboration}
\def\paperlicence{CC BY 4.0 licence}
\def\paperlicenceurl{https://creativecommons.org/licenses/by/4.0/}

\usepackage[top=1in, bottom=1.25in, left=1in, right=1in]{geometry}

\usepackage{microtype}
\usepackage{lineno}  
\usepackage{xspace} 
\usepackage{caption} 

\usepackage{graphicx}  
\usepackage{color}
\usepackage{colortbl}
\graphicspath{{./figs/}} 

\usepackage{amsmath} 
\usepackage{amssymb}
\usepackage{amsfonts}
\usepackage{upgreek} 

\newcommand*\patchAmsMathEnvironmentForLineno[1]{%
\expandafter\let\csname old#1\expandafter\endcsname\csname #1\endcsname
\expandafter\let\csname oldend#1\expandafter\endcsname\csname
end#1\endcsname
 \renewenvironment{#1}%
   {\linenomath\csname old#1\endcsname}%
   {\csname oldend#1\endcsname\endlinenomath}%
}
\newcommand*\patchBothAmsMathEnvironmentsForLineno[1]{%
  \patchAmsMathEnvironmentForLineno{#1}%
  \patchAmsMathEnvironmentForLineno{#1*}%
}
\AtBeginDocument{%
\patchBothAmsMathEnvironmentsForLineno{equation}%
\patchBothAmsMathEnvironmentsForLineno{align}%
\patchBothAmsMathEnvironmentsForLineno{flalign}%
\patchBothAmsMathEnvironmentsForLineno{alignat}%
\patchBothAmsMathEnvironmentsForLineno{gather}%
\patchBothAmsMathEnvironmentsForLineno{multline}%
\patchBothAmsMathEnvironmentsForLineno{eqnarray}%
}

\usepackage{hyperxmp}

\usepackage[pdftex,
            pdfauthor={\paperauthors},
            pdftitle={\paperasciititle},
            pdfkeywords={\paperkeywords},
            pdfcopyright={Copyright (C) \papercopyright},
            pdflicenseurl={\paperlicenceurl}]{hyperref}

\usepackage[colorinlistoftodos,textsize=scriptsize]{todonotes}

\usepackage[bottom,flushmargin,hang,multiple]{footmisc}

\usepackage[all]{hypcap} 

\usepackage{xspace} 
\usepackage{upgreek}

\def\lhcb   {\mbox{LHCb}\xspace}

\def\babar  {\mbox{BaBar}\xspace}
\def\belle  {\mbox{Belle}\xspace}

\def\cdf    {\mbox{CDF}\xspace}

\def\lhc    {\mbox{LHC}\xspace}

\def\rich   {RICH\xspace}

\def\MagUp {\mbox{\em Mag\kern -0.05em Up}\xspace}
\def\MagDown {\mbox{\em MagDown}\xspace}

\ifthenelse{\boolean{uprightparticles}}%
{

 \def\Pmu         {\ensuremath{\upmu}\xspace}                 
 \def\Pnu         {\ensuremath{\upnu}\xspace}                 
                  
 \def\Ppi         {\ensuremath{\uppi}\xspace}

 \def\PDelta      {\ensuremath{\Delta}\xspace}                 
 \def\PXi         {\ensuremath{\Xi}\xspace}                 
 \def\PLambda     {\ensuremath{\Lambda}\xspace}                 
 \def\PSigma      {\ensuremath{\Sigma}\xspace}                 
 \def\POmega      {\ensuremath{\Omega}\xspace}                 
 \def\PUpsilon    {\ensuremath{\Upsilon}\xspace}

 \def\PB      {\ensuremath{\mathrm{B}}\xspace}                 
                  
 \def\PD      {\ensuremath{\mathrm{D}}\xspace}

 \def\PK      {\ensuremath{\mathrm{K}}\xspace}

 \def\PP      {\ensuremath{\mathrm{P}}\xspace}

 \def\PW      {\ensuremath{\mathrm{W}}\xspace}

 \def\Pb      {\ensuremath{\mathrm{b}}\xspace}                 
 \def\Pc      {\ensuremath{\mathrm{c}}\xspace}                 
                  
 \def\Pe      {\ensuremath{\mathrm{e}}\xspace}

 \def\Ph      {\ensuremath{\mathrm{h}}\xspace}                 
 \def\Pi      {\ensuremath{\mathrm{i}}\xspace}

 \def\Pp      {\ensuremath{\mathrm{p}}\xspace}

 \def\Ps      {\ensuremath{\mathrm{s}}\xspace}                 
 \def\Pt      {\ensuremath{\mathrm{t}}\xspace}                 
 \def\Pu      {\ensuremath{\mathrm{u}}\xspace}

 \def\thebaroffset{0.0em}
}
{

 \def\Pmu         {\ensuremath{\mu}\xspace}                 
 \def\Pnu         {\ensuremath{\nu}\xspace}                 
                  
 \def\Ppi         {\ensuremath{\pi}\xspace}

 \mathchardef\PDelta="7101
 \mathchardef\PXi="7104
 \mathchardef\PLambda="7103
 \mathchardef\PSigma="7106
 \mathchardef\POmega="710A
 \mathchardef\PUpsilon="7107
                  
 \def\PB      {\ensuremath{B}\xspace}                 
                  
 \def\PD      {\ensuremath{D}\xspace}

 \def\PK      {\ensuremath{K}\xspace}

 \def\PP      {\ensuremath{P}\xspace}

 \def\PW      {\ensuremath{W}\xspace}

 \def\Pb      {\ensuremath{b}\xspace}                 
 \def\Pc      {\ensuremath{c}\xspace}                 
                  
 \def\Pe      {\ensuremath{e}\xspace}

 \def\Ph      {\ensuremath{h}\xspace}                 
 \def\Pi      {\ensuremath{i}\xspace}

 \def\Pp      {\ensuremath{p}\xspace}

 \def\Ps      {\ensuremath{s}\xspace}                 
 \def\Pt      {\ensuremath{t}\xspace}                 
 \def\Pu      {\ensuremath{u}\xspace}

 \def\thebaroffset{0.18em}
}
\newcommand{\offsetoverline}[2][\thebaroffset]{\kern #1\overline{\kern -#1 #2}}%

\makeatletter
\ifcase \@ptsize \relax
  \newcommand{\miniscule}{\@setfontsize\miniscule{4}{5}}
\or
  \newcommand{\miniscule}{\@setfontsize\miniscule{5}{6}}
\or
  \newcommand{\miniscule}{\@setfontsize\miniscule{5}{6}}
\fi
\makeatother

\DeclareRobustCommand{\optbar}[1]{\shortstack{{\miniscule (\rule[.5ex]{1.25em}{.18mm})}
  \\ [-.7ex] $#1$}}

\def\ep         {{\ensuremath{\Pe^+}}\xspace}

\def\mup        {{\ensuremath{\Pmu^+}}\xspace}
\def\mun        {{\ensuremath{\Pmu^-}}\xspace} 

\def\ellp       {{\ensuremath{\ell^+}}\xspace}

\def\neu        {{\ensuremath{\Pnu}}\xspace}

\def\neue       {{\ensuremath{\neu_e}}\xspace}

\def\neul       {{\ensuremath{\neu_\ell}}\xspace}

\def\W      {{\ensuremath{\PW}}\xspace}

\def\uquark    {{\ensuremath{\Pu}}\xspace}

\def\squark    {{\ensuremath{\Ps}}\xspace}

\def\cquark    {{\ensuremath{\Pc}}\xspace}

\def\bquark    {{\ensuremath{\Pb}}\xspace}

\def\tquark    {{\ensuremath{\Pt}}\xspace}

\def\hadron {{\ensuremath{\Ph}}\xspace}
\def\pion   {{\ensuremath{\Ppi}}\xspace}
\def\piz    {{\ensuremath{\pion^0}}\xspace}
\def\pip    {{\ensuremath{\pion^+}}\xspace}
\def\pim    {{\ensuremath{\pion^-}}\xspace}

\def\kaon    {{\ensuremath{\PK}}\xspace}

\def\KorKbar {\kern \thebaroffset\optbar{\kern -\thebaroffset \PK}{}\xspace}

\def\Kp      {{\ensuremath{\kaon^+}}\xspace}
\def\Km      {{\ensuremath{\kaon^-}}\xspace}

\def\Dbar    {{\ensuremath{\offsetoverline{\PD}}}\xspace}
\def\D       {{\ensuremath{\PD}}\xspace}

\def\DorDbar {\kern \thebaroffset\optbar{\kern -\thebaroffset \PD}\xspace}
\def\Dz      {{\ensuremath{\D^0}}\xspace}
\def\Dzb     {{\ensuremath{\Dbar{}^0}}\xspace}
\def\Dp      {{\ensuremath{\D^+}}\xspace}
\def\Dm      {{\ensuremath{\D^-}}\xspace}

\def\DpDm    {\ensuremath{\Dp {\kern -0.16em \Dm}}\xspace}
\def\Dstar   {{\ensuremath{\D^*}}\xspace}

\def\Dstarp  {{\ensuremath{\D^{*+}}}\xspace}
\def\Dstarm  {{\ensuremath{\D^{*-}}}\xspace}
\def\Dstarpm {{\ensuremath{\D^{*\pm}}}\xspace}

\def\theDstarp{{\ensuremath{\D^{*}(2010)^{+}}}\xspace}

\def\Dsp     {{\ensuremath{\D^+_\squark}}\xspace}

\def\B       {{\ensuremath{\PB}}\xspace}

\def\BorBbar {\kern \thebaroffset\optbar{\kern -\thebaroffset \PB}\xspace}
\def\Bz      {{\ensuremath{\B^0}}\xspace}

\def\Bd      {{\ensuremath{\B^0}}\xspace}

\def\BdorBdbar {\kern \thebaroffset\optbar{\kern -\thebaroffset \Bd}\xspace}

\def\Bs      {{\ensuremath{\B^0_\squark}}\xspace}

\def\BsorBsbar {\kern \thebaroffset\optbar{\kern -\thebaroffset \Bs}\xspace}

\def\Y#1S{\ensuremath{\PUpsilon{(#1S)}}\xspace}

\def\proton      {{\ensuremath{\Pp}}\xspace}

\def\LorLbar     {\kern \thebaroffset\optbar{\kern -\thebaroffset \PLambda}\xspace}

\newcommand{\decay}[2]{\ensuremath{#1\!\to #2}\xspace} 

\def\to                 {\ensuremath{\rightarrow}\xspace}

\newcommand{\tauDz}{{\ensuremath{\tau_{\Dz}}}\xspace}

\def\order   {{\ensuremath{\mathcal{O}}}\xspace}

\def\CP                {{\ensuremath{C\!P}}\xspace}
\def\CPT               {{\ensuremath{C\!PT}}\xspace}

\def\Vcs  {{\ensuremath{V_{\cquark\squark}^{\phantom{\ast}}}}\xspace}

\def\Vcb  {{\ensuremath{V_{\cquark\bquark}^{\phantom{\ast}}}}\xspace}

\def\Vuss  {{\ensuremath{V_{\uquark\squark}^\ast}}\xspace}

\def\Vubs  {{\ensuremath{V_{\uquark\bquark}^\ast}}\xspace}

\def\AT#1     {\ensuremath{A_{\mathrm{T}}^{#1}}\xspace}           

\def\C#1      {\ensuremath{\mathcal{C}_{#1}}\xspace}                       
\def\Cp#1     {\ensuremath{\mathcal{C}_{#1}^{'}}\xspace}                    
\def\Ceff#1   {\ensuremath{\mathcal{C}_{#1}^{\mathrm{(eff)}}}\xspace}        
\def\Cpeff#1  {\ensuremath{\mathcal{C}_{#1}^{'\mathrm{(eff)}}}\xspace}       
\def\Ope#1    {\ensuremath{\mathcal{O}_{#1}}\xspace}                       
\def\Opep#1   {\ensuremath{\mathcal{O}_{#1}^{'}}\xspace}                    

\newcommand{\bra}[1]{\ensuremath{\langle #1|}}             
\newcommand{\ket}[1]{\ensuremath{|#1\rangle}}              

\newcommand{\nospaceunit}[1]{\ensuremath{\text{#1}}}       
\newcommand{\aunit}[1]{\ensuremath{\text{\,#1}}}       

\newcommand{\tev}{\aunit{Te\kern -0.1em V}\xspace}
\newcommand{\gev}{\aunit{Ge\kern -0.1em V}\xspace}
\newcommand{\mev}{\aunit{Me\kern -0.1em V}\xspace}
\newcommand{\kev}{\aunit{ke\kern -0.1em V}\xspace}
\newcommand{\ev}{\aunit{e\kern -0.1em V}\xspace}
 
\newcommand{\mevc}{\ensuremath{\aunit{Me\kern -0.1em V\!/}c}\xspace}
\newcommand{\gevc}{\ensuremath{\aunit{Ge\kern -0.1em V\!/}c}\xspace}
\newcommand{\mevcc}{\ensuremath{\aunit{Me\kern -0.1em V\!/}c^2}\xspace}
\newcommand{\gevcc}{\ensuremath{\aunit{Ge\kern -0.1em V\!/}c^2}\xspace}

\def\cm   {\aunit{cm}\xspace}

\def\mm   {\aunit{mm}\xspace}

\def\mum  {\ensuremath{\,\upmu\nospaceunit{m}}\xspace}

\def\fb   {\ensuremath{\aunit{fb}}\xspace}
\def\invfb   {\ensuremath{\fb^{-1}}\xspace}

\def\order{{\ensuremath{\mathcal{O}}}\xspace}
\newcommand{\chisq}{\ensuremath{\chi^2}\xspace}
\newcommand{\chisqndf}{\ensuremath{\chi^2/\mathrm{ndf}}\xspace}
\newcommand{\chisqip}{\ensuremath{\chi^2_{\text{IP}}}\xspace}

\def\gsim{{~\raise.15em\hbox{$>$}\kern-.85em
          \lower.35em\hbox{$\sim$}~}\xspace}
\def\lsim{{~\raise.15em\hbox{$<$}\kern-.85em
          \lower.35em\hbox{$\sim$}~}\xspace}

\def\PDF {PDF\xspace}

\def\pt         {\ensuremath{p_{\mathrm{T}}}\xspace}

\def\ptot       {\ensuremath{p}\xspace}

\def\degrees{\ensuremath{^{\circ}}\xspace}

\def\mrad{\aunit{mrad}\xspace}
\def\rad{\aunit{rad}\xspace}

\def\evtgen     {\mbox{\textsc{EvtGen}}\xspace}

\def\geant      {\mbox{\textsc{Geant4}}\xspace}

\def\photos     {\mbox{\textsc{Photos}}\xspace}

\def\pythia     {\mbox{\textsc{Pythia}}\xspace}

\def\tell1  {TELL1\xspace}
\def\ukl1   {UKL1\xspace}

\newcommand{\vs}{\mbox{\itshape vs.}\xspace}

\usepackage{authblk}
\usepackage{enumitem}
\usepackage{placeins}

\usepackage{multirow}
\usepackage{booktabs}
\usepackage{dcolumn}
\usepackage{array}
\usepackage{rotating}

\usepackage[separate-uncertainty]{siunitx}
\usepackage{physics}
\usepackage{bm}

\def\agammaRSval  {\ensuremath{0.4}}
\def\agammaRSstat {\ensuremath{0.5}}

\def\agammaRSsys  {\ensuremath{0.2}}
\def\agammaKKval  {\ensuremath{2.3}}
\def\agammaKKstat {\ensuremath{1.5}}
\def\agammaKKsys  {\ensuremath{0.3}}
\def\agammaPPval  {\ensuremath{4.0}}
\def\agammaPPstat {\ensuremath{2.8}}
\def\agammaPPsys  {\ensuremath{0.4}}

\newcolumntype{C}{>{$}c<{$}}
\newcolumntype{L}{>{$}l<{$}}
\newcolumntype{R}{>{$}r<{$}}

\def\hp      {{\ensuremath{\hadron^+}}\xspace}
\def\hm      {{\ensuremath{\hadron^-}}\xspace}

\def\pisp    {{\ensuremath{\pion^{+}_{\mathrm{ \scriptscriptstyle tag}}}}\xspace}
\def\pism    {{\ensuremath{\pion^{-}_{\mathrm{ \scriptscriptstyle tag}}}}\xspace}

\def\KK {{\ensuremath{\Kp\Km  }}\xspace}
\def\PP {{\ensuremath{\pip\pim}}\xspace}
\def\RS {{\ensuremath{\Km\pip }}\xspace}
\def\HH {{\ensuremath{\hp\hm }}\xspace}

\def\DzKK {\decay{\Dz}{\Kp\Km}}
\def\DzPP {\decay{\Dz}{\pip\pim}}
\def\DzRS {\decay{\Dz}{\Km\pip}}

\def\DzbRS {\decay{\Dzb}{\Kp\pim}}

\def\theDstDzpip{\decay{\theDstarp}{\Dz\pip}}

\newcommand{\DY}[1]{{\ensuremath{\Delta Y_{#1}}}\xspace}
\newcommand{\agamma}[1]{{\ensuremath{A_{\Gamma}^{#1}}}\xspace}
\newcommand{\ycp}[1]{{\ensuremath{y_{\CP}^{#1}}}\xspace}

\def\phif   {{\ensuremath{\phi_{\lambda_{f}}}}\xspace}
\def\mhh    {{\ensuremath{m(\HH)}}\xspace}
\def\DstM   {{\ensuremath{m(\Dz\pisp)}}\xspace}

\def\af      {{\ensuremath{A_f}}\xspace}
\def\afb     {{\ensuremath{A_{\bar{f}}}}\xspace}
\def\abf     {{\ensuremath{\bar{A}_f}}\xspace}
\def\abfb    {{\ensuremath{\bar{A}_{\bar{f}}}}\xspace}
\newcommand{\Acpdec}[1]{{\ensuremath{a^{d}_{#1}}}\xspace}

\def\fdchisq {{\ensuremath{\chi^2_\text{FD}}}\xspace}

\newcommand{\cgev}{\ensuremath{\,c/\nospaceunit{Ge\kern -0.1em V}}\xspace}
\newcommand{\kevcc}{\ensuremath{\aunit{ke\kern -0.1em V\!/}c^2}\xspace}

\usepackage{cite} 
\usepackage{mciteplus}

\begin{document}

\renewcommand{\thefootnote}{\fnsymbol{footnote}}
\setcounter{footnote}{1}


\begin{titlepage}
\pagenumbering{roman}

\vspace*{-1.5cm}
\centerline{\large EUROPEAN ORGANIZATION FOR NUCLEAR RESEARCH (CERN)}
\vspace*{1.5cm}
\noindent
\begin{tabular*}{\linewidth}{lc@{\extracolsep{\fill}}r@{\extracolsep{0pt}}}
\ifthenelse{\boolean{pdflatex}}
{\vspace*{-1.5cm}\mbox{\!\!\!\includegraphics[width=.14\textwidth]{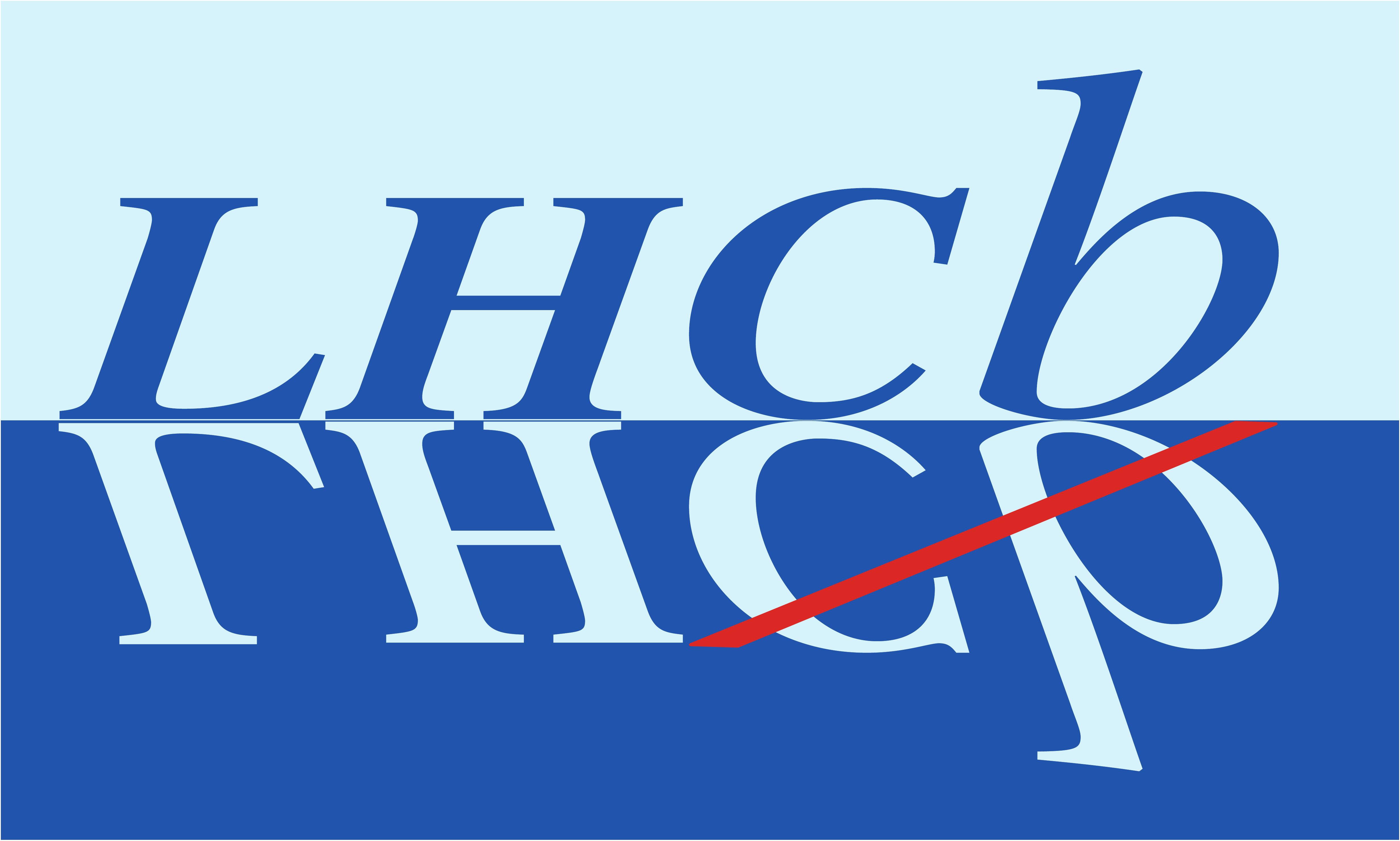}} & &}
{\vspace*{-1.2cm}\mbox{\!\!\!\includegraphics[width=.12\textwidth]{figs/lhcb-logo.eps}} & &}
\\
 & & CERN-EP-2021-060 \\
 & & LHCb-PAPER-2020-045 \\
 & & October 26, 2021 \\
 & & \\
\end{tabular*}

\vspace*{2.5cm}

{\normalfont\bfseries\boldmath\huge
\begin{center}
  \papertitle 
\end{center}
}

\vspace*{1.0cm}

\begin{center}
\paperauthors\footnote{Authors are listed at the end of this paper.}
\end{center}

\vspace{\fill}

\begin{abstract}
  \noindent
    A search for time-dependent violation of the charge-parity symmetry in \DzKK and \DzPP decays is performed at the \lhcb experiment using proton--proton collision data recorded from 2015 to 2018 at a centre-of-mass energy of $13\tev$, corresponding to an integrated luminosity of $6\invfb$.
    The \Dz meson is required to originate from a \theDstDzpip decay, such that its flavour at production is identified by the charge of the accompanying pion.
    The slope of the time-dependent asymmetry of the decay rates of \Dz and \Dzb mesons into the final states under consideration is measured to be
    \begin{align*}
        \DY{\KK} &= (-\agammaKKval \pm \agammaKKstat \pm \agammaKKsys )\times 10^{-4},\\
        \DY{\PP} &= (-\agammaPPval \pm \agammaPPstat \pm \agammaPPsys )\times 10^{-4},
    \end{align*}
    where the first uncertainties are statistical and the second are systematic.
    These results are compatible with the conservation of the charge-parity symmetry at the level of 2 standard deviations and improve the precision by nearly a factor of two.
\end{abstract}

\vspace*{2.0cm}

\begin{center}
  Published in \emph{Phys. Rev.} \textbf{D104} (2021) 072010
\end{center}

\vspace{\fill}

{\footnotesize 
\centerline{\copyright~\papercopyright. \href{\paperlicenceurl}{\paperlicence}.}}
\vspace*{2mm}

\end{titlepage}


\newpage
\setcounter{page}{2}
\mbox{~}

\renewcommand{\thefootnote}{\arabic{footnote}}
\setcounter{footnote}{0}

\cleardoublepage


\pagestyle{plain} 
\setcounter{page}{1}
\pagenumbering{arabic}


\section{Introduction}
\label{sect:introduction}

The breaking of the invariance of fundamental interactions under the combined charge conjugation ($C$) and parity ($P$) transformation, commonly named \CP violation, is a necessary condition to explain the much larger abundance of matter with respect to antimatter in the universe~\cite{Sakharov:1967dj}.
Within the standard model (SM) of particle physics, the weak interaction provides a source of \CP violation through a single complex phase in the Cabibbo--Kobayashi--Maskawa (CKM) matrix that governs the interaction of quarks with the \W boson~\cite{Cabibbo:1963yz,Kobayashi:1973fv}.
This CKM paradigm has been tested successfully in the decays of down-type quarks (\squark or \bquark) in \kaon and \B mesons.
However, the measured size of \CP violation is too small to explain the aforementioned matter--antimatter asymmetry~\cite{Dine:2003ax}, suggesting the existence of additional sources of \CP violation beyond the SM.

Hadrons containing charm quarks are the only ones where \CP violation and flavour-changing neutral currents (FCNC) involving up-type quarks (\uquark, \cquark or \tquark) can be studied, and provide a unique opportunity to detect new interactions beyond the SM that leave down-type quarks unaffected~\cite{Grossman:2006jg}.
Within the SM both \CP violation and FCNC for charm hadrons are predicted to be smaller than for kaons and beauty hadrons.
The Glashow--Iliopoulos--Maiani mechanism is more effective owing to the smaller mass of the beauty with respect to the top quark and to the smallness of the CKM matrix elements connecting the first two generations of quarks with the third.
Furthermore, the contributions from the strange and down quarks cancel in the $U$-spin limit, where $U$ spin is the $SU(2)$ subgroup of $SU(3)_\mathrm{F}$ relating the down and strange quarks.
In particular, the combination of CKM matrix elements responsible for \CP violation in charm decays in the SM is \mbox{$\Im(\Vcb\Vubs/\Vcs\Vuss)\approx -6\times 10^{-4}$}, corresponding to \CP asymmetries typically of the order of \mbox{$10^{-4}$} to \mbox{$10^{-3}$}~\cite{Grossman:2006jg}.

The \lhcb collaboration reported the first observation of \CP violation in the decay of \Dz mesons in 2019~\cite{LHCb-PAPER-2019-006}.
However, theoretical uncertainties on nonperturbative effects of the strong interaction do not allow a rigorous assessment of its compatibility with the SM~\cite{Grossman:2006jg,Li:2012cfa,Cheng:2012wr,Khodjamirian:2017zdu,Chala:2019fdb,Grossman:2019xcj}.
This has prompted a renewed interest of the theory community in the field~\cite{Buccella:2019kpn,Li:2019hho,Soni:2019xko,Cheng:2019ggx,Dery:2019ysp,Wang:2020gmn,Bause:2020obd,Dery:2021mll,Cheng:2021yrn}.
Complementary searches for time-dependent \CP violation in \Dz decays, which has not been observed so far, have the potential to clarify this picture~\cite{Kagan:2020vri}.

Cabibbo-suppressed \decay{\Dz}{f} decays, where the final state \mbox{$f = \Kp\Km$} or $\pip\pim$ is common to \Dz and \Dzb mesons, provide one of the most sensitive tests of time-dependent \CP violation through the measurement of the time-dependent asymmetry between the \Dz and \Dzb decay rates,
\begin{equation}
    \label{eq:acp}
    A_{\CP}(f,t) \equiv \frac{\Gamma(\decay{\Dz}{f},t) - \Gamma(\decay{\Dzb}{f},t)}
                             {\Gamma(\decay{\Dz}{f},t) + \Gamma(\decay{\Dzb}{f},t)},
\end{equation}
where \mbox{$\Gamma(\decay{\Dz}{f},t)$} indicates the decay rate of an initial \Dz meson decaying into the final state $f$ at time $t$.
The dependence of the asymmetry on decay time is due to the oscillation of \Dz into \Dzb mesons.
This process is parametrised through the mixing parameters $x_{12}$ and $y_{12}$, defined as \mbox{$x_{12} \equiv 2 \lvert M_{12} / \Gamma \rvert$} and \mbox{$y_{12} \equiv \lvert \Gamma_{12} / \Gamma \rvert$}~\cite{Grossman:2009mn}, where \mbox{$\bm{H}\equiv \bm{M} - \tfrac{i}{2}\bm{\Gamma}$} is the effective Hamiltonian governing the time evolution of the \Dz--\Dzb system and $\Gamma$ is the average decay width of the mass eigenstates.
Since both mixing parameters are smaller than 1\%~\cite{delAmoSanchez:2010xz,Lees:2012qh,Peng:2014oda,Staric:2015sta,LHCb-PAPER-2017-046,LHCb-PAPER-2018-038,LHCb-PAPER-2019-001,LHCb-PAPER-2021-009}, the asymmetry can be expanded to linear order in the mixing parameters as
\begin{equation}
    \label{eq:acp_expansion}
    A_{\CP}(f,t) \approx \Acpdec{f} + \DY{f} \frac{t}{\tauDz},
\end{equation}
where \Acpdec{f} is the \CP asymmetry in the decay, $\tauDz$ is the lifetime of the \Dz meson, and the \DY{f} parameter is approximately equal to~\cite{Kagan:2020vri}
\begin{equation}
    \label{eq:agamma_first_order}
    \DY{f} \approx - x_{12} \sin\phi^{M}_{f} + y_{12}\Acpdec{f}.
\end{equation}
Here, $\phi^{M}_{f}$ is defined as 
$\phi^{M}_{f} \equiv \arg \left(M_{12} \af /\abf\right)$, where \af (\abf) indicates the decay amplitude of a \Dz (\Dzb) meson into the final state $f$.
The parameter \DY{f} is approximately equal to the negative of the parameter \agamma{f} defined as the asymmetry of the effective decay widths of \Dz and \Dzb mesons into the final state $f$, as detailed in Appendix~\ref{app:delta_y}.

Within the SM, the value of \DY{f} is predicted to be of the order of $10^{-5}$ or less~\cite{Bigi:2011re,Bobrowski:2010xg,Kagan:2020vri,Li:2020xrz}, even though an enhancement up to the level of $10^{-4}$ by nonperturbative effects of the strong interaction is not excluded~\cite{Bobrowski:2010xg,Kagan:2020vri}.
At the current level of experimental precision, final-state dependent contributions to $\DY{f}$ can be safely neglected, as detailed in Appendix~\ref{app:delta_y}.
The measurements of \DY{\KK} and \DY{\PP} are thus expected to be consistent with each other and, under this assumption, they are collectively denoted as \DY{}.
Under the same approximation, the phase $\phi^{M}_{f}$ is equal to a dispersive mixing phase $\phi^{M}_{2}$ common to all \Dz decays, \mbox{$\DY{} \approx -x_{12}\sin\phi^{M}_{2}$}~\cite{Kagan:2020vri}.
The phase $\phi^{M}_{2}$ is defined as the phase of $M_{12}$ with respect to its $\Delta U = 2$ dominant contribution, hence the subscript ``2'', and coincides with the mixing phase $\phi_{12}$, defined as $\phi_{12} \equiv \arg(M_{12}/\Gamma_{12})$, in the superweak approximation~\cite{Wolfenstein:1964ks,Ciuchini:2007cw,Grossman:2009mn,Kagan:2009gb,Kagan:2020vri}.

Reducing the uncertainty on \DY{f} is also essential to determine the parameter \Acpdec{\KK} from the measurements of the time-integrated asymmetry of \DzKK decays~\cite{Aubert:2007if,Staric:2008rx,Aaltonen:2011se,LHCb-PAPER-2014-013,LHCb-PAPER-2016-035}, which is equal to
\begin{equation}
    A_\CP(\KK) \approx \Acpdec{\KK} + \DY{\KK}\frac{\langle t \rangle_{\KK}}{\tauDz},
\end{equation}
where \mbox{${\langle t \rangle}_{\KK}$} is the average measured decay time, which depends on the experimental environment.
In the most precise measurement to date, \mbox{$\langle t \rangle_{\KK}/\tauDz$} is equal to about 1.7~\cite{LHCb-PAPER-2016-035}.

The \DY{f} parameter has been measured by the \babar~\cite{Lees:2012qh}, \cdf~\cite{Aaltonen:2014efa}, \belle~\cite{Staric:2015sta} (which measures the parameter \agamma{f}) and \lhcb~\cite{LHCb-PAPER-2014-069,LHCb-PAPER-2016-063,LHCb-PAPER-2019-032} collaborations.
The world average, neglecting possible differences between the \KK and \PP final states, is \mbox{$\DY{} = (3.1 \pm 2.0)\times 10^{-4}$}~\cite{HFLAV18}.

This article presents a new measurement performed using proton--proton (\proton\proton) collision data collected by the \lhcb experiment at a centre-of-mass energy of $13\tev$ in 2015--2018, corresponding to an integrated luminosity of $6\invfb$.
Unlike in Ref.~\cite{LHCb-PAPER-2019-032}, the \Dz meson is required to originate from strong \decay{\theDstarp}{\Dz\pisp} decays, such that its initial flavour at production is identified by the charge of the tagging pion, \pisp.
The inclusion of charge-conjugate processes is implied throughout, except in the discussion of asymmetries.
Hereafter the \theDstarp meson is referred to as \Dstarp.

\section{Measurement overview}
\label{sect:analysis_strategy}

The measured raw asymmetry between the number of \Dz and \Dzb decays into the final state $f$ at time $t$,
\begin{equation}
    A_\text{raw}(f,t) \equiv \frac{N(\decay{\Dstarp}{\Dz(f,t)\pisp}) - N(\decay{\Dstarm}{\Dzb(f,t)\pism})}
                              {N(\decay{\Dstarp}{\Dz(f,t)\pisp}) + N(\decay{\Dstarm}{\Dzb(f,t)\pism})},
\end{equation}
is equal to
\begin{equation}
    \label{eq:A_raw}
    A_\text{raw}(f,t) \approx A_\CP(f,t) + A_\text{det}(\pisp) + A_\text{prod}(\Dstarp)
\end{equation}
up to corrections that are of third order in the asymmetries.
Here, \mbox{$A_\text{det}(\pisp)$} is the detection asymmetry due to different reconstruction efficiencies of positively and negatively charged tagging pions and \mbox{$A_\text{prod}(\Dstarp)$} is the production asymmetry of \Dstarpm mesons in \proton\proton collisions.
The measurement of \DY{f} from the slope of \mbox{$A_\text{raw}(f,t)$}, \textit{cf.} Eq.~(\ref{eq:acp_expansion}), is largely insensitive to time-independent asymmetries such as the detection and production asymmetries, which depend only on the kinematics of the particles.
However, the requirements used to select and reconstruct the decays introduce correlations between the kinematic variables and the measured decay time of the \Dz meson.
This causes an indirect time dependence of the production and detection asymmetries that needs to be accounted for.
These nuisance asymmetries are controlled with a precision better than $0.5 \times 10^{-4}$ by an equalisation of the kinematics of \Dstarp and \Dstarm candidates, as described in Sect.~\ref{sect:asymmetries}.
A further time dependence of \mbox{$A_\text{prod}(\Dstarp)$} arises if the \Dstarp meson is produced in the decay of a \B meson instead of in the \proton\proton collision.
The production asymmetry of these secondary \Dstarp mesons is different from that of \Dstarp mesons originating from the primary \proton\proton collision vertex (PV).
In addition, the measurement of the decay time of secondary \Dz mesons, which is performed with respect to the PV, is biased towards larger values.
The size of this background is assessed based on the distribution of the \Dz impact parameter with respect to its PV and its contribution to the asymmetry is subtracted as detailed in Sect.~\ref{sect:secondaries}.
Finally, \DY{f} is determined through a \chisq fit of a linear function to the time-dependent asymmetry, measured in 21 intervals of decay time in the range of 0.45 to 8\,\tauDz.

The analysis method is developed and validated using a sample of right-sign \mbox{\DzRS} decays --- thus named since the charges of the pions from the \Dstarp and \Dz decays have the same sign --- which consist mainly of Cabibbo-favoured decays of unmixed \Dz mesons.
This control sample has the same topology and kinematic distributions very similar to those of the signal channels, but its dynamical \CP asymmetry is known to be smaller than the current experimental uncertainty and can be neglected, as shown in Appendix~\ref{app:delta_y_rs}.
Therefore, the raw asymmetry between the number of \DzRS and \DzbRS decays is approximately equal to
\begin{equation}
    \label{eq:A_raw_RS}
    A_{\mathrm{raw}}(\RS,t) \approx A_\mathrm{det}(\pisp) + A_\mathrm{det}(\RS) + A_\mathrm{prod}(\Dstarp),
\end{equation}
where the right-hand side differs from that of Eq.~(\ref{eq:A_raw}) since it receives no contribution from dynamical \CP asymmetry, but contains an additional detection asymmetry from the \RS final state, $A_\mathrm{det}(\RS)$.
This asymmetry is removed by the kinematic equalisation described in Sect.~\ref{sect:asymmetries}, along with the other nuisance asymmetries.
The compatibility of the slope of the time-dependent asymmetry of \DzRS decays, \DY{\RS}, with zero is thus a useful cross-check of the analysis method.
Finally, the \DzRS sample is also used to estimate the size of the systematic uncertainties that are not expected to differ among the \Dz decay channels, allowing higher precision to be achieved than what would be possible by using the \DzKK and \DzPP samples.

To avoid experimenter's bias, the analysis method was developed without examining the values of \DY{\HH} of the signal channels, which were inspected only after the method had been finalised and the systematic uncertainties had been estimated.

\section{\lhcb detector \label{sect:detector}}

The \lhcb detector~\cite{LHCb-DP-2008-001,LHCb-DP-2014-002} is a single-arm forward spectrometer covering the pseudorapidity range $2<\eta <5$, designed for the study of particles containing \bquark or \cquark quarks.
The detector includes a high-precision tracking system consisting of a silicon-strip vertex detector surrounding the $pp$ interaction region, a large-area silicon-strip detector located upstream of a dipole magnet with a bending power of about $4{\mathrm{\,Tm}}$, and three stations of silicon-strip detectors and straw drift tubes placed downstream of the magnet.
The tracking system provides a measurement of the momentum, \ptot, of charged particles with a relative uncertainty that varies from 0.5\% at low momentum to 1.0\% at 200\gevc.
The minimum distance of a track to a PV, the impact parameter (IP), is measured with a resolution of $(15+29/\pt)\mum$, where \pt is the component of the momentum transverse to the beam, in\,\gevc.
The magnetic field deflects oppositely charged particles in opposite directions and this leads to detection asymmetries.
Therefore, its polarity is reversed around every two weeks throughout the data taking to reduce the effect.
Different types of charged hadrons are distinguished using information from two ring-imaging Cherenkov (\rich) detectors.
Photons, electrons and hadrons are identified by a calorimeter system consisting of scintillating-pad and preshower detectors, an electromagnetic and a hadronic calorimeter. Muons are identified by a system composed of alternating layers of iron and multiwire proportional chambers.

The online event selection is performed by a trigger, which consists of a hardware stage followed by a two-level software stage, which applies a full event reconstruction.
At the hardware-trigger stage, events are required to contain a muon with high \pt or a hadron, photon or electron with high transverse energy deposited in the calorimeters.
For hadrons, the transverse energy threshold is approximately $3.7\gev$.
In between the two software stages, an alignment and calibration of the detector is performed in near real-time~\cite{LHCb-PROC-2015-011} and updated constants are made available for the trigger, ensuring high-quality tracking and particle identification (PID) information.
The excellent performance of the online reconstruction
offers the opportunity to perform physics analyses directly using candidates reconstructed at the trigger level~\cite{LHCb-DP-2012-004,LHCb-DP-2016-001}, which the present analysis exploits.
The storage of only the triggered candidates enables a reduction in the event size by an order of magnitude.

Simulation is used to estimate the size of the background of secondary \Dstarp mesons from \B decays in Sect.~\ref{sect:secondaries}, and of three-body decays of charm mesons in Sect.~\ref{sect:systematics}.
In the simulation, $pp$ collisions are generated using \pythia~\cite{Sjostrand:2007gs,*Sjostrand:2006za} with a specific \lhcb configuration~\cite{LHCb-PROC-2010-056}.
Decays of unstable particles are described by \evtgen~\cite{Lange:2001uf}, in which final-state radiation is generated using \photos~\cite{davidson2015photos}.
The interaction of the generated particles with the detector, and its response, are implemented using the \geant toolkit~\cite{Allison:2006ve, *Agostinelli:2002hh} as described in Ref.~\cite{LHCb-PROC-2011-006}.

\section{Candidate selection}
\label{sect:selection}

\begin{figure}[bt]
  \begin{center}
    \includegraphics[width=0.4\linewidth]{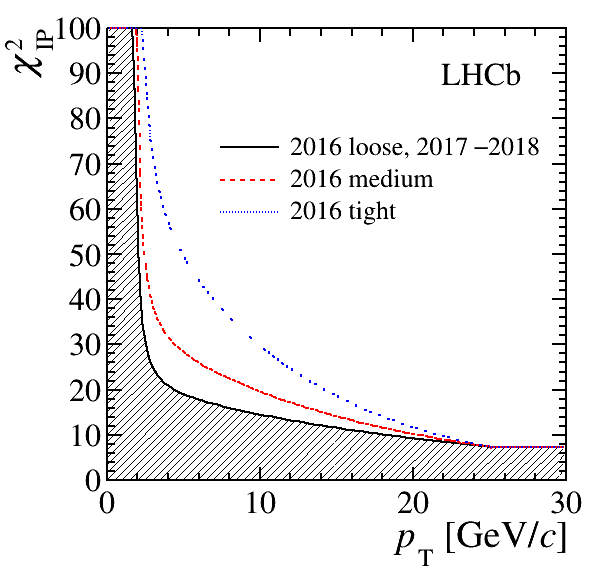}
  \end{center}
  \vspace*{-0.6cm}
  \caption{
    Requirement of the single-track selection of the first-stage software trigger in the \chisqip \textit{vs.} \pt plane, for the three configurations employed during data taking.
    The dashed region is excluded by the loosest configuration.
  }
  \label{fig:hlt1}
\end{figure}

The $\decay{\Dstarp}{\Dz\pisp}$ decay, where the \Dz meson subsequently decays into one of the following $\HH$ combinations, \RS, \KK, or \PP, is reconstructed at the trigger level.
No requirements on the type of hardware-trigger decision are applied, while at least one or both of the tracks from the \Dz decay are required to satisfy the single- or two-track selections of the first-stage software trigger.
The former requires the presence of at least one track with high \pt and large \chisqip with respect to all PVs, where the \chisqip\ is defined as the difference in the vertex-fit \chisq of a given PV reconstructed with and without the particle being considered.
The single-track requirement in the \chisqip \textit{vs.} \pt plane changed during the data taking, as illustrated in Fig.~\ref{fig:hlt1}.
In particular, the selection was tighter during 2016.
On the other hand, the two-track selection requires the presence of two high-\pt tracks forming a good-quality vertex that is significantly displaced from its associated PV, defined as the PV to which the IP of the two-track combination is the smallest.
In this case the selection is based on a bonsai boosted decision tree~\cite{BBDT} that takes as inputs the \chisq of the two-track vertex fit, the number of tracks with $\chisqip > 16$, the sum of the \pt of the two tracks and the significance of their flight-distance with respect to the associated PV, \fdchisq.
This is defined as the difference in the vertex-fit \chisq of the PV reconstructed including the two tracks, and the sum of the vertex-fit \chisq of the PV reconstructed without including them and of the two-track vertex-fit \chisq.
Also for the two-track selection, the requirements employed in 2016, in particular during the data taking with the magnetic field pointing upwards, were tighter with respect to the other years.

The second-stage software trigger combines pairs of oppositely charged tracks with distance of closest approach less than $0.1\mm$ to form \Dz candidates.
Both tracks are required to be of high quality based on the $\chisq$ per degree of freedom of their track fit  ($\chisqndf < 3$) and on the output of a multivariate classifier trained to identify fake tracks, by combining information from all of the tracking systems.
Furthermore, both tracks are required to have \mbox{$\ptot > 5 \gevc$} and to have a \chisqip\ with respect to all PVs in the event greater than 4.
The tracks are given a pion- or kaon-mass assignment, based on the information from the \rich detectors.
The \Dz decay vertex is required to be significantly displaced from the PV, and the angle between the \Dz momentum and the vector connecting the PV and the \Dz decay vertex is required to be less than $1\degrees$.
Finally, all remaining good-quality tracks of the event, as described above, which satisfy \mbox{$\ptot > 1\gevc$} and \mbox{$\pt > 200\mevc$}, are assigned a pion-mass hypothesis and are combined with the \Dz candidate to form a \Dstarp candidate, the vertex fit of which is required to be of good quality.

In the offline selection, the pseudorapidity of all \hp, \hm and \pisp tracks is required to lie in the range 2 to 4.2 to exclude candidates that traversed detector material corresponding to more than 0.3 interaction lengths between the \proton\proton interaction point and the end of the tracking system, as these candidates exhibit larger detection asymmetries~\cite{pajero:2021}.
The \Dz flight distance in the plane transverse to the beam is required to be less than $4\mm$ to remove \Dstarp candidates produced by hadronic interactions with the detector material, and the $z$ coordinate of the \Dz decay vertex is required to lie within $20\cm$ from the \proton\proton interaction point.\footnote{
    The \lhcb coordinate system is a right-handed system centred in the nominal \proton\proton collision point, with the $z$ axis pointing along the beam direction towards downstream of the detectors, the $y$ axis pointing vertically upwards, and the $x$ axis pointing in the horizontal direction.
}
The \HH invariant mass, \mhh, is required to lie in the range \mbox{$[1847.8,1882.6]$}, \mbox{$[1850.6,1879.9]$} and \mbox{$[1846.2,1884.2]\mevcc$} for the \DzRS, \DzKK and \DzPP candidates, respectively, corresponding to $\pm 2$ times the mass resolution around the known \Dz mass~\cite{PDG2020}.
Finally, to suppress the background from \decay{\Dz}{\Km\ep\neue} decays, the kaons from the \DzKK decay are required not to be identified as electrons or positrons, based on the output of a multivariate classifier combining information from all of the detectors.
This requirement is applied to both particles to avoid introducing different efficiencies for \Dz and \Dzb decays owing to possible PID asymmetries.

In order to improve the resolution on the \Dz decay time, a kinematic fit is performed in which the \Dstarp candidate is required to originate from the associated PV~\cite{Hulsbergen:2005pu}.
The resulting average decay-time resolution is $0.11\,\tauDz$.
At the same time, the resolution on the invariant mass of the \Dstarp candidates, \DstM, is improved by a factor of two.
However, the decay time of \Dz mesons coming from secondary \Dstarp mesons produced in the decay of \B mesons is overestimated.
The IP of these background \Dz mesons is, in general, greater than zero, contrary to signal candidates, whose IP is equal to zero within the experimental resolution.
The background from \B-meson decays is suppressed to the 4\% level by requiring that the \Dz IP is less than $60\mum$ and that its decay time is less than $8\,\tauDz$.
Finally, the \Dz decay time is required to be greater than $0.45\,\tauDz$ to exclude candidates with low reconstruction efficiency.

\begin{figure}[tb]
  \begin{center}
    \includegraphics[width=0.45\linewidth]{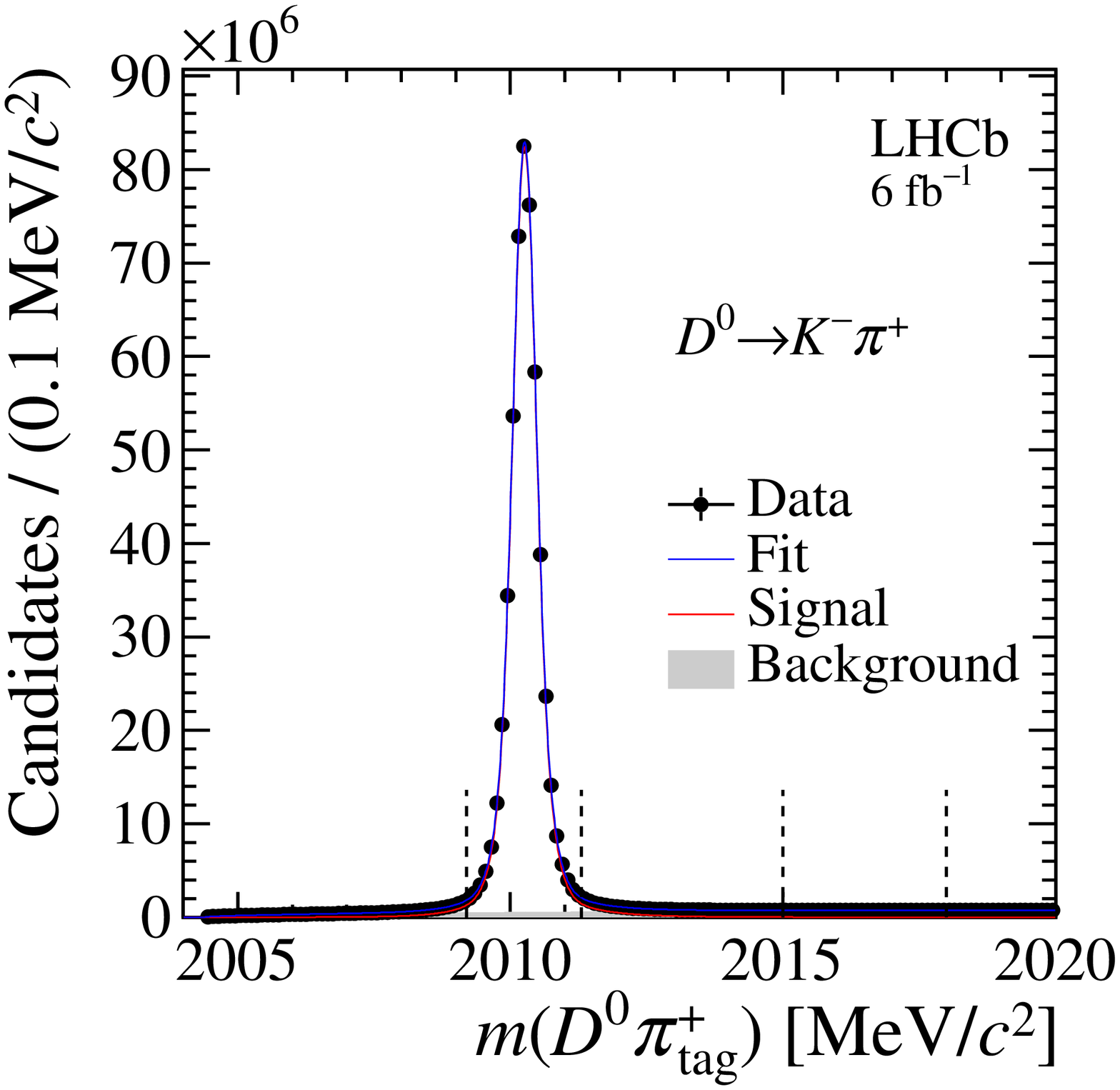}\\
    \includegraphics[width=0.45\linewidth]{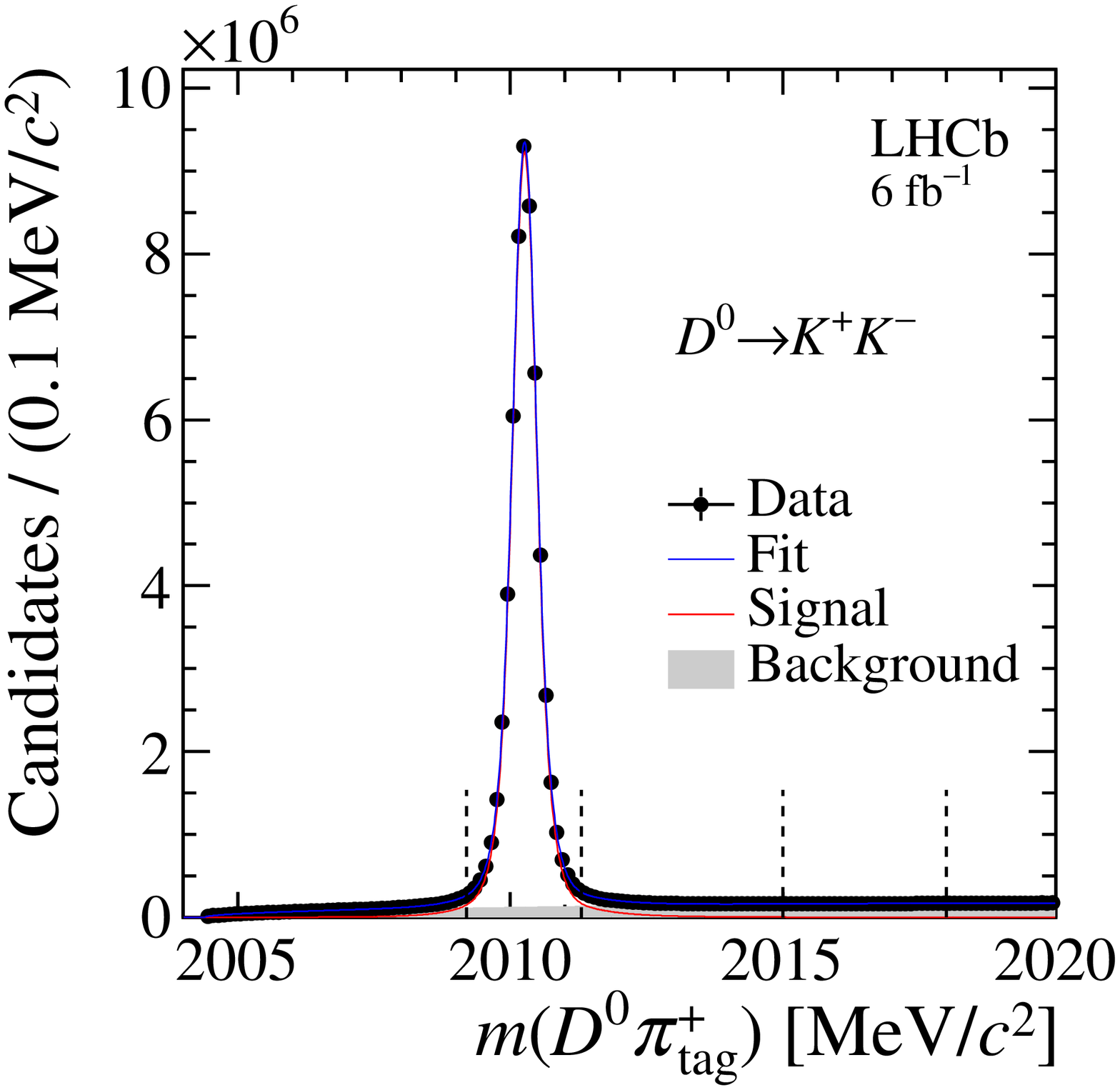}
    \includegraphics[width=0.45\linewidth]{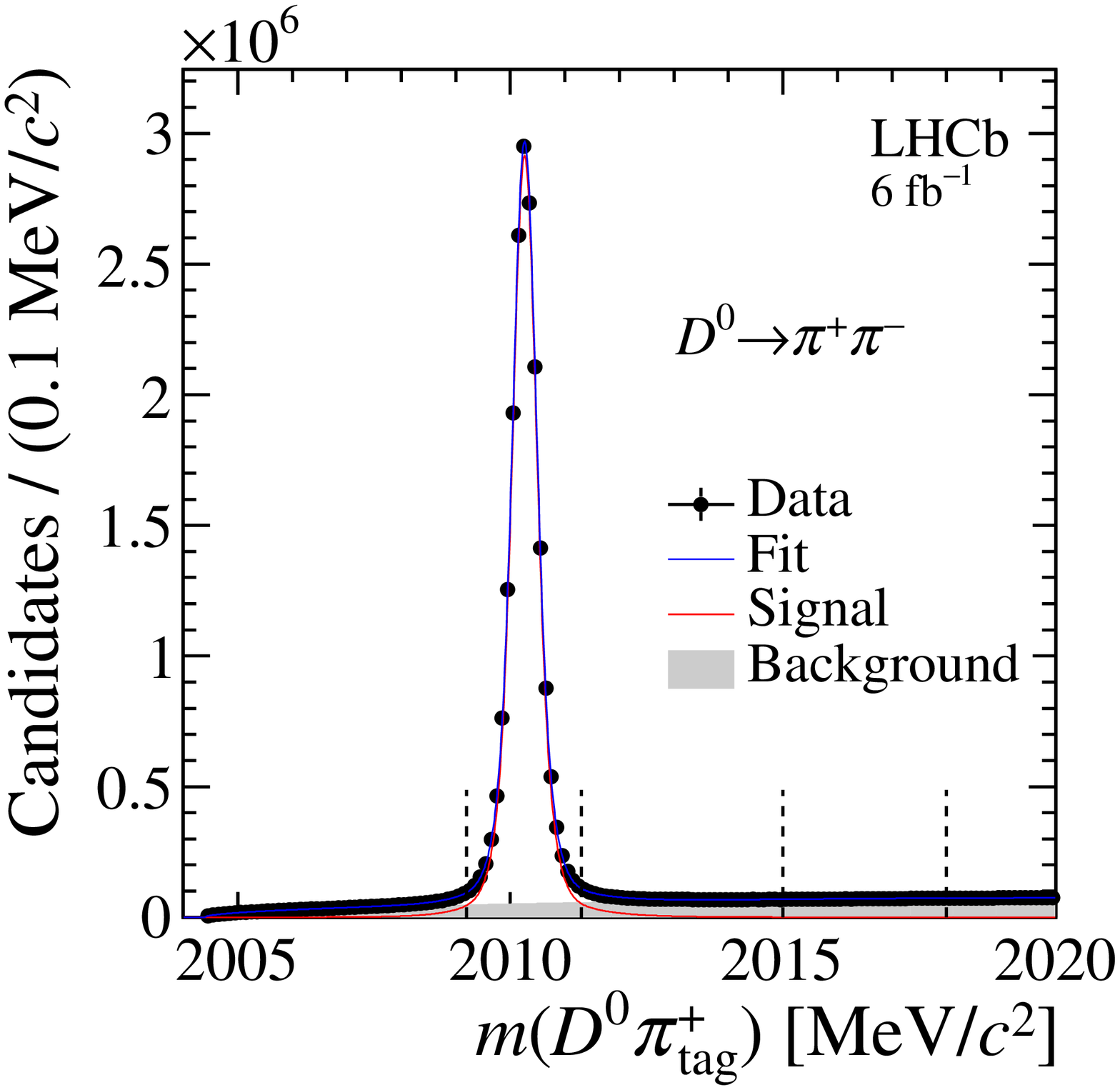}
  \end{center}
  \vspace*{-0.6cm}
  \caption{
    Distribution of \DstM for (top) \DzRS, (left) \DzKK and (right) \DzPP candidates. 
    The signal window and the lateral window employed to remove the combinatorial background (grey filled area) are delimited by the vertical dashed lines.
    Fit projections are overlaid.
  }
  \label{fig:dst_m}
\end{figure}

After these requirements, around 2.5\%, 4.7\% and 4.9\% of the \DzRS, \mbox{\DzKK} and \DzPP candidates are combined with more than one \pisp candidate to form a \Dstarp candidate.
In this case, one \Dstarp candidate per event is selected at random.
The distributions of \DstM of selected candidates are displayed for the three decay channels in Fig.~\ref{fig:dst_m}.
This quantity is calculated using the known \Dz mass in the determination of the \Dz energy.
This choice minimises the impact of the resolution of the invariant mass of the \Dz candidate.
The \DstM signal window is defined as \mbox{$[2009.2, 2011.3]\mevcc$} and retains about 96.9\% of the signal.
The purity within this window is 97.7\%, 95.5\% and 94.1\% for the \DzRS, \DzKK and \DzPP samples, respectively.
The residual background is dominated by real \Dz mesons associated with uncorrelated particles and is subtracted by using background candidates in the lateral mass window \mbox{$[2015, 2018]\mevcc$}, weighted with a suitable negative coefficient.
The coefficient is determined based on a binned maximum-likelihood fit to the \DstM distribution, which relies on an empirical model.
In particular, the signal probability density function (\PDF) is described by the sum of two Gaussian functions and a Johnson $S_U$ distribution~\cite{Johnson:1949zj},
\begin{equation}
    S_{U}(x;\mu,\sigma,\delta,\gamma) \propto
        \bigg[ 1 + \left(\frac{x - \mu}{\sigma}\right)^{2}\bigg]^{-\frac{1}{2}}
        \times \exp \left\{-\frac{1}{2}\left[ \gamma + \delta \sinh^{-1}\left(\frac{x - \mu}{\sigma}\right)\right]\right\},
\end{equation}
where the $\mu$ and $\sigma$ parameters are approximately equal to the mean and standard deviation of the Gaussian-like core, and the $\delta$ and $\gamma$ parameters describe its asymmetric tails.
The background \PDF, instead, is modelled by the function
\begin{equation}
    \sqrt{\DstM - m_0}\times\{1 + \alpha [\DstM - m_0] + \beta [\DstM - m_0]^2\},
\end{equation}
where $m_0$ is defined as the sum of the \Dz and \pip masses and the small parameters $\alpha$ and $\beta$ quantify the deviations from a square-root function.
The background subtraction is performed without distinguishing between \Dstarp and \Dstarm candidates, but separately in each decay-time interval.
The 21 intervals of decay time, which span the range $[0.45,8]\,\tauDz$, are chosen to be equally populated, except the last two intervals, which contain half the number of candidates with respect to the others.

\begin{figure}[tb]
  \begin{center}
    \includegraphics[width=0.45\linewidth]{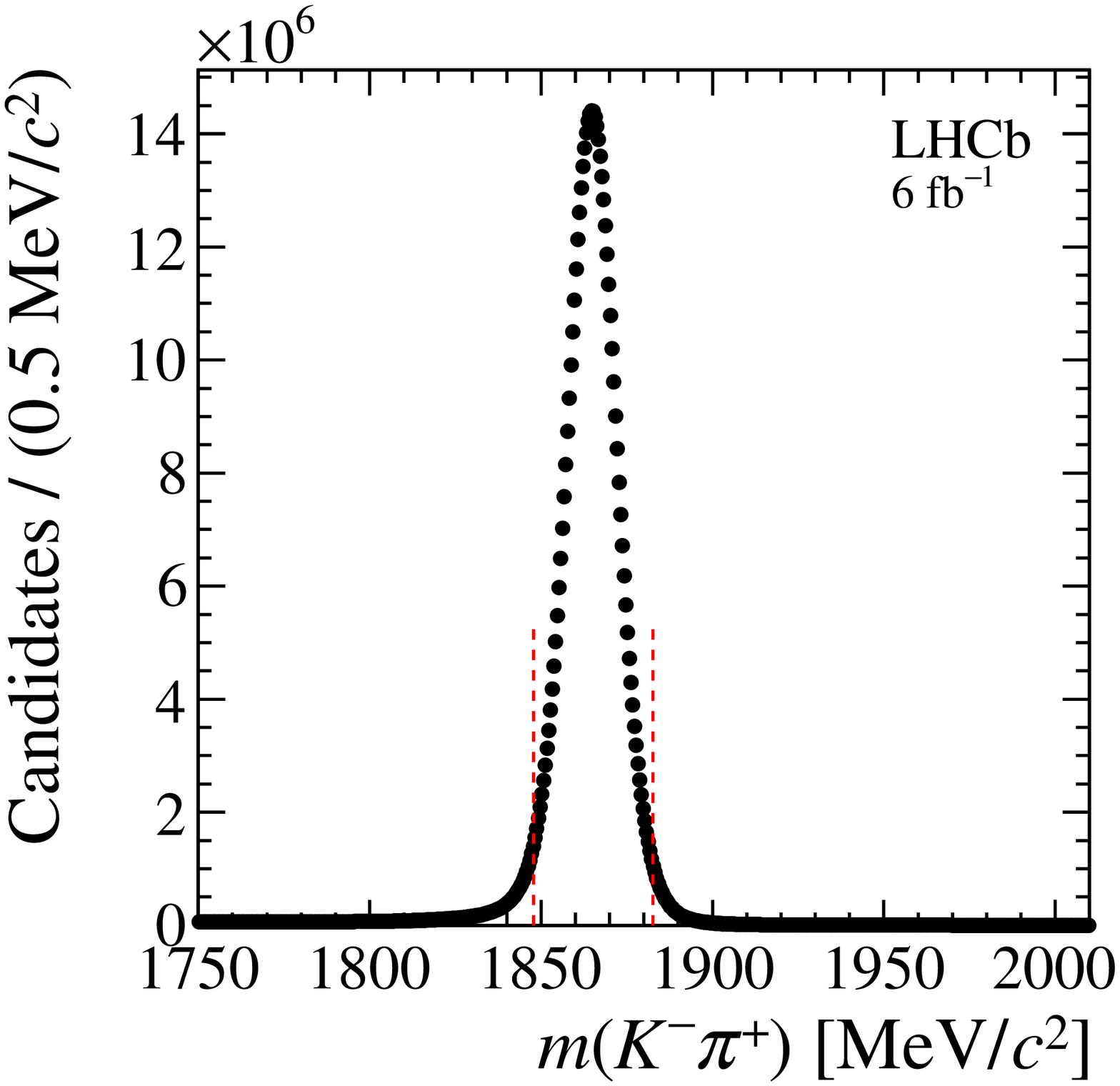}\\
    \includegraphics[width=0.45\linewidth]{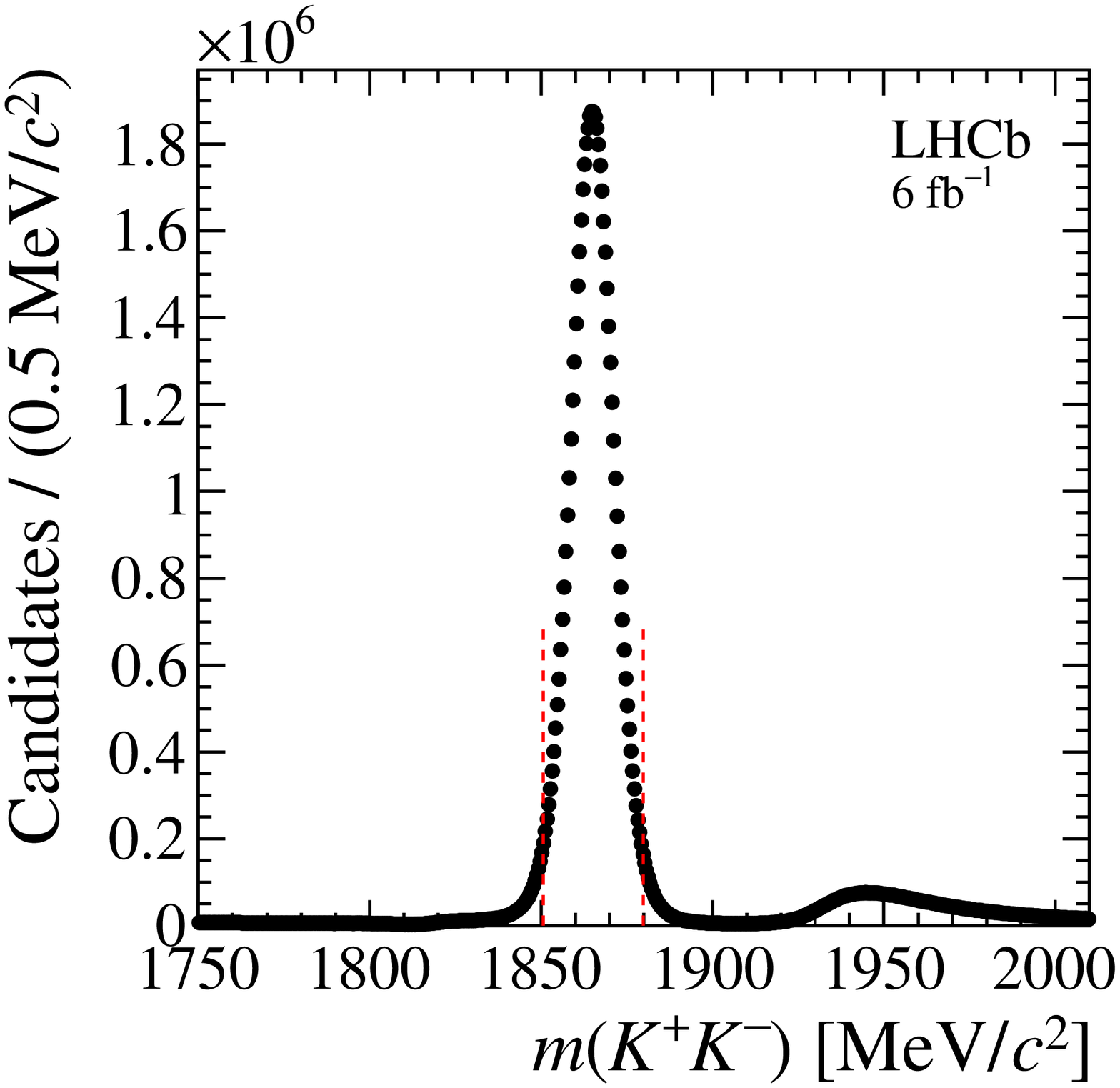}
    \includegraphics[width=0.45\linewidth]{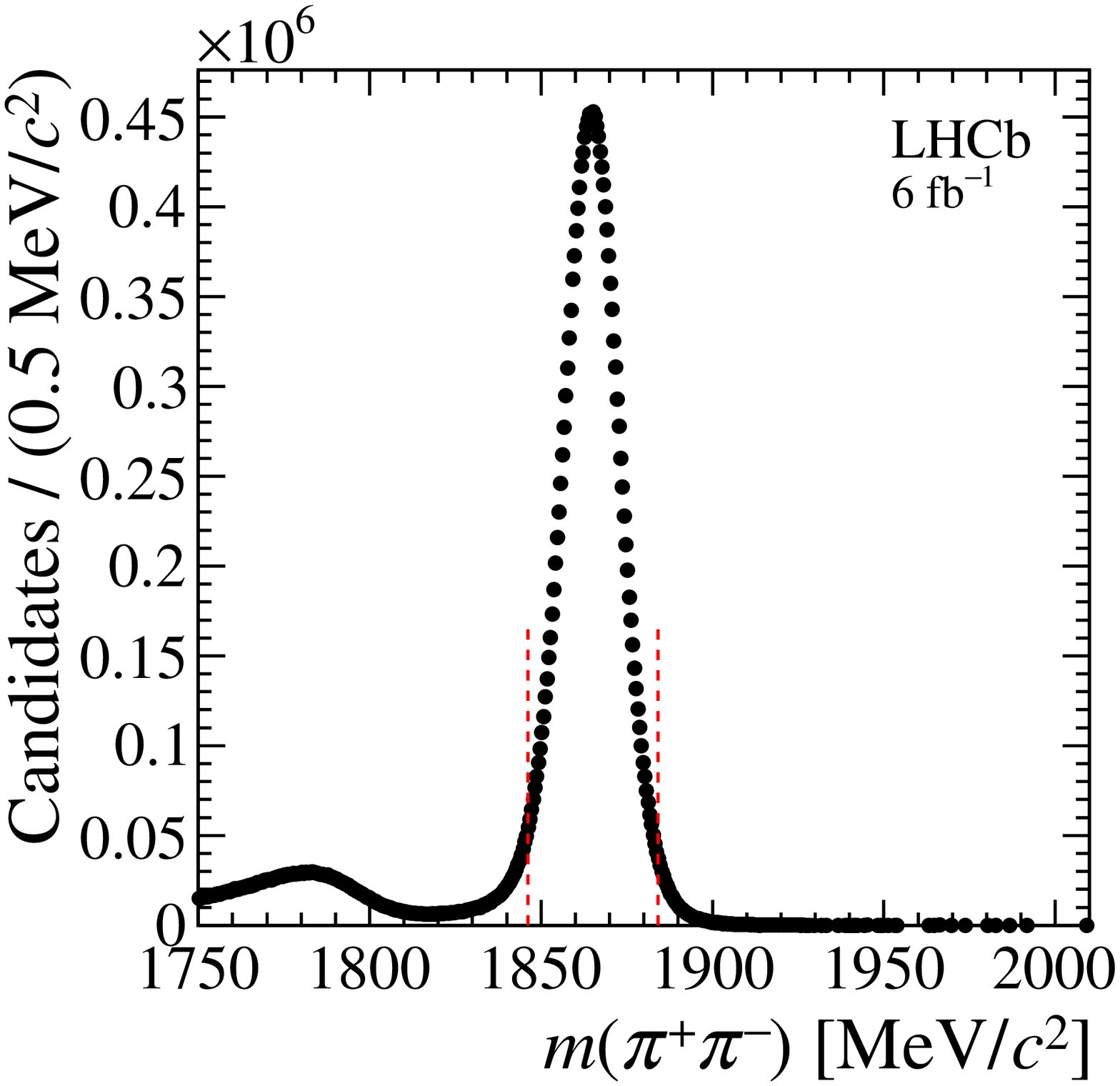}
  \end{center}
  \vspace*{-0.6cm}
  \caption{
    Distribution of \mhh for (top) \DzRS, (left) \DzKK and (right) \DzPP background-subtracted candidates. 
    The signal window is delimited by the vertical dashed lines.
  }
  \label{fig:mhh}
\end{figure}

The \mhh distributions after the removal of the \DstM background are displayed in Fig.~\ref{fig:mhh}.
The number of candidates in the signal region is 519, 58 and 18 million for the \RS, \KK and \PP decay channels, respectively.
The number of candidates per integrated luminosity is by a factor of 3.4 larger than that of the measurement with the data collected in 2011--2012~\cite{LHCb-PAPER-2016-063}, owing to the increased charm-quark production cross-section at the higher centre-of-mass energy~\cite{LHCb-PAPER-2012-041,LHCb-PAPER-2015-041}, to the increased trigger rate allowed by the real-time reconstruction of the events~\cite{LHCb-DP-2012-004,LHCb-DP-2016-001}, and to the implementation of the two-track selection in the first stage of the software trigger, which increases the selection efficiency at low decay times.

\section{Momentum-dependent asymmetries}
\label{sect:asymmetries}

\begin{figure}[tb]
  \begin{center}
    \includegraphics[width=0.47\linewidth]{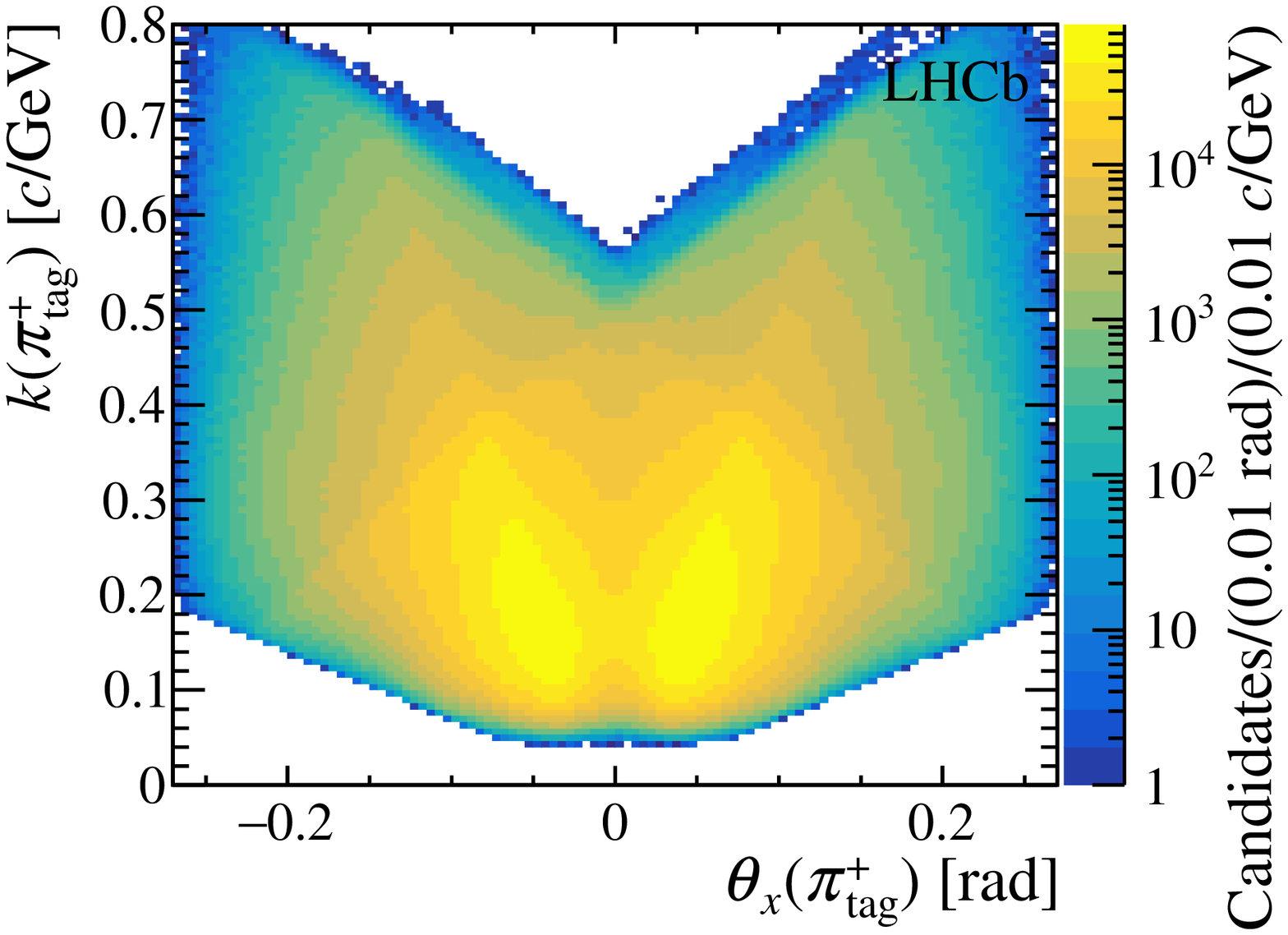}
    \includegraphics[width=0.47\linewidth]{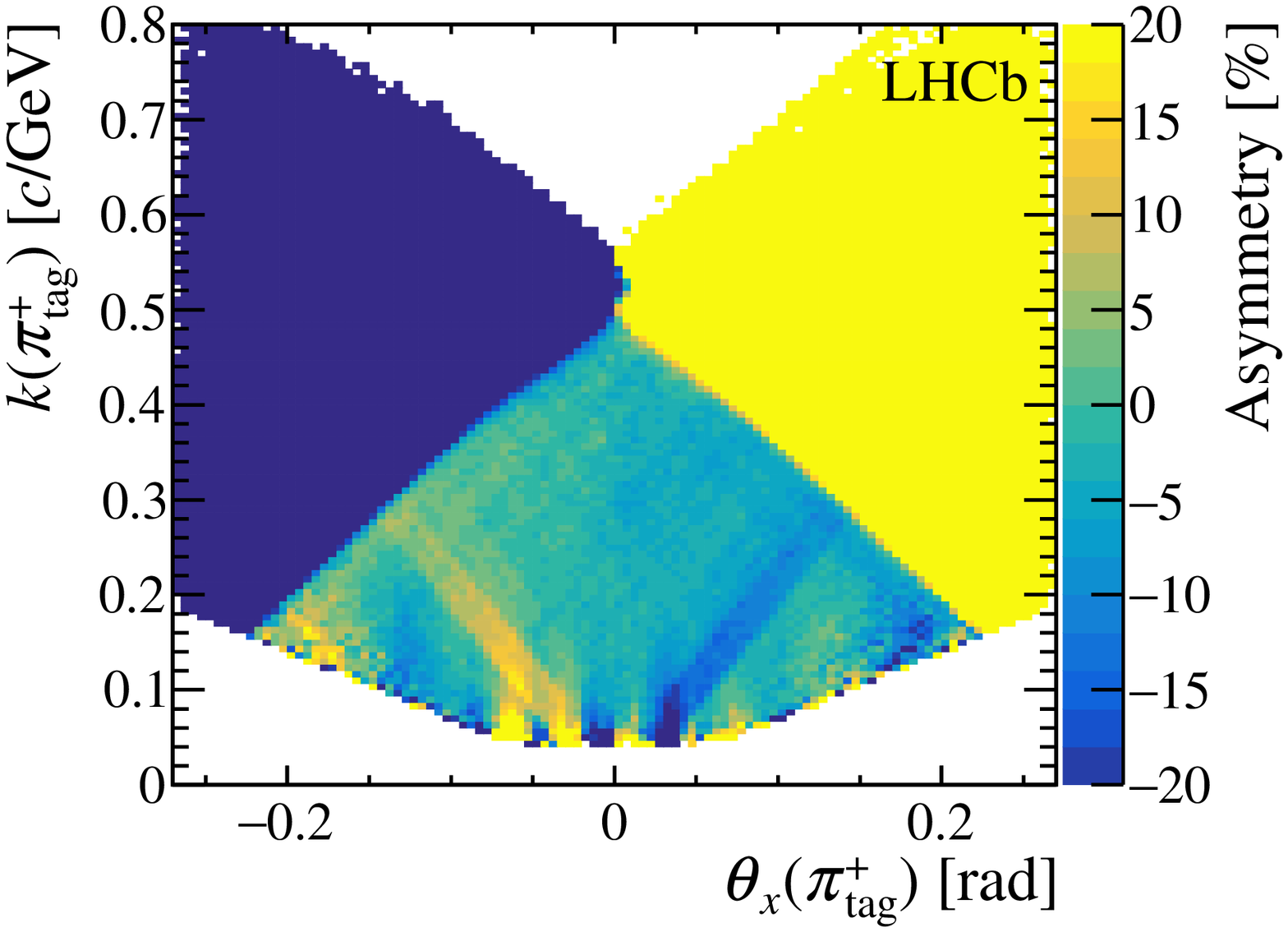}
    \includegraphics[width=0.47\linewidth]{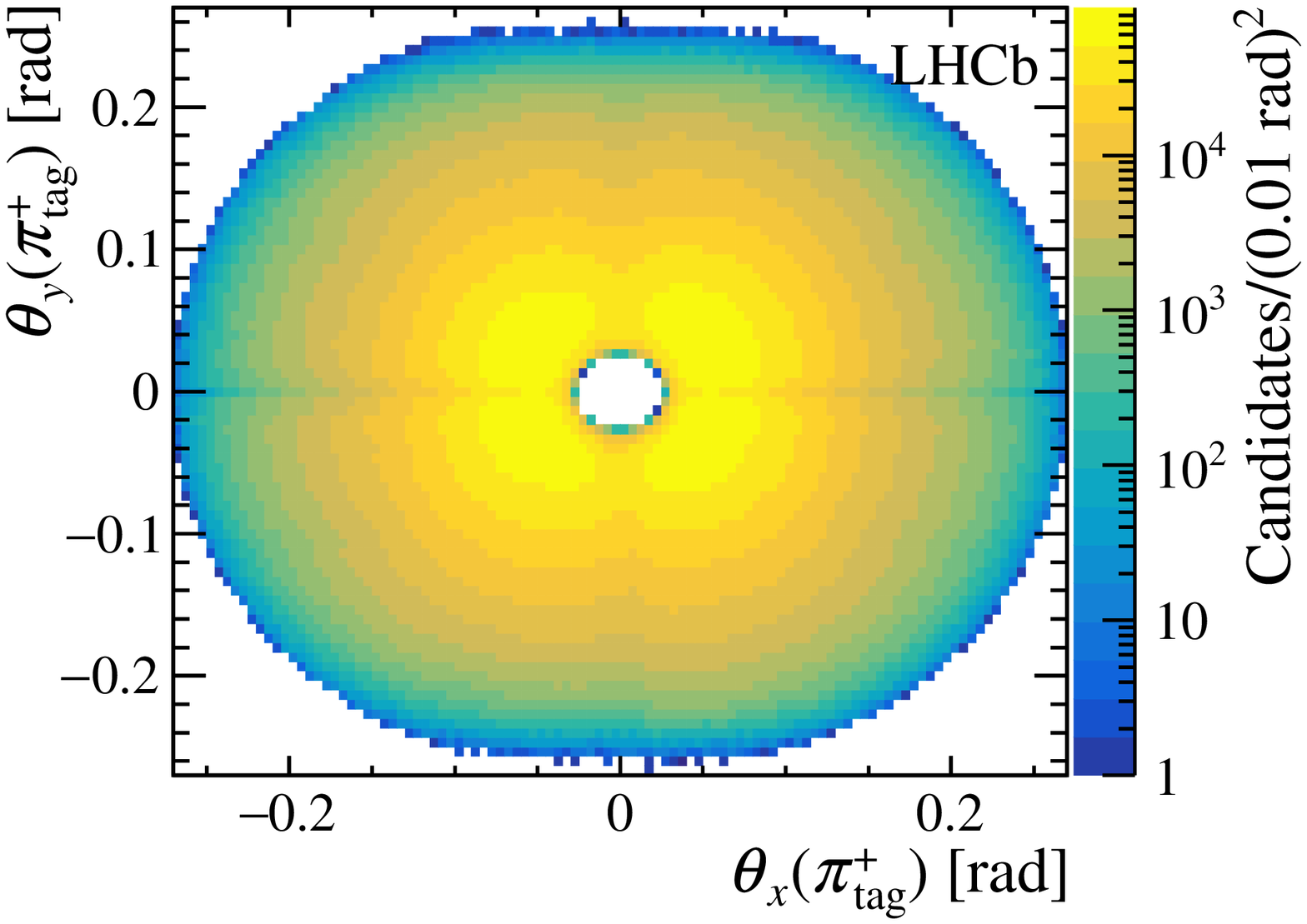}
    \includegraphics[width=0.47\linewidth]{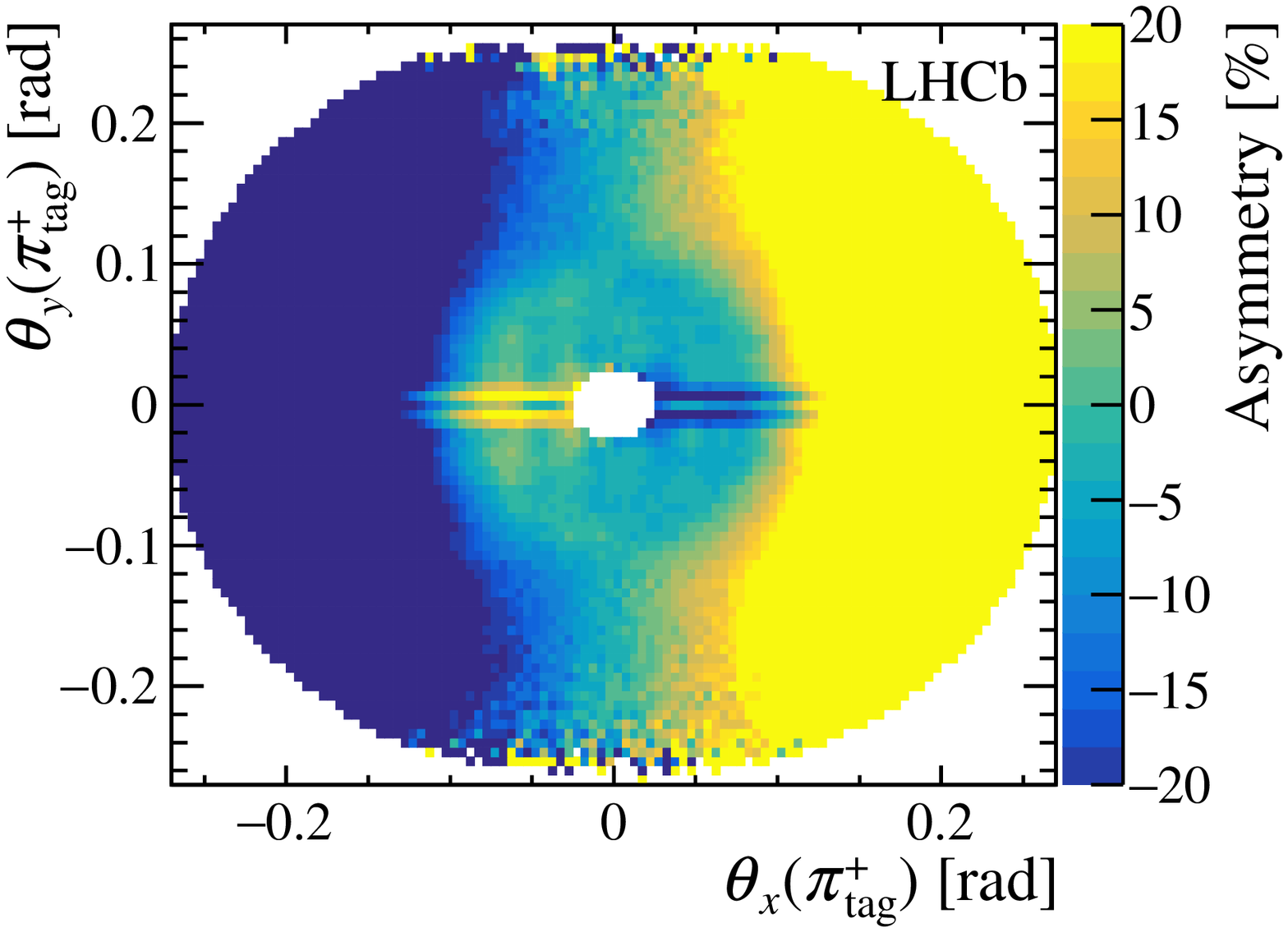}
  \end{center}
  \vspace*{-0.6cm}
  \caption{
    (Left) sum and (right) asymmetry of the distributions of the momentum of \pisp and \pism candidates, projected on the (top) $k$ \vs $\theta_x$ and (bottom) $\theta_y$ \vs $\theta_x$ planes, for the \DzRS candidates collected in 2017 with the \MagUp polarity.
    The angles $\theta_{x(y)}$ and the curvature $k$ are defined in the text.
    The asymmetries during the other data-taking years are similar, and opposite in sign for the data collected with the \MagDown polarity.
    The regions of the distributions with asymmetries whose magnitude is larger than 20\% are plotted with the same colour as $\pm 20\%$.
    These regions are discarded from the data sample after the kinematic weighting.
  }
  \label{fig:kinematics}
\end{figure}

The data sample is affected by momentum-dependent nuisance asymmetries.
The largest of these arise from the \pisp meson and are caused by the vertical magnetic field, which bends oppositely charged particles in opposite directions.
For a given magnet polarity, low-momentum particles of one charge at large or small emission angles in the horizontal plane may be deflected out of the detector or into the uninstrumented \lhc beam pipe, whereas particles with the opposite charge are more likely to remain within the acceptance.
This is shown in Fig.~\ref{fig:kinematics}, where the momentum of the \pisp meson is parametrised through its emission angles in the bending and vertical planes, \mbox{$\theta_{x(y)} \equiv {\arctan}(p_{x(y)} / p_z)$}, and its curvature in the magnetic field, \mbox{$k \equiv 1/\sqrt{p_x^2 + p_z^2}$}.
These asymmetries cancel to a large extent in the average between the samples collected with the magnet polarities pointing upwards (\MagUp) and downwards (\MagDown).
Smaller residual asymmetries remaining after the averaging are due to right--left misalignment of detector elements and to the nonzero $x$ coordinate of the collision point, to different beam--beam crossing angles for the \MagUp and \MagDown polarities, and to variations of the detection efficiency over time.
Other momentum-dependent asymmetries, which are independent of the magnet polarity, are the \Dstarp production asymmetry and the track-efficiency asymmetry.
The latter is caused by the higher occupancy of the detector part downstream of the magnet towards which the negatively charged particles are bent, owing to the large number of electrons produced in the particles interaction with the detector material.
Finally, for the \DzRS decay channel, the asymmetry caused by the different interaction cross-section of positively and negatively charged kaons and pions with matter (with the latter being much smaller) is independent of the magnet polarity.

\begin{figure}[tb]
  \begin{center}
    \includegraphics[width=0.35\linewidth]{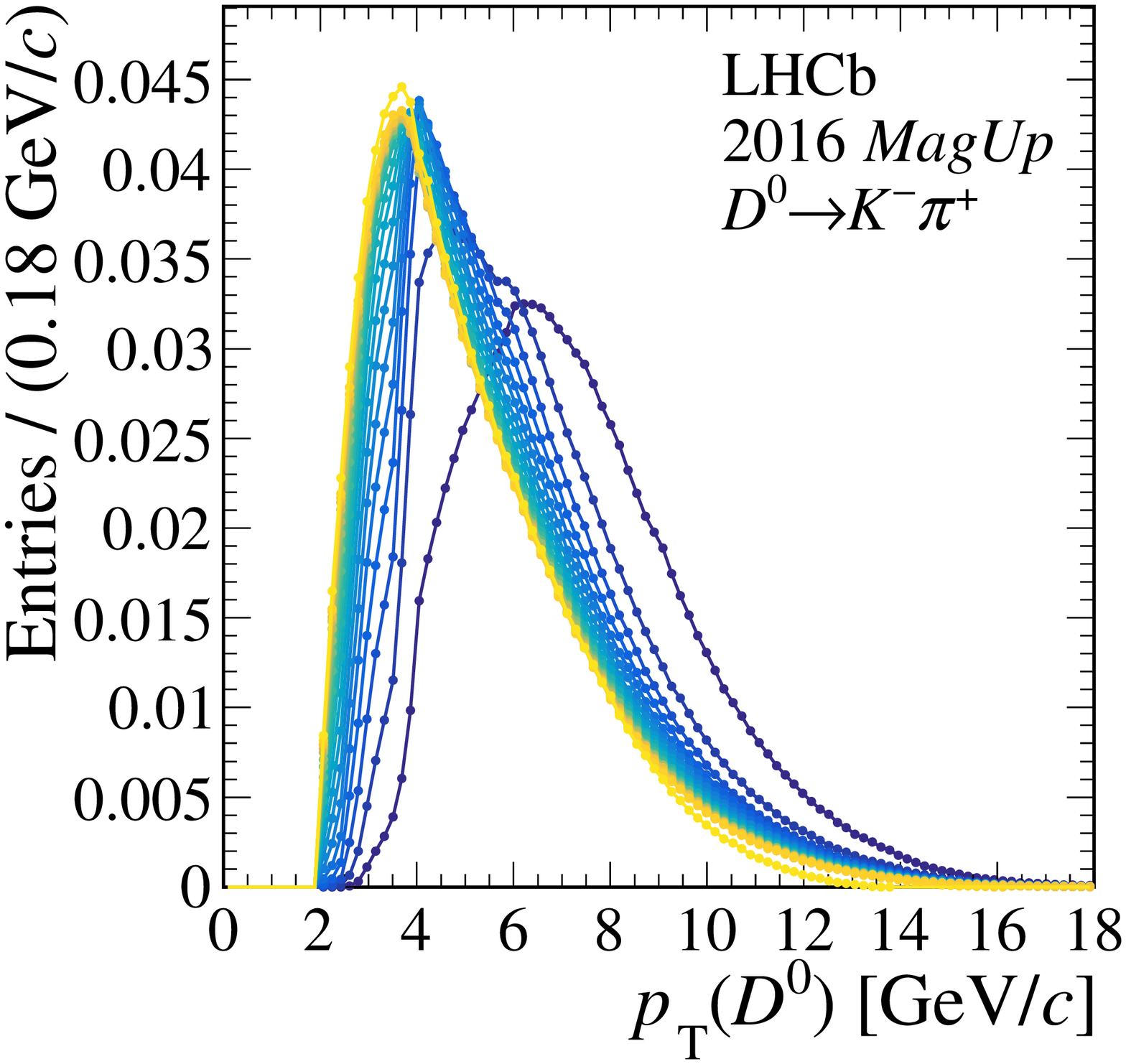}
    \includegraphics[width=0.35\linewidth]{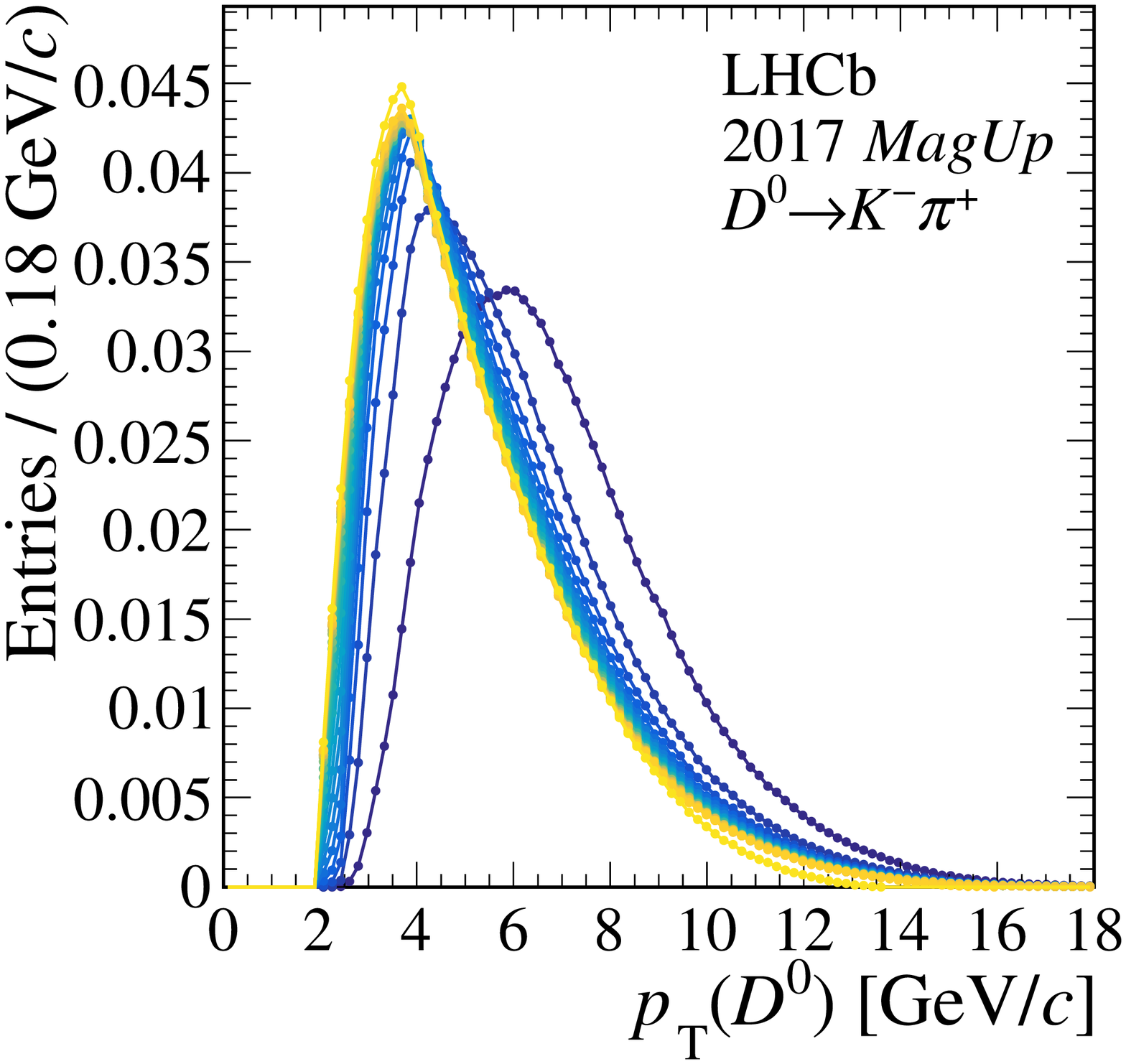}\\
    \includegraphics[width=0.35\linewidth]{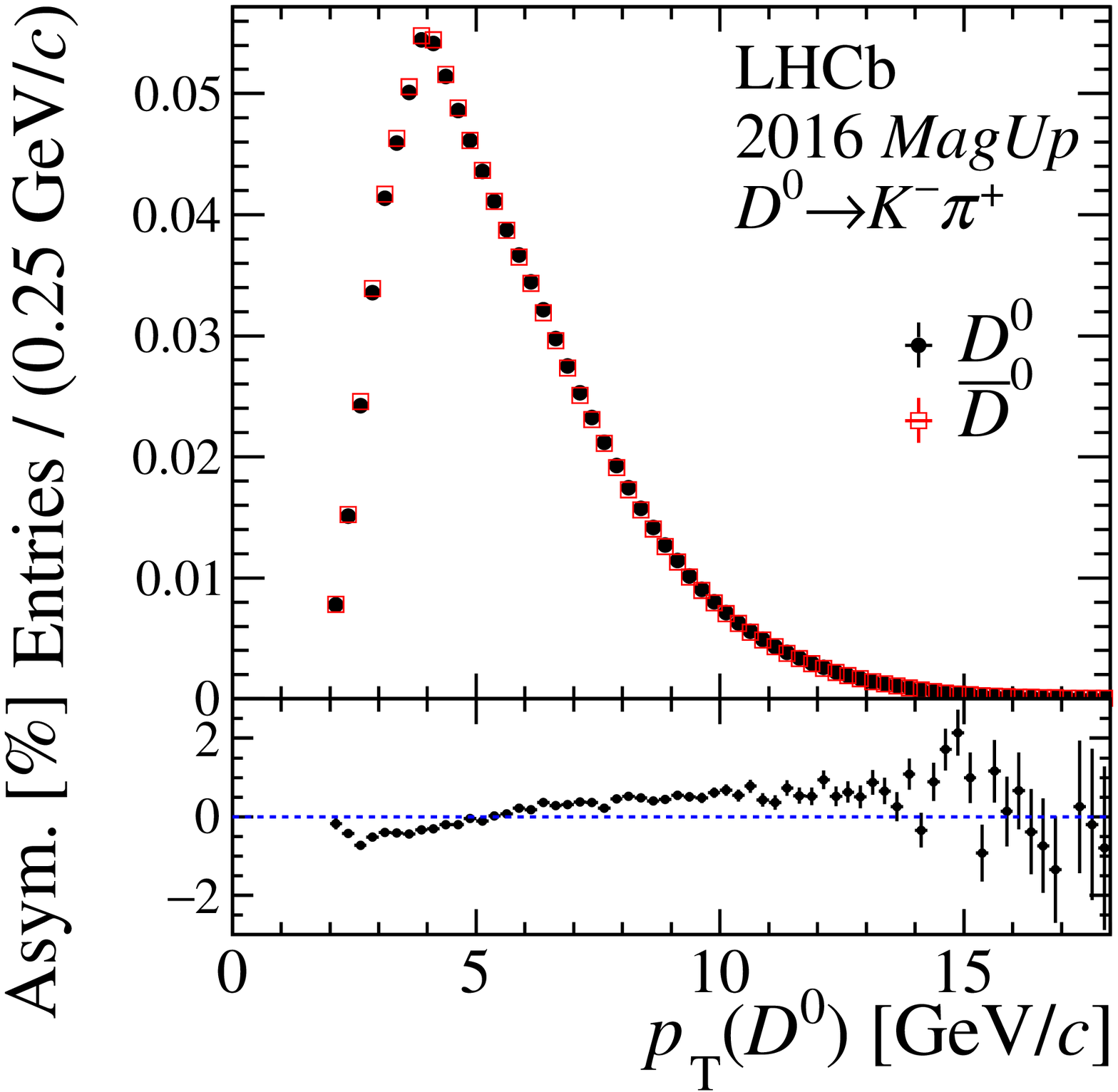}
    \includegraphics[width=0.35\linewidth]{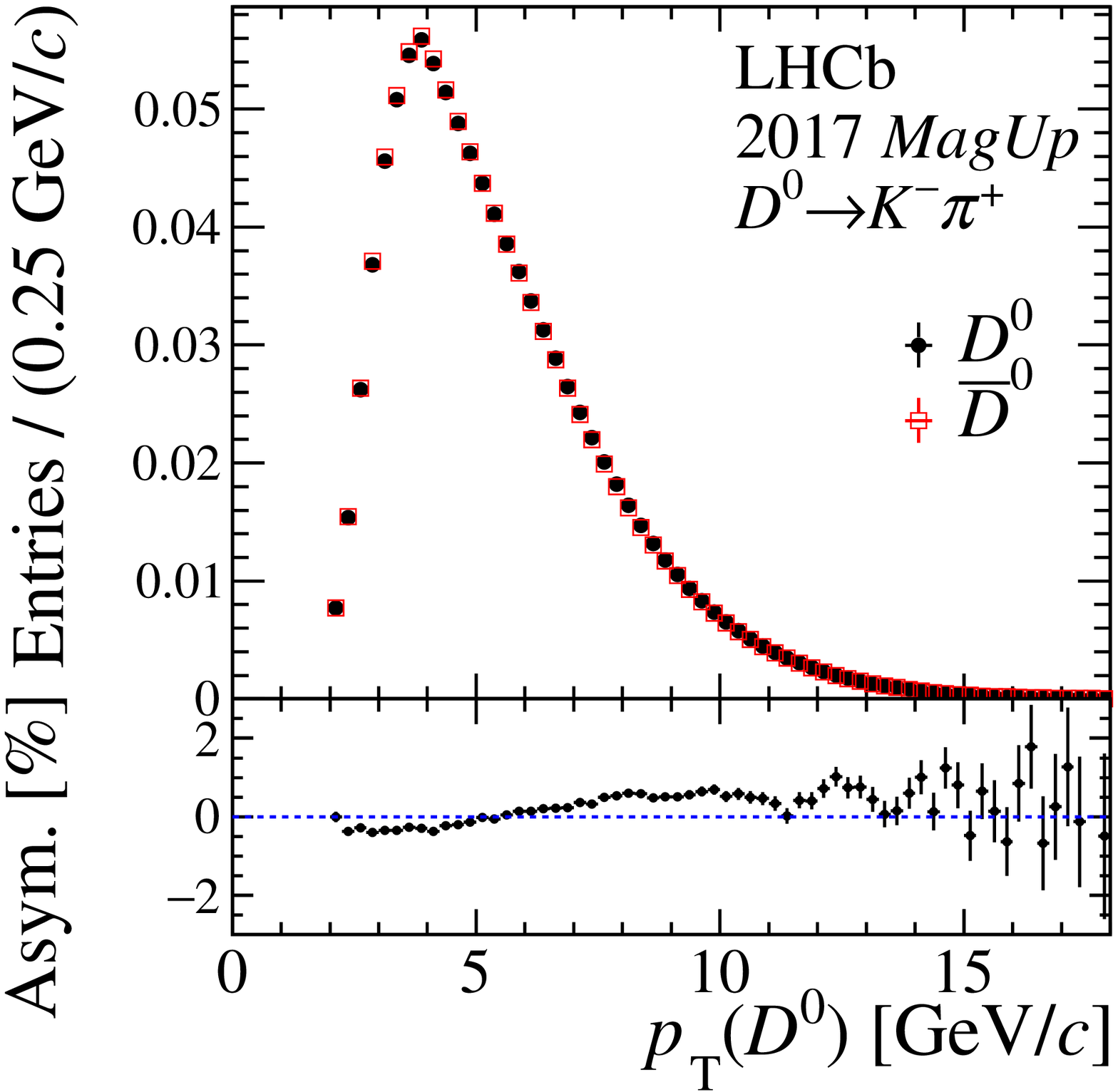}\\
    \includegraphics[width=0.35\linewidth]{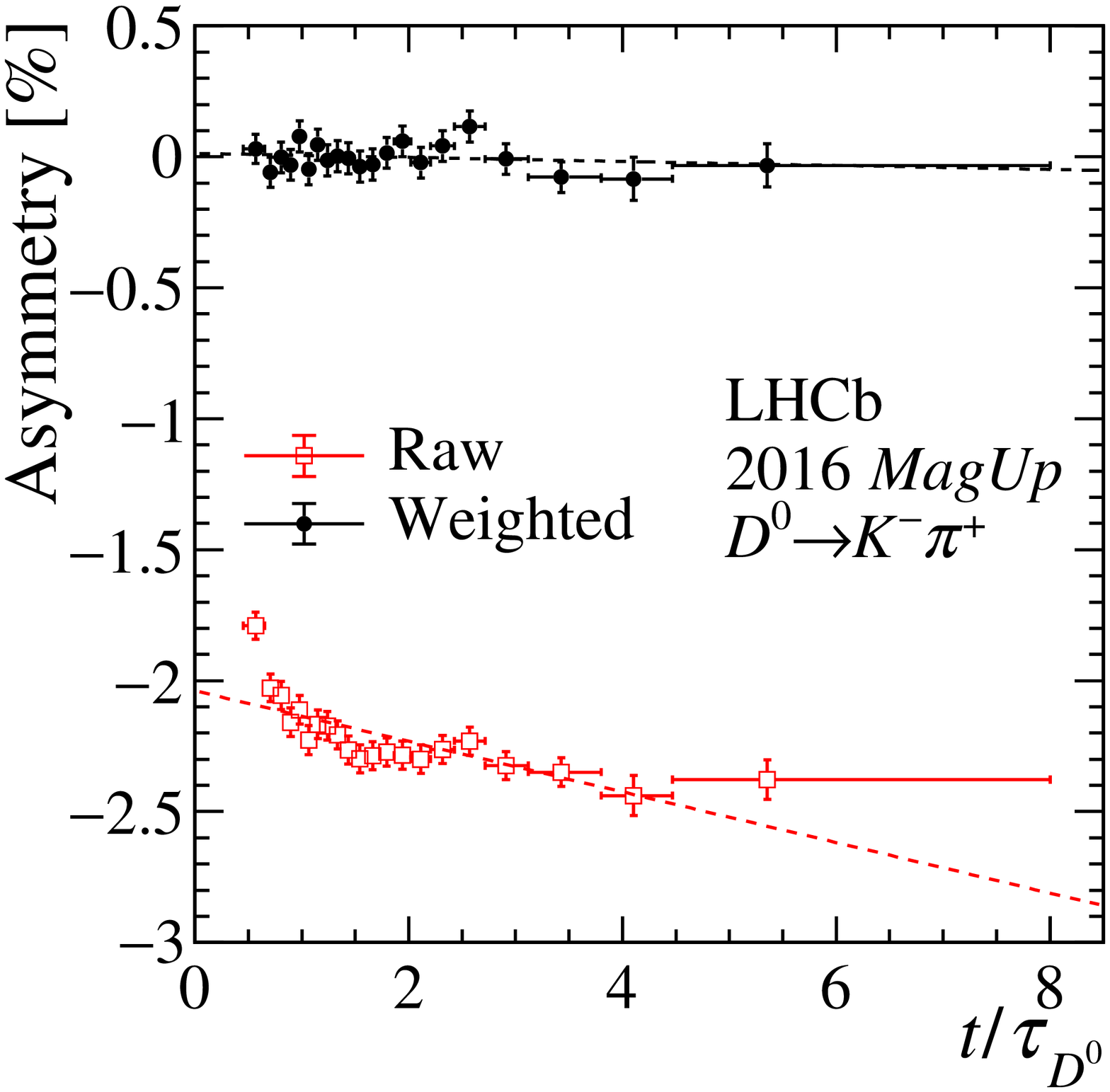}
    \includegraphics[width=0.35\linewidth]{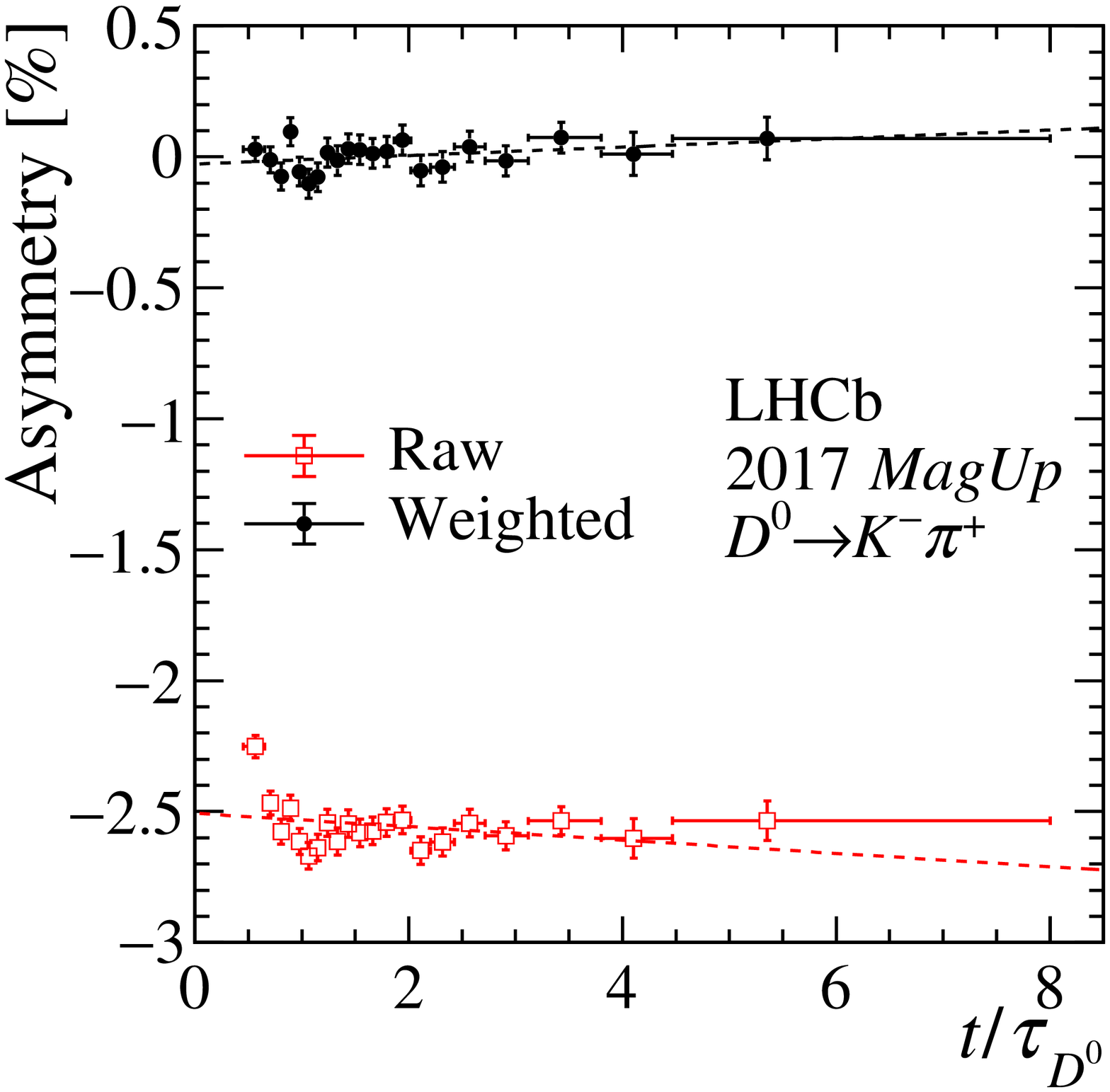}
  \end{center}
  \vspace*{-0.6cm}
  \caption{
    (Top) Normalised distributions of the \Dz transverse momentum, in different colours for each decay-time interval.
    Decay time increases from blue to yellow colour.
    (Centre) Asymmetry between the normalised $\pt$ distributions of \Dz and \Dzb mesons.
    (Bottom)
    Linear fit to the time-dependent asymmetry
    (red) before and (black) after the kinematic weighting.
    All plots correspond to \DzRS candidates recorded with the \MagUp polarity in (left) 2016 and (right) 2017.
  }
  \label{fig:time_dep_asymm}
\end{figure}

Since the $Q$ value of the \Dstarp decay is small with respect to the pion mass, both the magnitude and the direction of the momenta of the \Dstarp, \pisp and \Dz mesons are highly correlated.
As a consequence, all aforementioned asymmetries reflect into momentum-dependent asymmetries of the \Dz candidate, with all being of similar size.
These asymmetries would not bias the measurement of \DY{\HH} if they did not depend on the \Dz decay time.
However, even if the momentum of the \Dz meson is uncorrelated with its decay time, the selection requirements introduce correlations between their measured values.
For example, due to the requirement on the $\chi^2_\text{FD}$ of the \Dz candidate, low decay-time values are measured only if the \Dz momentum is sufficiently large.
The largest correlations concern the \Dz transverse momentum, the normalised distribution of which is plotted for each decay-time interval in Fig.~\ref{fig:time_dep_asymm} (top).
The raw asymmetry of the data collected with the \MagUp polarity, of the order of 1\%, increases as a function of transverse momentum, and correspondingly decreases as a function of decay time, as shown in Fig.~\ref{fig:time_dep_asymm} (centre) and (bottom).
As a result, the dependence on decay time is not linear.
The data collected in 2016 present a much larger slope of the time-dependent asymmetry, even if their momentum-dependent asymmetries are similar to those of data collected in 2017, since the correlations induced by the first-stage software trigger during 2016 are larger.
The asymmetry slopes for the data collected with the \MagDown polarity are considerably smaller, as a result of smaller observed momentum-dependent asymmetries.

These nuisance asymmetries are removed by equalising the kinematics of \pisp and \pism candidates and of \Dz and \Dzb candidates.
This is obtained by weighting their kinematic distributions to their average.
The weighting is performed with a binned approach in two steps.
The first equalises the \mbox{$(\theta_x, \theta_y, k)$} distributions of \pisp and \pism candidates to remove the largest acceptance and detector asymmetries, and employs
36 intervals in the range $[-0.27, 0.27]\rad$ for $\theta_x$,
27 intervals in the range $[-0.27, 0.27]\rad$ for $\theta_y$, and
40 intervals in the range $[0, 0.8]\cgev$ for $k$.
For each variable, all intervals have the same width.
In addition, intervals with fewer than 40 \pisp or \pism candidates, or where the asymmetry between the number of \pisp and \pism candidates is greater than 20\% in magnitude, are removed by setting the corresponding weights to zero.
This avoids weights whose value would be prone to large statistical fluctuations or very different from unity.
The effect of these requirements is very similar to the application of the fiducial requirements used to remove phase-space regions characterised by large detector asymmetries in Ref.~\cite{LHCb-PAPER-2019-006}, but removes fewer candidates from the data sample.

Even after this weighting procedure, residual asymmetries of about 0.5\% and dependent on the \Dz momentum and pseudorapidity are observed~\cite{pajero:2021}.
These asymmetries are removed by the second step of the weighting, which  considers the tridimensional distribution of $(\pt(\Dz), \eta(\Dz), \eta(\pisp))$.
The first two variables have the largest correlation with decay time, while $\eta(\pisp)$ is included to avoid that the weighting of the \Dz kinematics spoils that of the \pisp meson.
This second weighting employs
32 intervals in the range $[2, 18]\gevc$ for $\pt(\Dz)$,
25 intervals in the range $[2, 4.5]$ for $\eta(\Dz)$ and
22 intervals in the range $[2, 4.2]$ for $\eta(\pisp)$.
All intervals in each variable have the same width; the same limits on the minimum number of candidates and on the maximum asymmetry per interval, as in the first weighting, are applied.

While the impact on the result of the second step of the weighting is smaller than that of the first, the corresponding size of the shift in \DY{\HH} is of the same order as that of the final statistical uncertainty.
In particular, the second step is essential to remove the asymmetries of the momentum distribution of the \Dz meson.
For the \RS decay channel, these receive a contribution from the detection asymmetries of the \RS final state, which are not eliminated by a dedicated weighting.
They are instead removed by the second step of the baseline weighting, as are the other asymmetries affecting the \Dz momentum.

Since the detection asymmetries and the correlations induced by the trigger depend on the data-taking conditions, the weighting is performed separately in eight subsamples, divided according to the year and to the magnet polarity.
Furthermore, since the asymmetries are different between the \RS final state and the signal decay channels, the weighting is performed independently for each decay channel.
The weighting slightly modifies the combined momentum distribution of \Dz and \Dzb candidates and, consequently, also the \DstM distribution.
Therefore, the fits performed to calculate the coefficients to subtract the \DstM background are repeated after each step of the weighting and the coefficients are updated accordingly.

\begin{figure}[tb]
  \begin{center}
    \includegraphics[width=0.49\linewidth]{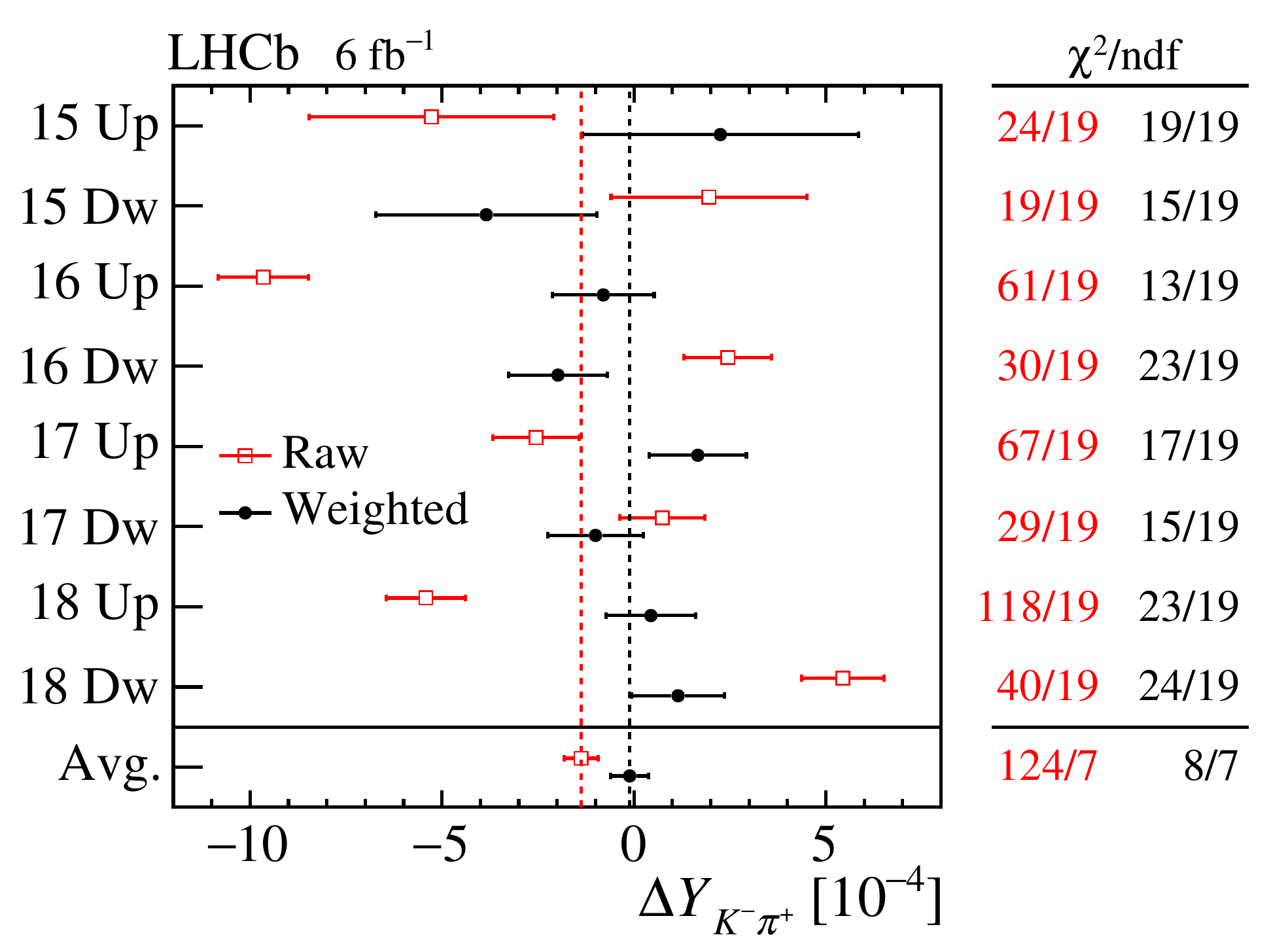}\\
    \includegraphics[width=0.49\linewidth]{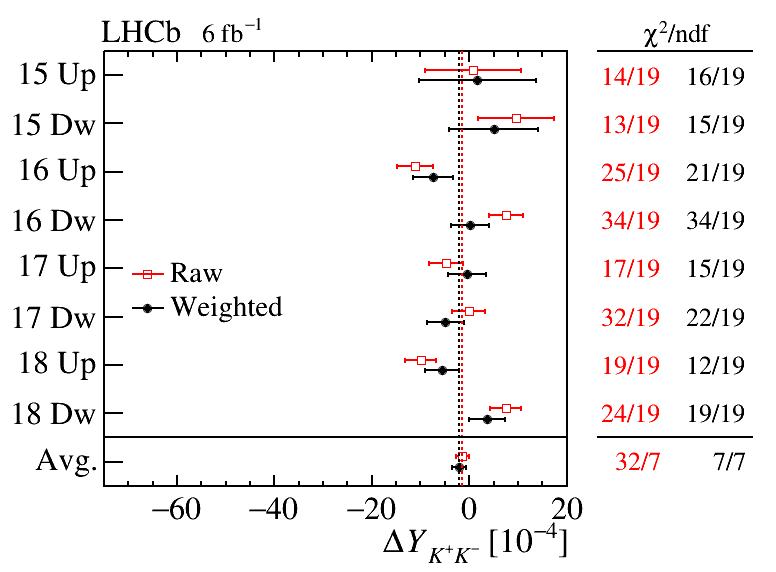}
    \includegraphics[width=0.49\linewidth]{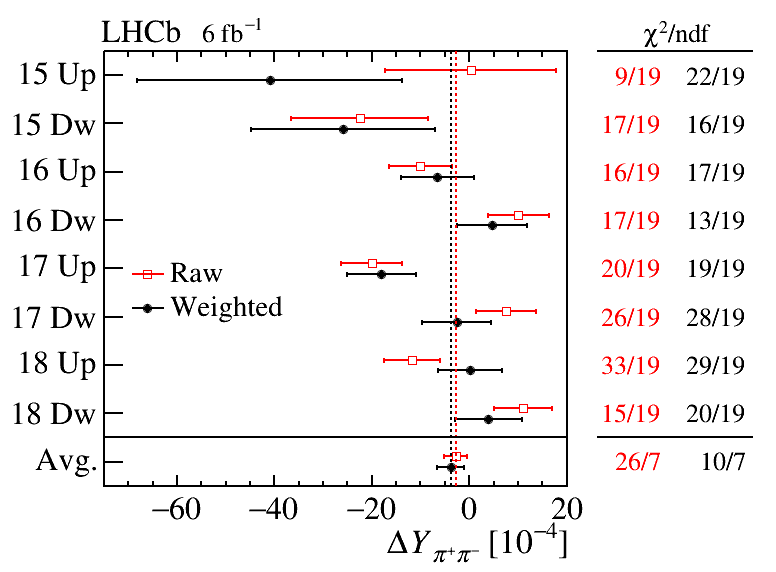}
  \end{center}
  \vspace*{-0.6cm}
  \caption{
    Results of the fits to the time-dependent asymmetry for each subsample, for (top) \RS, (bottom left) \KK and (bottom right) \PP final states.
    In the $y$-axis labels, the data-taking year is abbreviated with the last two digits only and the magnet polarity \MagUp (\MagDown) is abbreviated as ``Up'' (``Dw''), while ``Avg.'' denotes the weighted average.
    The fit \chisqndf are reported on the right of each plot.
    Empty red squares and solid black dots correspond to data before and after the equalisation of the \pisp and \Dz kinematics, respectively, and prior to subtracting the contribution of secondary decays described in Sect.~\ref{sect:secondaries}.
  }
  \label{fig:weighting_summary}
\end{figure}

The measured values of \DY{\HH} for the three decay channels are displayed in Fig.~\ref{fig:weighting_summary} before and after the kinematic weighting.
After the weighting, the time dependence of the asymmetry in each sample is well described by a linear function, as confirmed by the fact that all fits have a \chisqndf compatible with unity.
In addition, the measurements of \DY{\HH} are compatible among different years and magnet polarities.
On the other hand, the compatibility of \DY{\RS} with zero should be confirmed only after the subtraction of the contribution to the asymmetry of secondary \Dstarp mesons from \B decays, which is described in Sect.~\ref{sect:secondaries}.
The agreement among the measured values of \DY{\RS} in the eight subsamples and the compatibility of their average with zero after the aforementioned subtraction confirm the effectiveness of the weighting, which removes even the larger detection asymmetries of this decay channel with a precision three times better than that of the signal samples.

Due to the correlation between the measured decay time and momentum of the \Dz meson, a possible time-dependent asymmetry due to a nonzero value of \DY{\HH} would reflect into momentum-dependent asymmetries and would be partially cancelled by the kinematic weighting.
This would cause a dilution of the true value of \DY{\HH}.
The size of this dilution is measured by introducing an artificial value of \DY{\RS} in the \DzRS raw sample, obtained by filtering the candidates according to an efficiency that changes linearly with decay time, with opposite slopes for \Dz and \Dzb candidates.
The kinematic weighting is then applied and the measured value of \DY{\RS} is compared to the introduced one.
This procedure is repeated for different values of \DY{\RS}, up to values as large as 100 times the statistical uncertainty of the final measurement.
The dilution is found to have a linear effect on the measured value of \DY{\RS}, which is equal to \mbox{$(96.9\pm0.1)\%$} of the introduced one.
As a cross-check, the same study is performed also for the signal channels, obtaining compatible dilution factors, although less precise.
In Fig.~\ref{fig:weighting_summary} and in the following, the results of all decay channels are corrected to account for this dilution factor, using the value measured in the \RS channel.

\section{Removal of \texorpdfstring{\B}{B} decays}
\label{sect:secondaries}

The background from \B decays produces a biasing contribution to the asymmetry, $A(t)$, even after the removal of the nuisance asymmetries described in Sect.~\ref{sect:asymmetries}.
In fact, such asymmetry is equal to
\begin{equation}
    \label{eq:secondaries}
    A(t) = A_\text{sig}(t) + f_{\B}(t)[A_{\B}(t) - A_\mathrm{sig}(t)],
\end{equation}
where \mbox{$A_\text{sig}(t)$} and \mbox{$A_{\B}(t)$} are the asymmetries of signal and secondary decays from \B mesons, and \mbox{$f_{\B}(t)$} is the fraction of secondary decays among the \Dz candidates at given decay time $t$.
Since the flight distance of \Dz candidates is measured with respect to their associated PV, the decay time of the secondary background from \B decays is overestimated and the fraction $f_{\B}(t)$ increases as a function of decay time.
Moreover, the asymmetries of signal and secondary decays differ, mainly because of the different production asymmetries of \Dstarp and \B mesons, and of the different asymmetries of their selection efficiency at the hardware-level trigger.
As a consequence, secondary decays will introduce a bias on the measurement of \DY{\HH}.

\begin{figure}[tb]
  \begin{center}
    \includegraphics[width=0.45\linewidth]{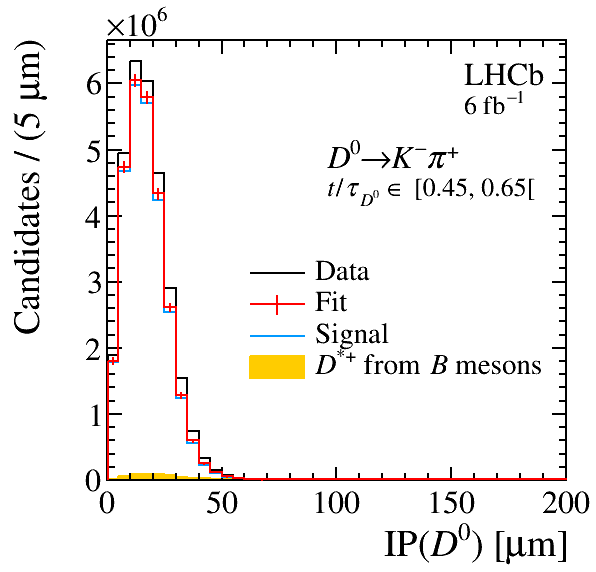}
    \includegraphics[width=0.45\linewidth]{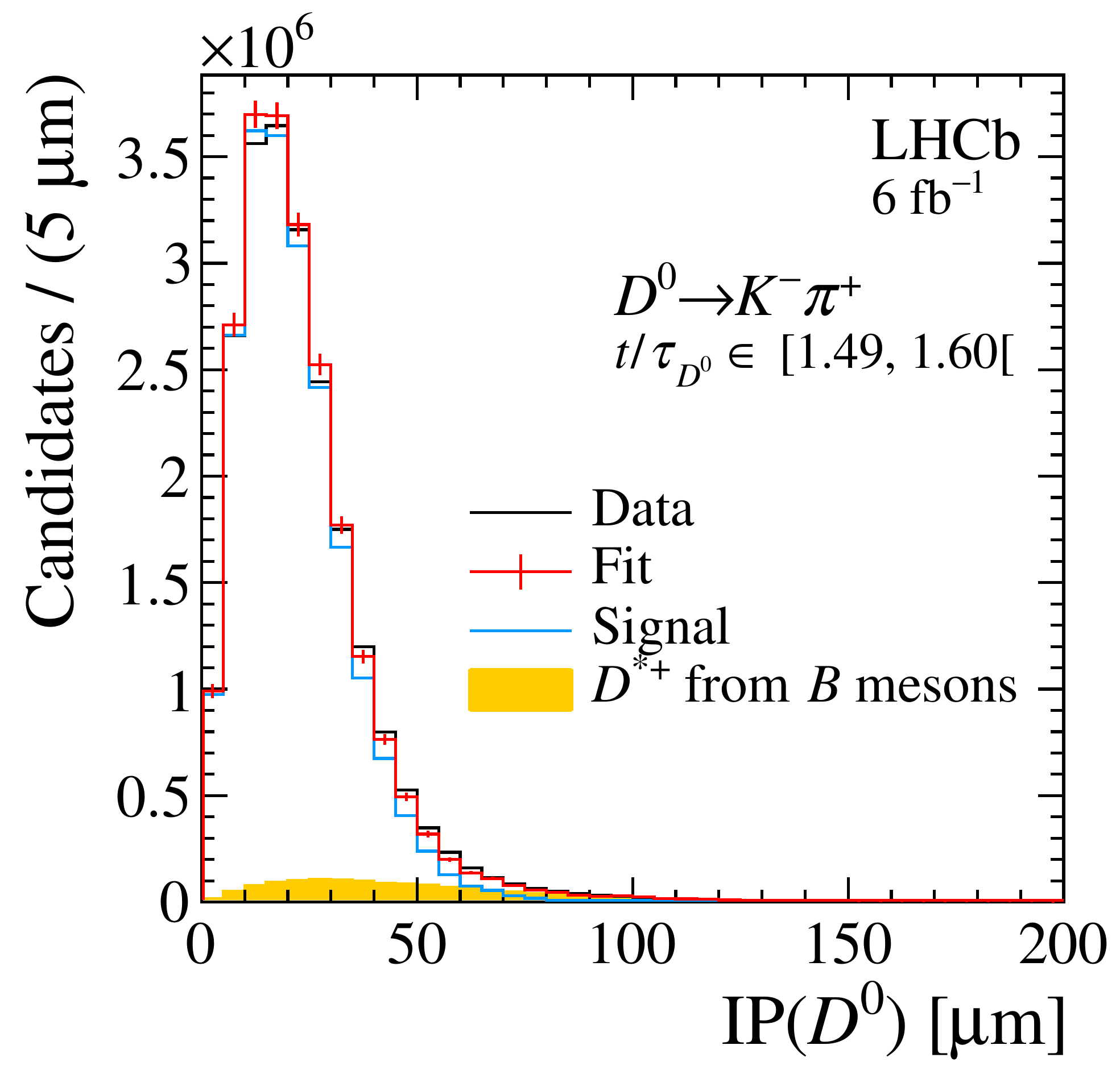}
    \includegraphics[width=0.45\linewidth]{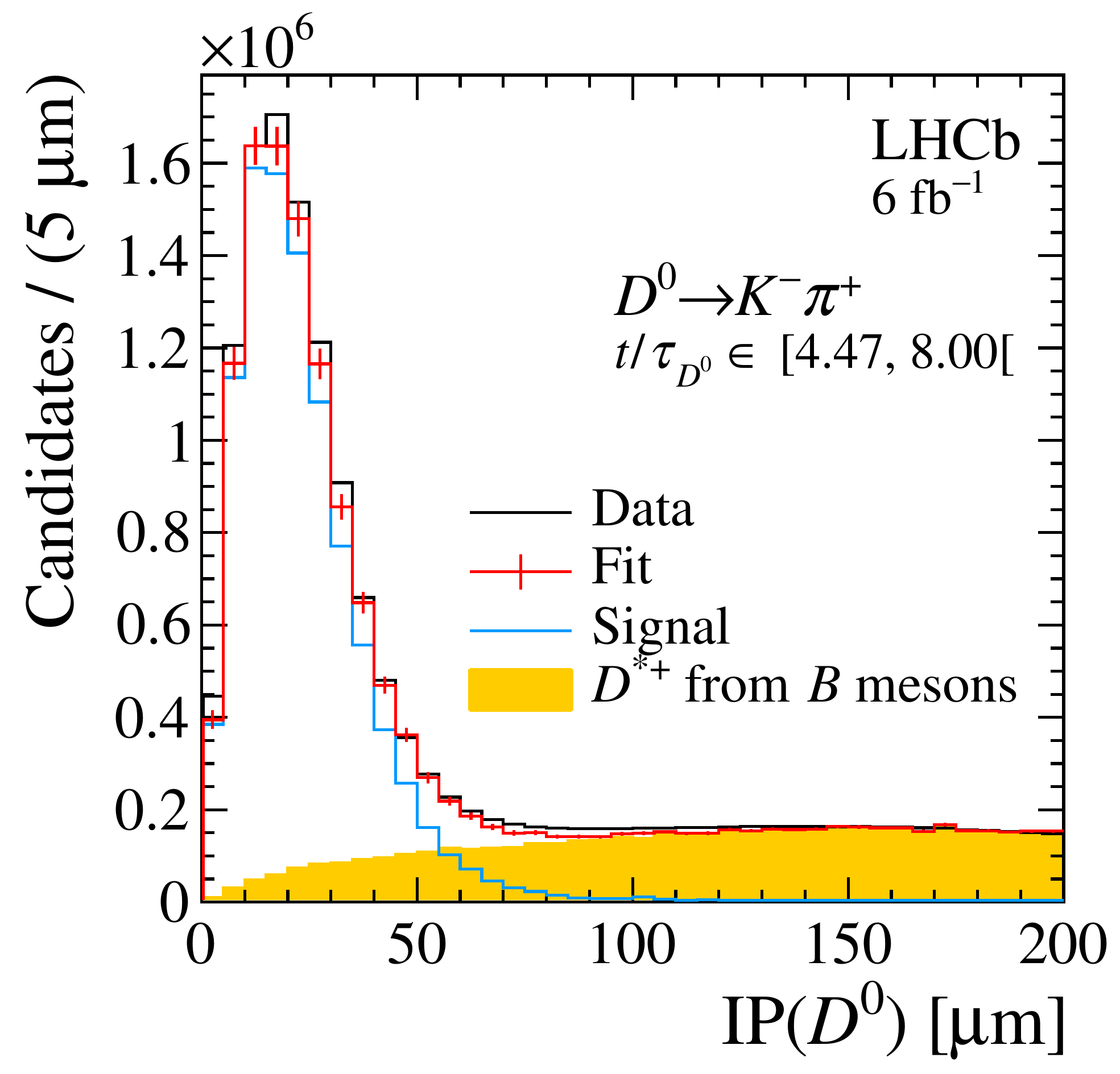}
  \end{center}
  \vspace*{-0.6cm}
  \caption{
    Impact parameter distribution of the \DzRS candidates for the (top) first, (left) middle and (right) last decay-time interval.
    The projections of the two-dimensional template fit are superimposed.
  }
  \label{fig:ip_fit}
\end{figure}

\begin{figure}[tb]
  \begin{center}
    \includegraphics[width=0.45\linewidth]{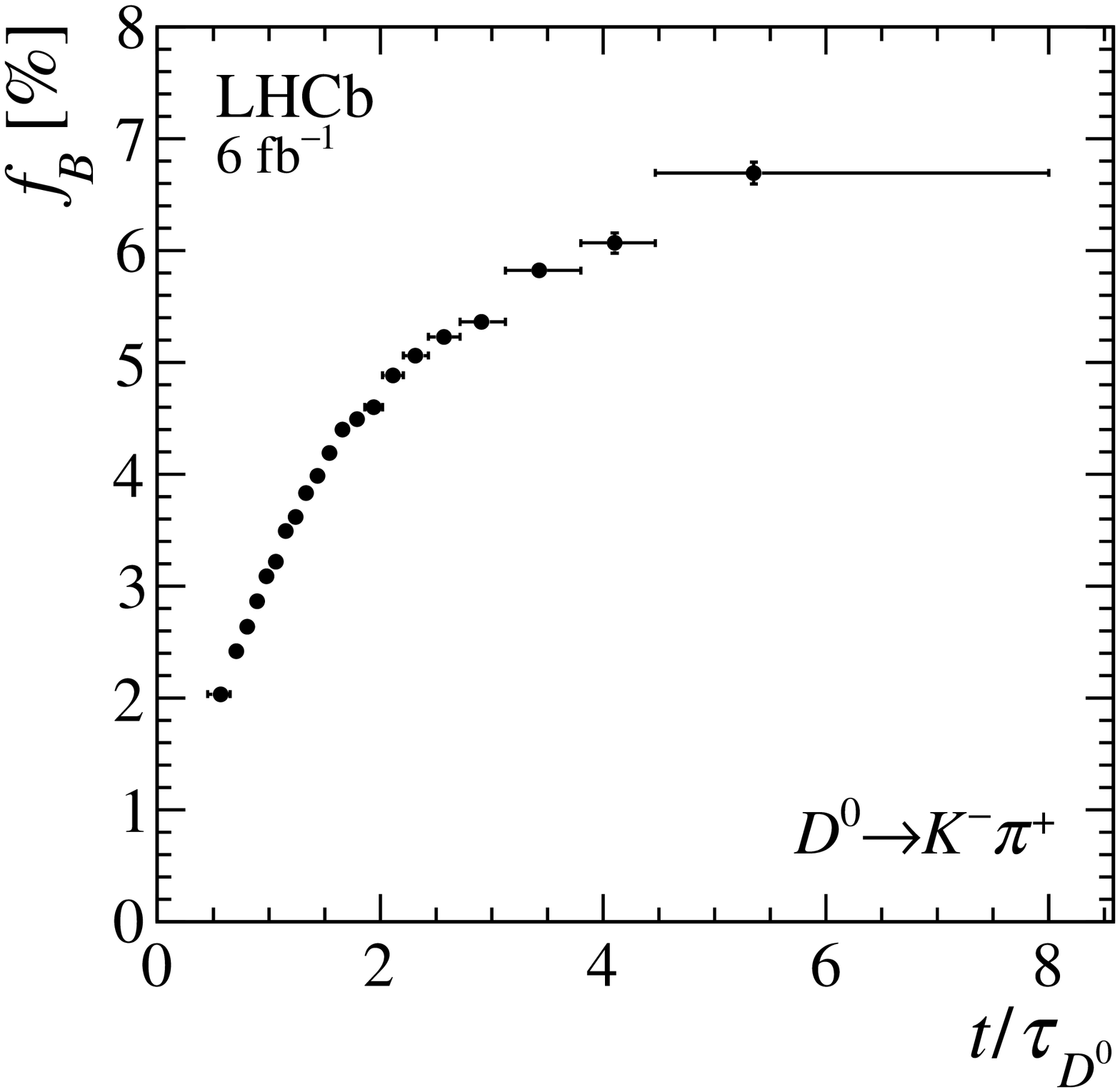}
    \includegraphics[width=0.45\linewidth]{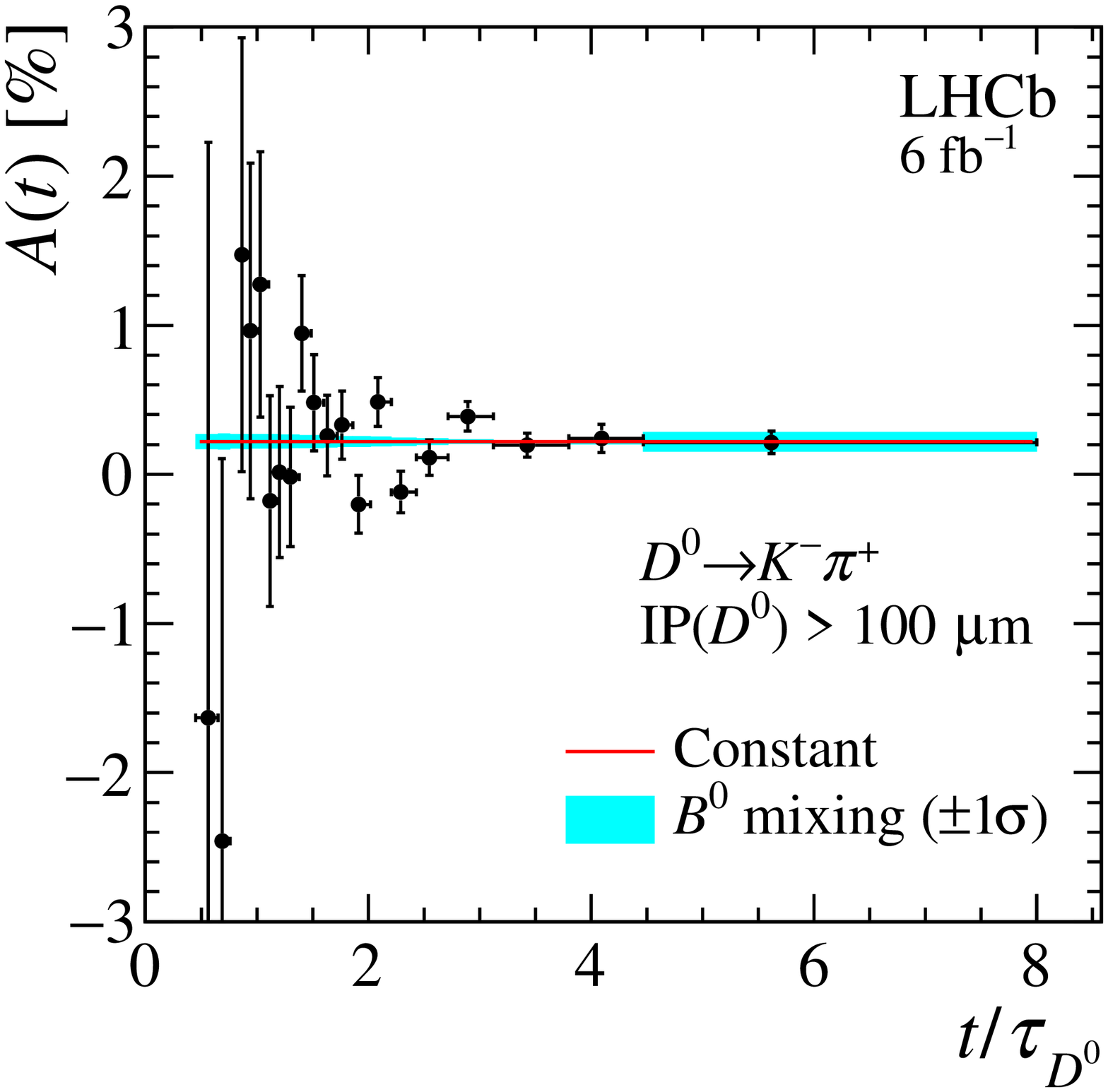}
  \end{center}
  \vspace*{-0.7cm}
  \caption{
    (Left) Fraction of secondary decays from \B mesons measured in the template fit.
    Only statistical uncertainties are displayed.
    (Right) Measured asymmetry, after the kinematic weighting, of data with \mbox{$\mathrm{IP}(\Dz) > 100\mum$} in the enlarged \DstM signal window.
    The fit of a constant function is shown in red, while the $1\sigma$ band corresponding to the possible impact of \Bz mixing is plotted in cyan.
  }
  \label{fig:f_sec}
\end{figure}

The fraction $f_{\B}(t)$ is determined through a binned maximum-likelihood fit to the $\mathrm{IP}(\Dz)~\vs~t(\Dz)$ bidimensional distribution of \DzRS candidates, for \Dz and \Dzb samples combined.
In the fit, the selection requirement on the \mbox{$\mathrm{IP}(\Dz)$} range is loosened from $[0,60]$ to $[0, 200]\mum$ to increase the discriminating power between signal and secondary decays and have a better handle on the latter category.
For the same reason, the \DstM signal window is enlarged to \mbox{$[2007.5, 2011.3]\mevcc$}.
In fact, for secondary decays the PV constraint biases the measured value of the angle between the \Dz and \pisp momentum, and consequently the \DstM invariant mass, to lower values.
The template {\PDF}s are taken from a simplified simulation of signal and secondary decays only, whereas all other particles produced in the \proton\proton collision are discarded to minimise the usage of computing resources.
To reduce small discrepancies between data and simulation, the kinematics of the \Dz meson is weighted to match that of data~\cite{Rogozhnikov:2016bdp}.
The weighting coefficients are calculated using data with \mbox{$\mathrm{IP}(\Dz) < 60\mum$} ($>100 \mum$) for signal (secondary) decays.
In the fit, the two-dimensional template {\PDF}s are determined from simulation, and only the time-integrated fraction of secondary decays is left free to vary.
The ratio of the fit projections, which are shown for three decay-time intervals in Fig.~\ref{fig:ip_fit}, to data agrees with unity within \mbox{10--20\%}, with the largest discrepancies being due to the accuracy of the simulation of the trigger requirements at low decay times.
The impact of these discrepancies, which affect similarly signal and secondary decays and cancel to good extent in the calculation of the fraction \mbox{$f_{\B}(t)$}, is estimated as a systematic uncertainty in Sect.~\ref{sect:systematics}.
The dependence of \mbox{$f_{\B}(t)$} on decay time is displayed for the baseline $\mathrm{IP}(\Dz)$ and \DstM requirements in Fig.~\ref{fig:f_sec} (left).
The fraction increases with decay time from around 2\% to 7\%, corresponding to a time-integrated value of around 4\%.

The difference in asymmetry of secondary and signal decays, entering Eq.~(\ref{eq:secondaries}), is measured from \DzRS candidates satisfying $\mathrm{IP}(\Dz) > 100\mum$, where the fraction of secondary decays is about 95\%.
By construction, the weighting of Sect.~\ref{sect:asymmetries} sets to zero the asymmetry of the candidates satisfying $\mathrm{IP}(\Dz) < 60\mum$, which are signal decays apart from the 4\% contamination of secondary \Dstarp decays.
Thus, the size of the time-integrated asymmetry of signal decays after the kinematic weighting is negligible with respect to that of secondary decays, which is equal to the asymmetry at $\mathrm{IP}(\Dz) > 100\mum$ up to a dilution of around 5\%.
This asymmetry, which is shown in Fig.~\ref{fig:f_sec} (right), is compatible with being independent of decay time and amounts to \mbox{$(2.2 \pm 0.4) \times 10^{-3}$}.
The constant behaviour of the asymmetry difference, \mbox{$A_{\B}(t) - A_\mathrm{sig}(t)$}, is in agreement with expectations.
The nuisance time-dependent asymmetries of Sect.~\ref{sect:asymmetries} cancel to good extent in the difference between secondary and signal decays even before the kinematic weighting, since the kinematics of the two categories of decays are similar.
Moreover, both the difference of the production and of the selection asymmetries, where the latter is mainly due to particles other than the \Dstarp decay products responsible for the hardware-trigger decision, are expected to depend weakly on momenta.
Therefore, the asymmetry difference is not expected to depend on decay time before the kinematic weighting to first order.
Since the weighting does not modify the asymmetry difference to first order, this assumption holds after the kinematic weighting as well.
In particular, it is verified explicitly in data that the fraction \mbox{$f_{\B}(t)$} and the asymmetry difference are not changed to first order by the kinematic weighting.
Therefore, the order in which the weighting and the subtraction of secondary decays are performed does not affect the results.
Finally, the dependence of the production asymmetry on the decay time due to \Bz mixing is found to be negligible within the experimental uncertainty, as is the contribution from \CP violation; see Sect.~\ref{sect:systematics}.

\begin{figure}[tb]
  \begin{center}
    \includegraphics[width=0.6\linewidth]{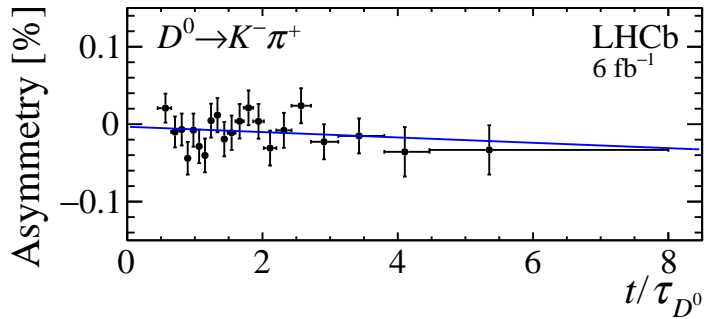}
  \end{center}
  \vspace*{-0.6cm}
  \caption{Asymmetry, $A_\mathrm{sig}(t)$, of \DzRS candidates as a function of decay time.
  A linear fit is superimposed.
  The \chisqndf of the fit is \mbox{$17/19$}.
  }
  \label{fig:asymmetry_fit_RS}
\end{figure}

The asymmetry of signal decays is calculated in each interval of decay time by subtracting the term \mbox{$f_{\B}(t)(A_{\B} - A_\text{sig})$} from the measured asymmetry, $A(t)$, using the fitted values of $f_\text{\B}(t)$ and of the time-independent asymmetry difference above.
The results for the \DzRS control sample are plotted in Fig.~\ref{fig:asymmetry_fit_RS}.
The shift of the \DY{\RS} value with respect to that of Fig.~\ref{fig:weighting_summary}, where the contribution from \B-meson decays is not subtracted, is approximately equal to $-0.26 \times 10^{-4}$.

Since no differences are expected in $f_\text{\B}(t)$ and in the asymmetry difference among different \Dz decay channels, their estimates for the \RS control channel are employed to correct the signal samples as well, to minimise the statistical fluctuations on the values of the shift.
The results are reported in Sect.~\ref{sect:results}.

\section{Systematic uncertainties}
\label{sect:systematics}

The main systematic uncertainties on \DY{f} are due to the subtraction of the combinatorial background under the \Dstarp mass peak, the asymmetry of the time-dependent shifts of the peak position for \Dstarp and \Dstarm mesons, and uncertainties in the subtraction of the contribution of the background from \B-meson decays.
Minor contributions are related to limitations in the removal of the nuisance asymmetries described in Sect.~\ref{sect:asymmetries}, as well as to the background of misidentified \D-meson decays under the \Dz mass peak.
Whenever they are not expected to depend on the decay channel, the systematic uncertainties are evaluated relying on the \RS final state to minimise the statistical uncertainty on their estimated value.

The removal of the background under the \Dstarp mass peak relies on the assumption that the kinematics and the asymmetry of the background are the same in the signal and in the lateral window used for the background subtraction.
A systematic uncertainty on this assumption is assigned by repeating the measurement of \DY{\RS} using three alternative windows, namely $[2004.5,2008.5]\mevcc$, $[2013,2015]\mevcc$ and $[2018,2020]\mevcc$.
No systematic trends are spotted, and additional studies of the background properties do not reveal any significant differences among the four lateral windows.
Therefore, the root mean square of the deviations, \mbox{$0.10 \times 10^{-4}$}, is employed as a conservative estimate of the systematic uncertainty.
For the \mbox{\KK (\PP)} channel, instead, the systematic uncertainty is calculated by scaling this value by the ratio of the signal-to-background ratios in the \RS and in the \mbox{\KK (\PP)} channels, yielding \mbox{$0.20\times 10^{-4}$ ($0.28 \times 10^{-4}$)}.
Uncertainties in the determination of the coefficient used for the background subtraction can cause a bias.
A systematic uncertainty on this effect is estimated by repeating the measurement fitting the combinatorial-background distribution using the alternative {\PDF}s employed in Refs.~\cite{LHCb-PAPER-2019-006,LHCb-PAPER-2016-063} instead of the baseline background model.
The maximum deviations from the baseline result, which amount to \mbox{$0.01 \times 10^{-4}$}, \mbox{$0.04 \times 10^{-4}$} and \mbox{$0.05 \times 10^{-4}$} for the \RS, \KK and \PP decay channels, respectively, are assigned as systematic uncertainties.
Finally, the impact of possible differences in the background \PDF for \pisp and \pism mesons is estimated with the \RS channel by repeating the fits separately for \Dstarp and \Dstarm candidates.
The fits are performed separately for different years and magnet polarities.
The deviation from the baseline result, \mbox{$0.07 \times 10^{-4}$}, is taken as systematic uncertainty.
The corresponding systematic uncertainties for the \KK and \PP samples are calculated by scaling this value to account for different signal-to-background ratios as before, yielding \mbox{$0.14\times 10^{-4}$} and \mbox{$0.19 \times 10^{-4}$}, respectively.
All the systematic uncertainties on the subtraction of the combinatorial background are summed in quadrature, giving the final values \mbox{$0.12\times 10^{-4}$}, \mbox{$0.24\times 10^{-4}$} and \mbox{$0.34 \times 10^{-4}$} for the \RS, \KK and \PP channels, respectively.

\begin{figure}[tb]
  \begin{center}
    \includegraphics[width=0.45\linewidth]{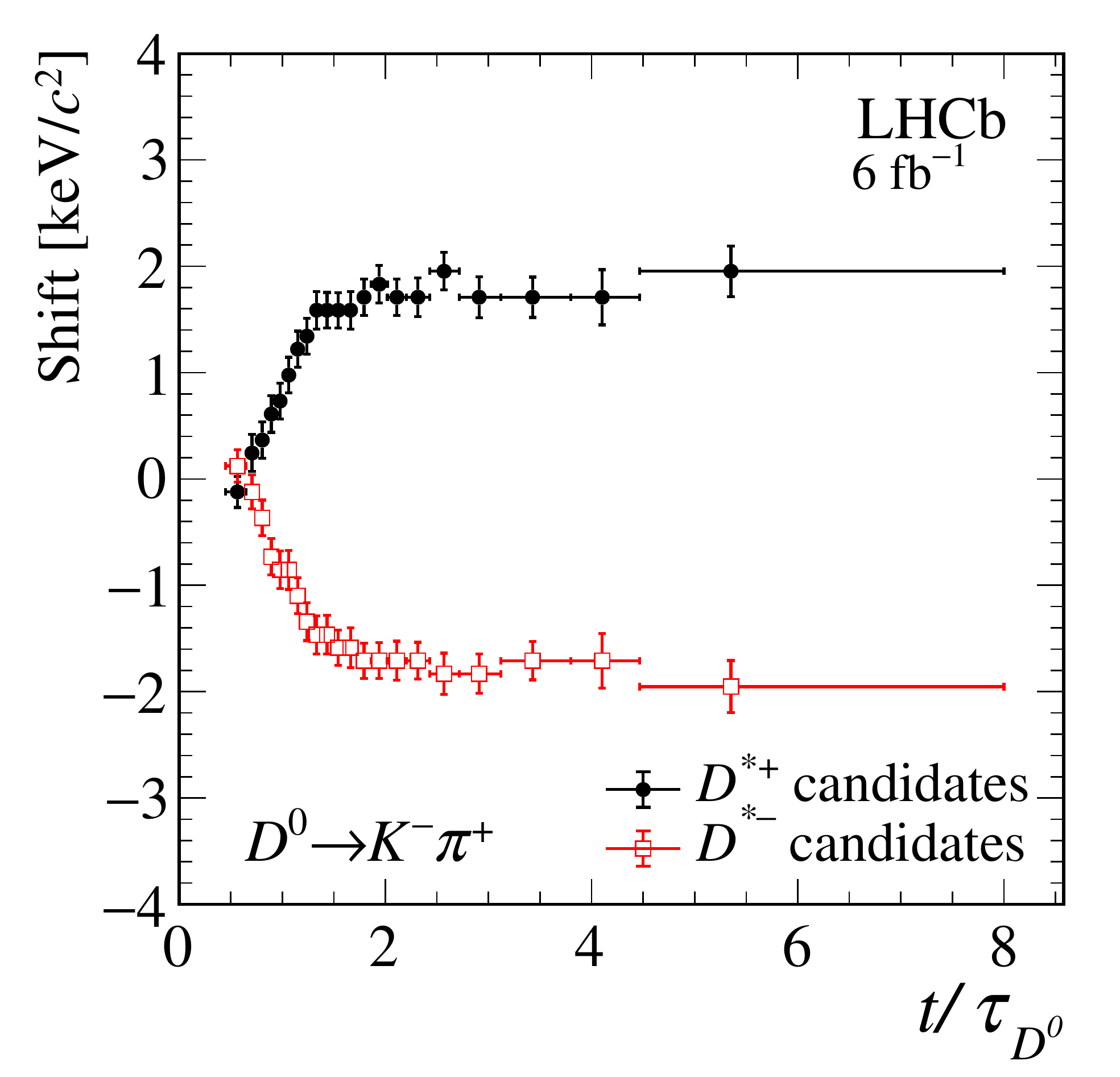}
  \end{center}
  \vspace*{-0.7cm}
  \caption{
    Relative shift of the \Dstarp and \Dstarm mass peaks as a function of decay time for the \RS decay channel.
  }
  \label{fig:dst_m_shift}
\end{figure}

The usage of a fixed signal window, \mbox{$[2009.2, 2011.3]\mevcc$}, for the \DstM variable can bias the measurement of \DY{\HH} if the signal {\PDF}s of \Dstarp and \Dstarm candidates are shifted with respect to each other and the size of the shift changes as a function of time.
In each time interval, the size of the shift is estimated by comparing the \Dstarp and \Dstarm signal distributions.
The shift, which is displayed in Fig.~\ref{fig:dst_m_shift}, is compatible with zero at small decay times, and increases up to $\pm 2\kevcc$ at large decay times.
The impact of this variation is estimated by repeating the measurement of \DY{\RS} using a time- and flavour-dependent \DstM signal window, defined in each time interval by shifting the baseline window for \Dstarp (\Dstarm) candidates by plus half (minus half) the measured shift.
The deviation of \DY{\RS} from its baseline value, \mbox{$0.14 \times 10^{-4}$}, is taken as systematic uncertainty for all decay channels.

The subtraction of the contribution of \B-meson decays from the asymmetry, see Eq.~(\ref{eq:secondaries}), relies on the correct determination of both their fraction as a function of decay time and of their asymmetry difference with respect to signal decays.
The impact of the finite precision of the asymmetry difference is equal to \mbox{$0.04 \times 10^{-4}$} and is taken as systematic uncertainty.
The time dependence of the asymmetry difference owing to the \Bz mixing is estimated in simulation, and the obtained template \PDF is added to the constant function in the fit in Fig.~\ref{fig:f_sec} (right).
However, its normalisation is compatible with zero and its impact on the measurement is estimated to be less than \mbox{$0.04 \times 10^{-4}$}.
Since this value is equal to the systematic uncertainty on the precision on the asymmetry difference in the constant hypothesis, no additional systematic uncertainty is assigned to avoid double counting.
The uncertainty in the determination of the fraction $f_{\B}$ as a function of time receives two separate contributions.
The first comes from the finite size of the simulation sample used to produce the template {\PDF}s.
The second, larger contribution, is due to possible discrepancies of the two-dimensional $t(\Dz)~\vs~\mathrm{IP}(\Dz)$ distribution between simulation and data.
These are estimated using the data subsample where the \Dstarp meson forms a good-quality vertex with a \mun, which provides a pure sample of \B-meson decays.
The measured differences between data and simulation are found to be of the same order as those observed in the results of the fit to the $t(\Dz)~\vs~\mathrm{IP}(\Dz)$ distribution, whose projections are displayed in Fig.~\ref{fig:ip_fit}.
Taking into account the sum of these two contributions, the absolute uncertainty on the fraction at high and low decay times, which drives the impact of \B-meson decays on the measurement, is equal to 1.0\% and 0.7\%, respectively.
The corresponding uncertainty on the subtraction of the contribution of the asymmetry of \B-meson decays is \mbox{$0.07\times 10^{-4}$}.
The two systematic uncertainties arising due to the uncertainty of the asymmetry and fraction of \B-meson decays are summed in quadrature, yielding a total systematic uncertainty equal to \mbox{$0.08\times 10^{-4}$}.
As a cross-check, the asymmetry difference is measured also for the signal channels; the results are compatible with that obtained for the \RS channel, but less precise.
Furthermore, the fraction of \B-meson decays of the signal channels is checked in simulation to be compatible with that of the \RS channel within an uncertainty smaller than that with which the fraction is known in data.

Another source of background contributing to the systematic uncertainty is that of multibody decays of \D mesons, where one daughter particle is not reconstructed.
If one of the final-state particles is misidentified, a wrong mass assignment can compensate for the underestimation of the invariant mass due to the unreconstructed particle.
For \Dz mesons produced in the decay of a \Dstarp meson, these background contributions appear as a peak in the \DstM distribution, albeit with a larger width with respect to the signal.
Therefore, unlike pure \HH combinatorial background, they are not removed by the subtraction of the \DstM combinatorial background.
The same applies also to \decay{\Dz}{\HH} decays, where one of the daughter particles is misidentified.
Even if these misidentified decays mostly lie outside of the \mhh signal region, they need to be taken into account to determine correctly the contribution of the other background decays.
The same observation applies to \decay{\Dsp}{\Kp\Km\pip} decays, where the \Dsp meson is produced at the PV and the pion is assigned as \pisp.
Since their reconstructed \mhh and \DstM masses are anticorrelated, they are not removed by the subtraction of the \DstM combinatorial background, which instead causes the appearance of a dip in their \mhh distribution.

\begin{figure}[tb]
  \begin{center}
    \includegraphics[width=0.45\linewidth]{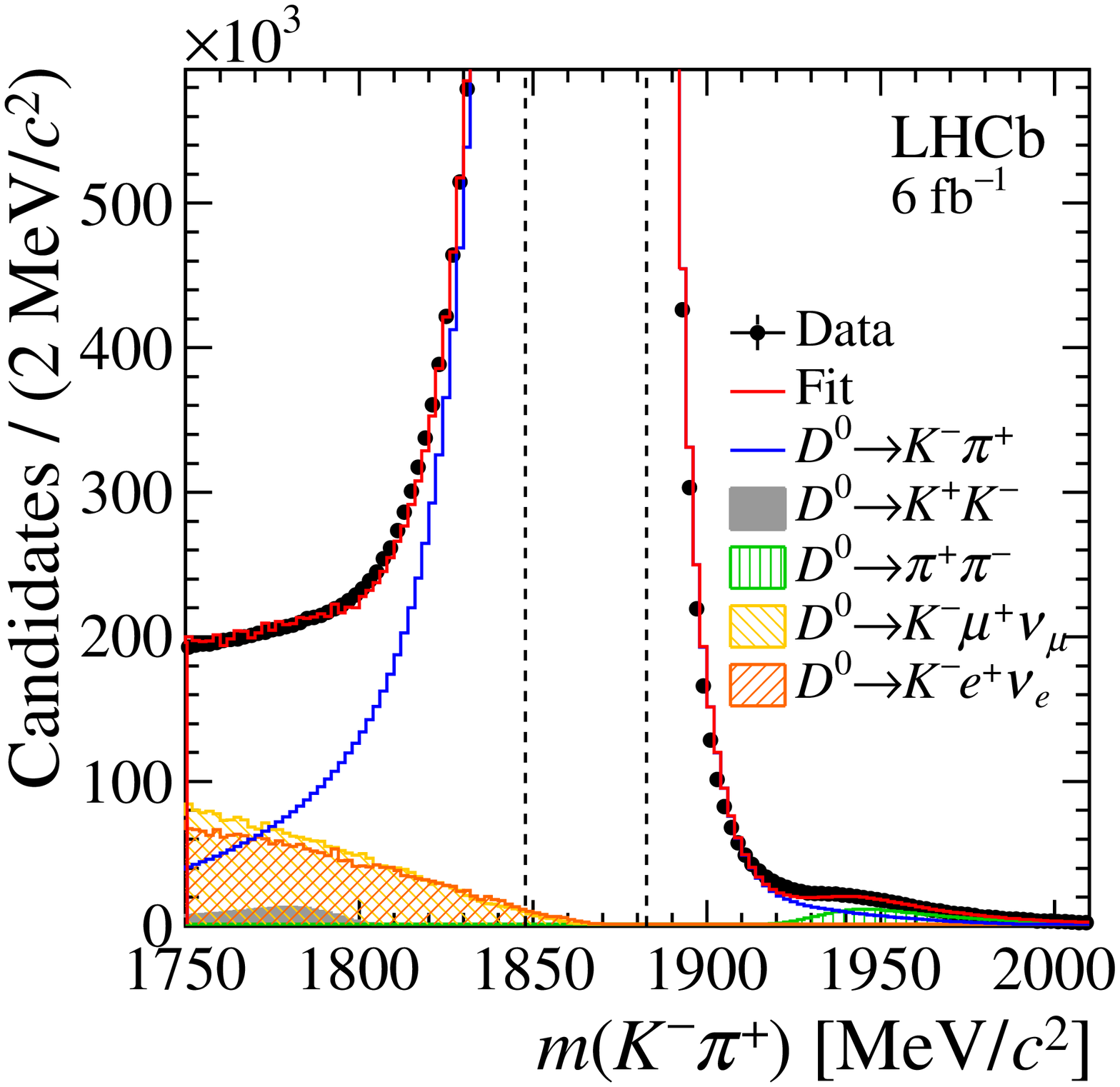} \\
    \includegraphics[width=0.45\linewidth]{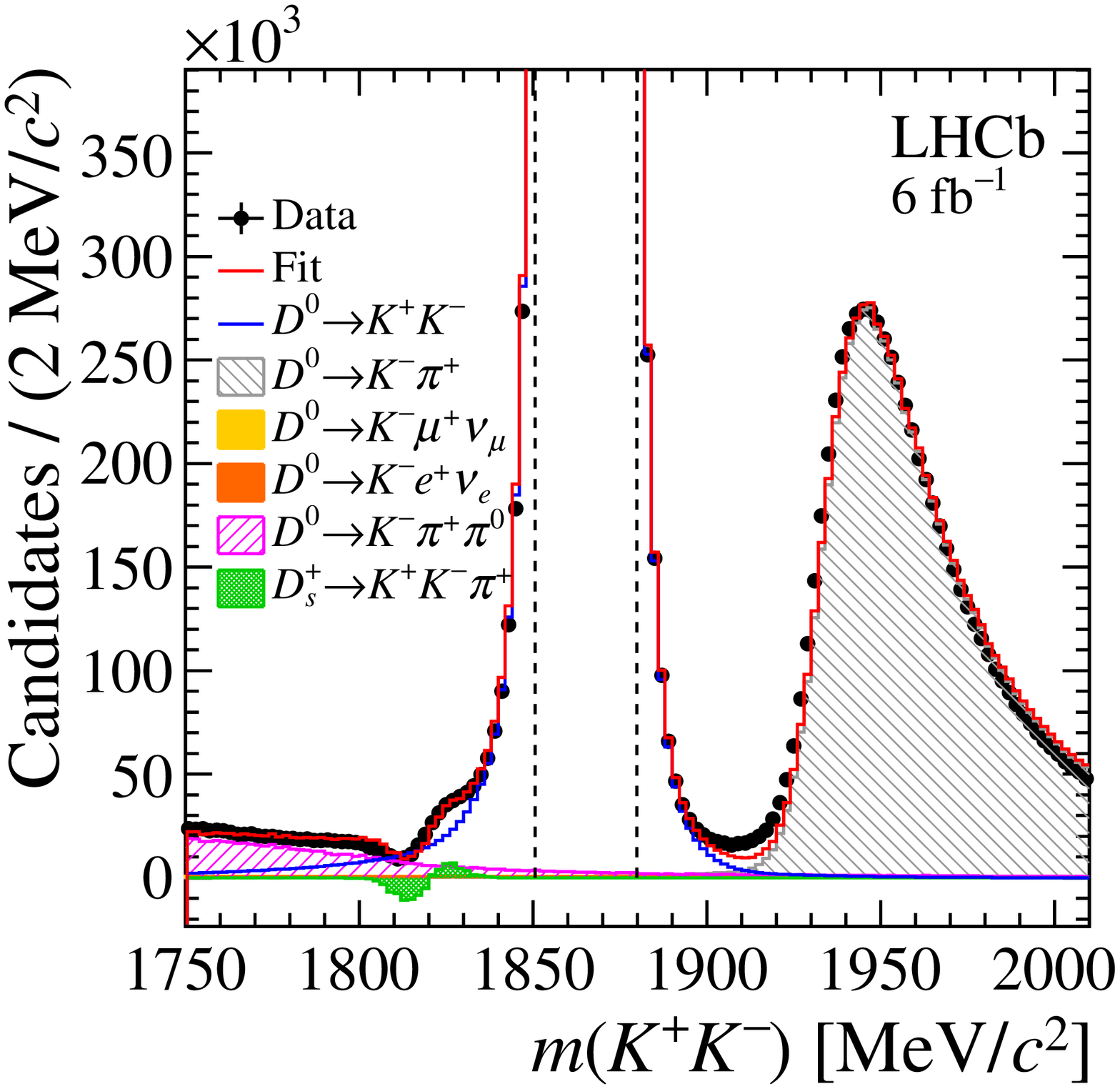}
    \includegraphics[width=0.45\linewidth]{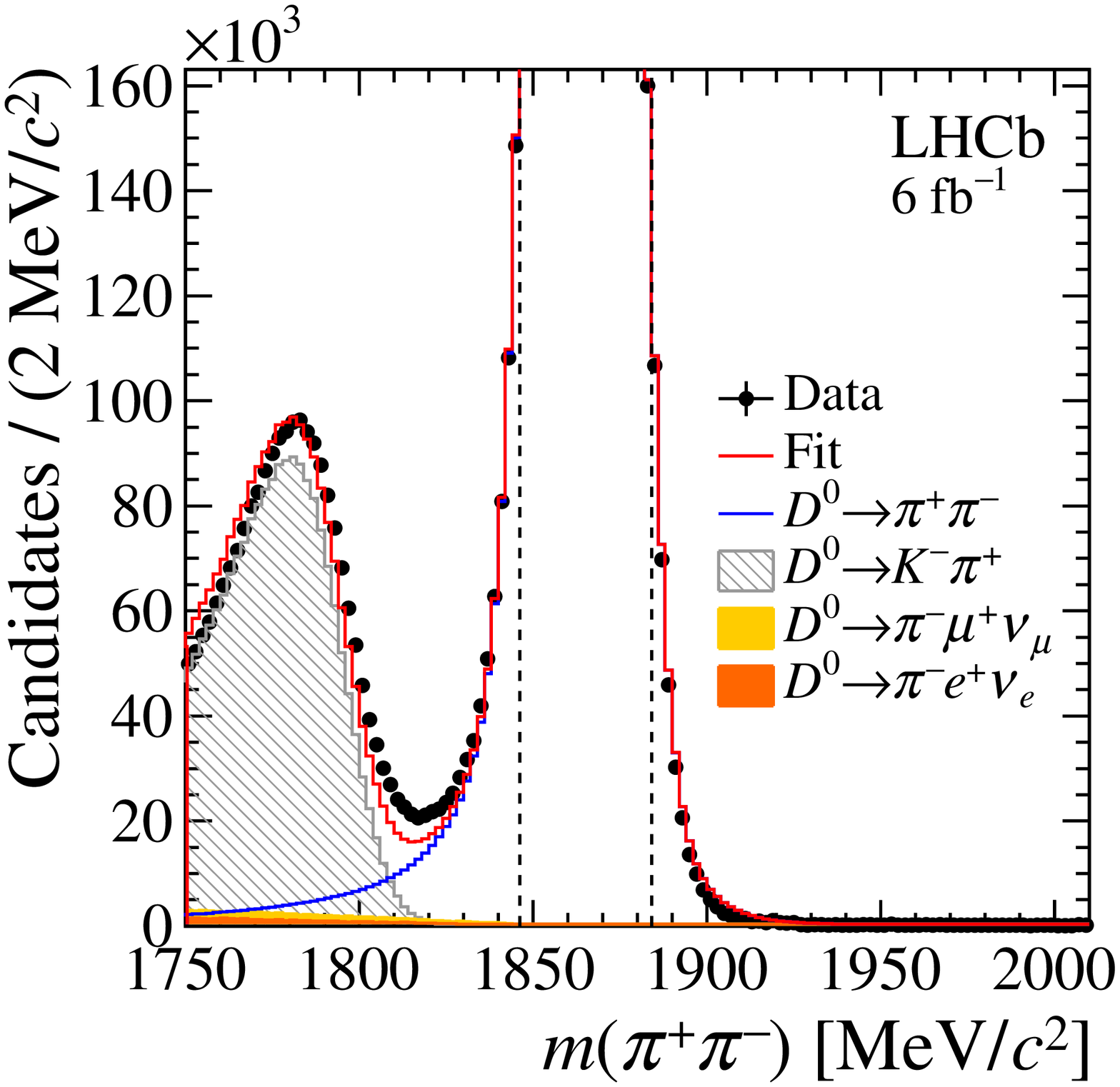}
  \end{center}
  \vspace*{-0.5cm}
  \caption{
    Distributions of \mhh of (top) \RS, (left) \KK and (right) \PP final states, with the results superimposed.
    The background components are stacked; the \mhh template \PDF of the \decay{\Dsp}{\Kp\Km\pip} decay has a negative contribution to the left of the known \Dz mass due to the subtraction of the \DstM background.
    The vertical dashed lines delimit the signal region.
  }
  \label{fig:mhh_bkgs}
\end{figure}

All the background components are studied using a simplified simulation~\cite{Cowan:2016tnm}, where the decays of unstable particles are described by \evtgen~\cite{Lange:2001uf}, final-state radiation (FSR) is generated using \photos~\cite{davidson2015photos} and the acceptance and the momentum, vertex and IP resolutions are simulated in a parametric way.
The same simulation is used to determine the FSR distribution of the signal decays with better precision than what would be possible by using the smaller simulated sample described in Sect.~\ref{sect:detector}.
However, while the FSR distribution of signal decays is fixed to the results of the simplified simulation, the signal mass resolution and the \PDF tails due to decays in flight of pions and kaons into muons are fixed to those measured in the simulation described in Sect.~\ref{sect:detector}.
For all decays, simulated events are weighted to reproduce the effect of the PID requirements on the particles reconstructed as coming from the \Dz final state.
The weights are calculated with a data-driven method by employing large calibration samples~\cite{LHCb-PUB-2016-021,LHCb-DP-2018-001}, and are parametrised as a function of momentum and pseudorapidity.
The background contamination in the signal region is estimated through template fits to the \mhh data distribution.
Only the signal and background yields and the resolution of the signal component are varied, the latter to correct for $\order{10\%}$ discrepancies in the resolution between data and simulation, whereas the background template {\PDF}s and the signal FSR and tails due to decays in flight are fixed to the simulation results.
The results of the fit are displayed in Fig.~\ref{fig:mhh_bkgs}.
The fitted ratio of the relative normalisation of the background components with respect to the signal agrees with expectations within 15\%, except for \decay{\Dsp}{\Kp\Km\pip} decays, for which the discrepancy is at the 35\% level.
While the agreement with data is not perfect, the projections capture all the main features of the \mhh distributions and allow an estimate of the size of the background contamination under the \Dz mass peak with a precision sufficient to assess a systematic uncertainty.

For the \RS decay channel, the largest contamination in the signal window is due to \decay{\Dz}{\Km\ellp\neul} decays, where \ellp stands for \ep or \mup, and amounts to \mbox{$(2.5 \pm 0.1) \times 10^{-4}$} of the \DzRS yield.
The time-dependent asymmetry of this background is estimated in the $[1750,1780]\mevcc$ sideband, after subtracting the contribution from signal decays.
The total estimated bias on \DY{\RS} is less than $0.01 \times 10^{-4}$.
For the \KK final state, the largest background fractions are \mbox{$(8.2 \pm 0.8)\times 10^{-4}$} for \decay{\Dz}{\Km\pip\piz}, \mbox{$(3.7 \pm 0.2)\times 10^{-4}$} for \decay{\Dz}{\Km\ellp\neul} and \mbox{$(2.3 \pm 0.2)\times 10^{-4}$} for \decay{\Dz}{\Km\pip} decays.
Their asymmetries are estimated in the $[1750,1780]\mevcc$ and $[1920,1970]\mevcc$ sidebands for the \decay{\Dz}{\Km\pip\piz} and \decay{\Dz}{\Km\pip} decay channels, respectively.
Measuring the asymmetry of \decay{\Dz}{\Km\ellp\neul} decays is particularly challenging owing to the tiny fraction of these decays in data.
Therefore, its size is conservatively assigned to the maximum value of the asymmetries measured for all other background channels in all of the \decay{\Dz}{\HH} decay channels.
The total bias on the \DY{\KK} value due to the background components in the $m(\KK)$ signal window is estimated to be \mbox{$0.06 \times 10^{-4}$}.
Finally, for the \PP decay channel the only relevant background contribution is due to \decay{\Dz}{\pim\ellp\neul} decays, whose fraction in the signal region is \mbox{$(2.5 \pm 0.2) \times 10^{-4}$}.
Their asymmetry difference with respect to the signal is estimated in the same way as for \decay{\Dz}{\Km\ellp\neul} decays for the \KK final state, and provokes a bias on \DY{\PP} less than \mbox{$0.03 \times 10^{-4}$}.

The kinematic equalisation of the momentum distribution of \pisp and \pism and of \Dz and \Dzb mesons is performed through a binned approach.
While the choice of the concerned variables is optimised to remove the kinematic asymmetries, the intervals size has to be kept large enough to avoid large statistical fluctuations.
Therefore, detector-induced, time-dependent asymmetries might not be completely removed by the kinematic weighting if they vary considerably within the intervals. 
The size of the residual asymmetries is estimated in the \DzRS sample by reducing progressively the size of the intervals until the measured value of \DY{\RS} does not change within the statistical uncertainty.
A systematic uncertainty of \mbox{$0.05\times 10^{-4}$} is estimated as the difference between the value of \DY{\RS} measured with the baseline scheme and its asymptotic value.
As a cross-check of the effectiveness of the kinematic weighting in removing the nuisance asymmetries, alternative configurations of the kinematic weighting acting on different variables have been employed, including that described in Ref.~\cite{LHCb-CONF-2019-001}.
The baseline configuration minimises the residual asymmetries of all kinematic variables of the \Dz and \pisp mesons after the weighting.
However, all weighting procedures that remove satisfactorily the asymmetries of the \Dz momentum provide measurements of \DY{\RS} within \mbox{$0.13 \times 10^{-4}$} from the baseline value.

\begin{table}[tb]
  \caption{Summary of the systematic uncertainties, in units of $10^{-4}$.
  The statistical uncertainties are reported for comparison.}
  \vspace*{-0.5cm}
  \begin{center}
  \begin{tabular}{l >{$}c<{$} >{$}c<{$}}
    \toprule
    Source&			            \DY{\KK} [10^{-4}]&	\DY{\PP}[10^{-4}]\\
    \midrule
    Subtraction of the \DstM background&    0.2&   0.3\\
    Flavour-dependent shift of \Dstar-mass peak&   0.1&   0.1\\
    \Dstarp from \B-meson decays&   0.1&   0.1\\
    \mhh background&    0.1&   0.1\\
    Kinematic weighting&    0.1&   0.1\\
    \midrule
    Total systematic uncertainty&   \agammaKKsys&   \agammaPPsys\\
    Statistical uncertainty&    \agammaKKstat&  \agammaPPstat\\
    \bottomrule
  \end{tabular}
  \end{center}
\label{tab:systematics}
\end{table}
All systematic uncertainties are summarised in Table~\ref{tab:systematics}.
The slope of the time-dependent asymmetry of the control sample is measured to be \mbox{$\DY{\RS} = (-\agammaRSval \pm \agammaRSstat \pm \agammaRSsys )\times 10^{-4}$}, where the first uncertainty is statistical and the second is systematic, and is compatible with zero as expected.
Note, however, that this measurement was not performed blindly.
Additional robustness tests are performed to check that the measured value of \DY{\HH} does not display unexpected dependencies on various observables, including
the selections that are satisfied by the \Dz candidate at the hardware and at the first software stage of the trigger;
the momentum, the transverse momentum and the pseudorapidity of the \Dz and \pisp mesons;
the \Dz flight distance in the plane transverse to the beam;
the position of the PV along the beamline;
and the number of PVs in the event.
No significant dependencies of \DY{\HH} on any of these variables are found.
The measurement is repeated for the signal channels, assigning a zero weight in the weighting procedure of Sect.~\ref{sect:asymmetries} only to the candidates in the tridimensional-space intervals for which the corresponding intervals of the \RS sample have fewer than 40 candidates or an asymmetry greater than 20\%.
In this way, the choice of the zero weights is made independent of the value of \DY{\hp\hm}.
The stability of the measurement is further checked as a function of the threshold of the minimum number of candidates and of the maximum asymmetry per interval.
The results of all these tests are compatible with the baseline one within the statistical uncertainty.
Finally, possible biases due to the decay-time resolution, approximately $0.11\,\tauDz$, are determined in simulation to be less than $0.01 \times 10^{-4}$, and thus are neglected.

\section{Results}
\label{sect:results}

\begin{figure}[tb]
  \begin{center}
    \includegraphics[width=0.6\linewidth]{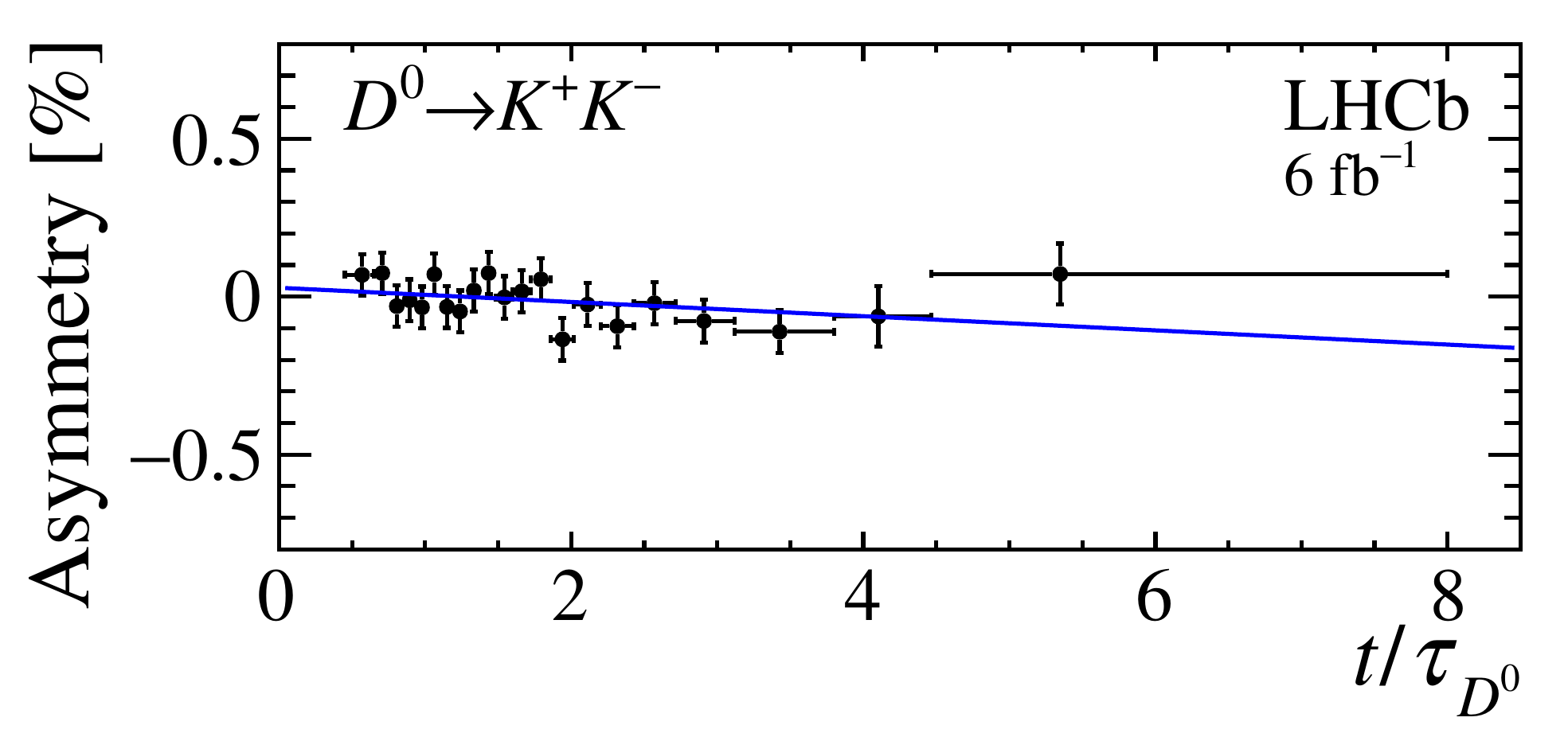}
    \includegraphics[width=0.6\linewidth]{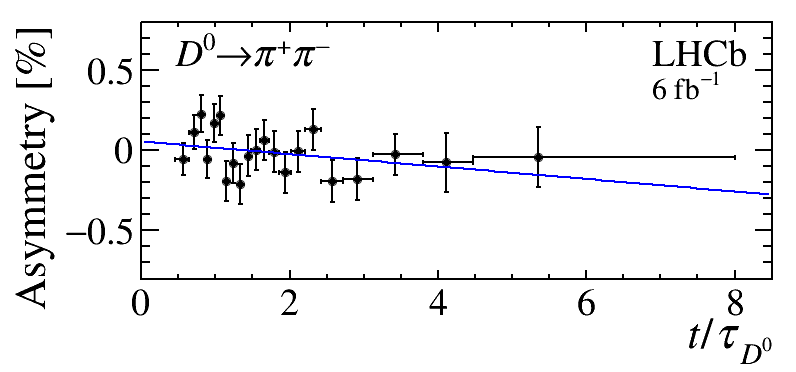}
  \end{center}
  \vspace*{-0.6cm}
  \caption{Asymmetry, $A_\mathrm{sig}(t)$, as a function of decay time for (top) \DzKK and (bottom) \DzPP candidates.
  A linear fit is superimposed.
  The \chisqndf of the fits are \mbox{$15/19$} and \mbox{$21/19$}, respectively.
  }
  \label{fig:asymmetry_fit}
\end{figure}

The time-dependent asymmetries of the \DzKK and \DzPP channels, after the kinematic weighting and the subtraction of the contribution from \B-meson decays, are displayed in Fig.~\ref{fig:asymmetry_fit}.
Linear fits are superimposed, and the resulting slopes are
\begin{align*}
        \DY{\KK} &= (-\agammaKKval \pm \agammaKKstat \pm \agammaKKsys )\times 10^{-4},\\
        \DY{\PP} &= (-\agammaPPval \pm \agammaPPstat \pm \agammaPPsys )\times 10^{-4},
\end{align*}
where the first uncertainties are statistical and the second are systematic.
Assuming that all systematic uncertainties are 100\% correlated, except those on the $m(\hp\hm)$ background, which are taken to be uncorrelated, the difference of \DY{f} between the two final states is equal to
\begin{equation*}
    \DY{\KK} - \DY{\PP} = (1.7 \pm 3.2 \pm 0.1)\times 10^{-4},
\end{equation*}
and is consistent with zero.
Neglecting final-state dependent contributions to \DY{f}, the two values are combined using the best linear unbiased estimator~\cite{Lyons:1988rp,Nisius:2020jmf}.
The result,
\begin{equation*}
    \DY{} = (-2.7 \pm 1.3 \pm 0.3)\times10^{-4},
\end{equation*}
is consistent with zero within two standard deviations, and both its statistical and systematic uncertainties are improved by more than a factor of two with respect to the previous most precise measurement~\cite{LHCb-PAPER-2016-063}.

These results are combined with previous \lhcb measurements~\cite{LHCb-PAPER-2014-069,LHCb-PAPER-2016-063,LHCb-PAPER-2019-032}, with which they are consistent, yielding the \lhcb legacy results with the 2011--2012 and 2015--2018 data samples,
\begin{equation*}
\begin{aligned}
    \DY{\KK} &= (-0.3 \pm 1.3 \pm 0.3) \times 10^{-4},\\
    \DY{\PP} &= (-3.6 \pm 2.4 \pm 0.4) \times 10^{-4},\\
    \DY{} &= (-1.0 \pm 1.1 \pm 0.3) \times 10^{-4},\\
    \DY{\KK} - \DY{\PP} &= (+3.3 \pm 2.7 \pm 0.2) \times 10^{-4}.
\end{aligned}
\end{equation*}
Finally, the arithmetic average of \DY{\KK} and \DY{\PP}, which would allow final-state dependent contributions to be suppressed by a factor of $\epsilon$~\cite{Kagan:2020vri}, where $\epsilon$ is the parameter quantifying the breaking of the $U$-spin symmetry in these decays, is
\[
    \tfrac{1}{2}(\DY{\KK} + \DY{\PP}) = (-1.9 \pm 1.3 \pm 0.4) \times 10^{-4}.
\]
These results are consistent with no time-dependent \CP violation in \DzKK and \DzPP decays, and improve by nearly a factor of two on the precision of the previous world average~\cite{HFLAV18}.

\section*{Acknowledgements}
\noindent We express our gratitude to our colleagues in the CERN
accelerator departments for the excellent performance of the LHC. We
thank the technical and administrative staff at the LHCb
institutes.
We acknowledge support from CERN and from the national agencies:
CAPES, CNPq, FAPERJ and FINEP (Brazil); 
MOST and NSFC (China); 
CNRS/IN2P3 (France); 
BMBF, DFG and MPG (Germany); 
INFN (Italy); 
NWO (Netherlands); 
MNiSW and NCN (Poland); 
MEN/IFA (Romania); 
MSHE (Russia); 
MICINN (Spain); 
SNSF and SER (Switzerland); 
NASU (Ukraine); 
STFC (United Kingdom); 
DOE NP and NSF (USA).
We acknowledge the computing resources that are provided by CERN, IN2P3
(France), KIT and DESY (Germany), INFN (Italy), SURF (Netherlands),
PIC (Spain), GridPP (United Kingdom), RRCKI and Yandex
LLC (Russia), CSCS (Switzerland), IFIN-HH (Romania), CBPF (Brazil),
PL-GRID (Poland) and OSC (USA).
We are indebted to the communities behind the multiple open-source
software packages on which we depend.
Individual groups or members have received support from
AvH Foundation (Germany);
EPLANET, Marie Sk\l{}odowska-Curie Actions and ERC (European Union);
A*MIDEX, ANR, Labex P2IO and OCEVU, and R\'{e}gion Auvergne-Rh\^{o}ne-Alpes (France);
Key Research Program of Frontier Sciences of CAS, CAS PIFI,
Thousand Talents Program, and Sci. \& Tech. Program of Guangzhou (China);
RFBR, RSF and Yandex LLC (Russia);
GVA, XuntaGal and GENCAT (Spain);
the Royal Society
and the Leverhulme Trust (United Kingdom).

\section*{Appendices}
\appendix
\section{Formalism}
\label{app:delta_y}
The theoretical parametrisation of the \Dz decay rates is introduced in Ref.~\cite{Kagan:2020vri}, where the following phases are defined for the decays into Cabibbo-suppressed \CP eigenstates:
\begin{equation}
    \phi_f^M \equiv \arg\left(M_{12}\frac{\af}{\abf}\right),\qquad
    \phi_f^\Gamma \equiv \arg\left(\Gamma_{12}\frac{\af}{\abf}\right).
\end{equation}
Here, $M_{12}$ and $\Gamma_{12}$ are the off-diagonal elements of the two Hermitian matrices defined by $\bm{H}\equiv \bm{M} - \tfrac{i}{2}\bm{\Gamma}$, where $\bm{H}$ is the $2 \times 2$ effective Hamiltonian governing the evolution of the \Dz--\Dzb system, and $\af \equiv \bra{f}\mathcal{H}\ket{\Dz}$ ($\abf \equiv \bra{f}\mathcal{H}\ket{\Dzb}$) is the decay amplitude of a \Dz (\Dzb) meson into the final state $f$, with $\mathcal{H}$ the $\lvert \Delta C \rvert = 1$ effective Hamiltonian.
The time-dependent decay rates into the final state $f$ can be parametrised to second order in the mixing parameters as
\begin{equation}
    \label{eq:decay_widths_SCS}
    \begin{aligned}
    \Gamma(\decay{\Dz}{f,t}) &\equiv
            e^{-\tau}\lvert\af\rvert^{2}
            \left(1 + c^+_f \tau + c^{\prime +}_f \tau^2\right),\\
    \Gamma(\decay{\Dzb}{f,t}) &\equiv
            e^{-\tau}\lvert\abf\rvert^{2}
            \left(1 + c^-_f \tau + c^{\prime -}_f \tau^2\right),
    \end{aligned}
\end{equation}
where $\tau$ is defined as $\tau \equiv \Gamma t$, a normalisation factor common to the two equations is implicit, and the coefficients $c^\pm_f$ and $c^{\prime \pm}_{f}$ are equal to
\begin{align}
    \begin{split}
    c^\pm_f &=\left[
            \mp x_{12}\sin\phi_{f}^{M} -y_{12}\cos\phi_{f}^{\Gamma}\right]\left\lvert\frac{\abf}{\af}\right\rvert^{\pm 1}\\
            &\approx
            \mp x_{12}\sin\phi_{f}^{M} -y_{12}(1 \mp a^{d}_{f}), \label{eq:c_SCS}
    \end{split}\\
    \begin{split}
    c^{\prime \pm}_{f} &=
            \tfrac{1}{4}\Big[y_{12}^2 - x_{12}^2 + (y_{12}^2 + x_{12}^2 \pm 2x_{12}y_{12}\sin\phi_{12})\left\lvert\frac{\abf}{\af}\right\rvert^{\pm 2} \Big] \\
            &\approx
            \tfrac{1}{2}\left[y_{12}^2 \pm x_{12}y_{12}\sin\phi_{12} \mp (x_{12}^2 + y_{12}^2)a^{d}_{f} \right],
    \end{split}
\end{align}
where \mbox{$\phi_{12} \equiv \arg(M_{12}/\Gamma_{12}) = \phi^{M}_{f} - \phi^{\Gamma}_{f}$}.
In the approximate expressions, the relation
\begin{equation}
    a_{f}^{d} \equiv \frac{\lvert \af \rvert^{2} - \lvert \abf \rvert^{2}}
                          {\lvert \af \rvert^{2} + \lvert \abf \rvert^{2}} \approx 1 - \left\lvert\frac{\abf}{\af}\right\rvert
\end{equation}
has been used, and all terms have been expanded to first order in the \CP-violation parameters $a_{f}^{d}$, $\sin\phi_{f}^{M}$ and $\sin\phi_{f}^{\Gamma}$.
Both phases $\phi_{f}^{M}$ and $\phi_{f}^{\Gamma}$ are measured to be approximately equal to zero rather than $\pi$ with a significance greater than $5$ standard deviations~\cite{Kagan:2020vri,HFLAV18,LHCb-PAPER-2021-009}.

The \DY{f} parameter is defined as~\cite{Kagan:2020vri}
\begin{equation}
    \label{eq:Yf_SCS}
    \Delta Y_{f} \equiv \frac{c^{+}_{f} - c^{-}_{f}}{2}
                 \approx (- x_{12} \sin\phi^{M}_{f} + y_{12} a^{d}_{f}),
\end{equation}
and has first been measured (although with a relative minus sign in the definition of \DY{}) in Ref.~\cite{Lees:2012qh} as
\begin{equation}
    \DY{f} \approx - \frac{\hat{\Gamma}_{\decay{\Dz}{f}} - \hat{\Gamma}_{\decay{\Dzb}{f}}}
         {2\hat{\Gamma}_{\decay{\Dz}{\RS}}},
\end{equation}
by modelling the time distributions of \decay{\Dz}{f} and \DzRS decays, see Eq.~(\ref{eq:decay_widths_SCS}), with an exponential function, \mbox{$\textnormal{exp}(-\hat{\Gamma}\tau )$}, and assuming that the effective decay width $\hat{\Gamma}$ is equal to unity for \DzRS decays.
This method neglects the contributions to the effective decay widths from \mbox{$c^{\prime \pm}_{f}$}, assuming that $\hat{\Gamma}_{\decay{\Dz/\Dzb}{f}} = 1 - c^{\pm}_{f}$.
The \agamma{} observable, which has been used as alternative to \DY{f} in Refs.~\cite{LHCb-PAPER-2011-032,LHCB-PAPER-2013-054,Staric:2015sta}, is similarly defined as the asymmetry of the effective decay widths of \Dz and \Dzb mesons into the final state $f$,
\begin{equation}
    \agamma{f} \equiv \frac{\hat{\Gamma}_{\decay{\Dz}{f}} - \hat{\Gamma}_{\decay{\Dzb}{f}}}
         {\hat{\Gamma}_{\decay{\Dz}{f}} + \hat{\Gamma}_{\decay{\Dzb}{f}}},
\end{equation}
and is related to \DY{f} as
\begin{equation}
    \label{eq:DY_vs_ycp}
    \agamma{f} = -\frac{\DY{f}}{1 + \ycp{f}},
\end{equation}
where the \ycp{f} parameter is defined as \mbox{$\ycp{f} \equiv - (c_f^+ + c_f^-)/2$}
and is equal to $y_{12}$ up to second order in the \CP-violation parameters defined above.
However, as the statistical precision improves, approximating the time-dependent decay widths with the effective ones might not be a good approximation any longer, since \CP-even corrections to the exponential decay rate quadratic in the mixing parameters might be of the same order as the \CP-odd first-order ones.

On the contrary, the definition of $A_{\CP}(t)$ in Eq.~(\ref{eq:acp}) employed in Refs.~\cite{Aaltonen:2014efa,LHCb-PAPER-2014-069,LHCb-PAPER-2016-063,LHCb-CONF-2019-001,LHCb-PAPER-2019-032} and in the present article is always dominated by the first-order terms, since the \CP-even second-order terms cancel in the difference in the numerator.
In particular, the coefficient of the linear expansion of $A_{\CP}(t)$ in Eq.~(\ref{eq:acp_expansion}) is equal to \DY{f} up to a multiplicative factor of \mbox{$4\lvert\af\rvert^2\lvert\abf\rvert^2/(\lvert\af\rvert^2 + \lvert\abf\rvert^2)^2$}, whose difference with unity is approximately equal to $(\Acpdec{f})^2 / 2 \lesssim 10^{-6}$~\cite{LHCb-PAPER-2019-006,LHCb-PAPER-2016-035}.
This coefficient has been denoted as $-\agamma{f}$ in Refs.~\cite{Aaltonen:2014efa,LHCb-PAPER-2014-069,LHCb-PAPER-2016-063,LHCb-CONF-2019-001,LHCb-PAPER-2019-032}, neglecting the 1\% correction due to \ycp{f} in Eq.~(\ref{eq:DY_vs_ycp}).

The final-state dependent contributions to \DY{f} in Eq.~(\ref{eq:Yf_SCS}) can be isolated by defining $\phi^{M}_f \equiv \phi^{M}_2 + \delta\phi_f$,
where $\phi^{M}_2$ is the intrinsic \CP-violating mixing phase of \Dz mesons, defined as the argument of the dispersive mixing amplitude $M_{12}$ with respect to its dominant $\Delta U = 2$ component,
and $\delta\phi_f$ is the relative weak phase of the subleading amplitude responsible for \CP violation in the decay with respect to the dominant decay amplitude~\cite{Kagan:2020vri}.
By defining $\delta_f$ the strong-phase analogue of $\delta\phi_f$, and using $\delta\phi_f = - \Acpdec{f} \cot\delta_{f}$, Eq.~(\ref{eq:Yf_SCS}) can be written as
\begin{equation}
    \label{eq:agamma_phiM}
    \DY{f} \approx - x_{12} \sin\phi^{M}_2 + y_{12} \Acpdec{f} \left(1 + \frac{x_{12}}{y_{12}}\cot\delta_{f} \right),
\end{equation}
where the first term is universal and the second encloses the final-state dependence.
The term \mbox{$y_{12} \lvert \Acpdec{f} \rvert$} is estimated to be less than $0.1\times 10^{-4}$ by using available experimental data~\cite{LHCb-PAPER-2019-006,HFLAV18} and the minimal assumption that \Acpdec{\KK} and \Acpdec{\PP} have opposite signs, which is motivated by $U$-spin symmetry arguments as well as by experimental evidence~\cite{LHCb-PAPER-2019-006,LHCb-PAPER-2017-046}.
The factor \mbox{$\tfrac{x_{12}}{y_{12}}\cot\delta_f$} can enhance the dependence on the final state, even though the phase $\delta_f$ is expected to be of $\order{1}$ due to large rescattering at the charm mass scale.
On the other hand, the SM predictions for $\phi^{M}_2$ are of the order $2 \mrad$ or less~\cite{Bigi:2011re,Bobrowski:2010xg,Kagan:2020vri,Li:2020xrz}, even though enhancements up to one order of magnitude due to low-energy nonperturbative strong interactions cannot be excluded~\cite{Bobrowski:2010xg,Kagan:2020vri}.

An alternative parametrisation of \CP violation and mixing is based on the explicit expansion of the mass eigenstates of $\bm{H}$ in terms of the flavour eigenstates, \mbox{$\ket{D_{1,2}} \equiv p\ket{\Dz} \pm q\ket{\Dzb}$}, with \mbox{$\lvert p\rvert^2 + \lvert q\rvert^2 = 1$} (\CPT invariance is assumed).
The corresponding mixing parameters are defined as \mbox{$x \equiv (m_2-m_1)/\Gamma$} and \mbox{$y \equiv (\Gamma_2-\Gamma_1)/(2\Gamma)$}, where \mbox{$m_{1,2}$} and \mbox{$\Gamma_{1,2}$} are the masses and decay widths of the mass eigenstates.
Adopting the convention that $\ket{D_1}$ ($\ket{D_2}$) is the approximately \CP-odd (\CP-even) eigenstate, the following relations hold, $x_{12} \approx x$ and $y_{12} \approx y$, up to corrections quadratic in the \CP violation parameter $\sin\phi_{12}$~\cite{Grossman:2009mn,Kagan:2009gb,Kagan:2020vri}.
In this parametrisation, the parameter \DY{f} defined in Eq.~(\ref{eq:Yf_SCS}) is equal to
\begin{equation}
    \label{eq:hflav}
    \DY{f} = \frac{1}{2}
                 \left[ \left( \left\lvert\frac{q}{p}\right\rvert \left\lvert\frac{\abf}{\af}\right\rvert + \left\lvert\frac{p}{q}\right\rvert \left\lvert\frac{\af}{\abf}\right\rvert\right)x \sin\phif
                      -\left( \left\lvert\frac{q}{p}\right\rvert \left\lvert\frac{\abf}{\af}\right\rvert - \left\lvert\frac{p}{q}\right\rvert \left\lvert\frac{\af}{\abf}\right\rvert\right)y \cos\phif
                 \right],
\end{equation}
where \phif is defined as \mbox{$\phif \equiv \arg[-(q\abf)/(p\af)]$}.
Neglecting terms of order higher than one in the \CP-violation parameters $(\lvert q/p \rvert -1)$, $\sin\phif$ and $\Acpdec{f}$, Eq.~(\ref{eq:hflav}) can be written as
\begin{equation}
    \label{eq:agamma_first_order_hflav}
    \DY{f} \approx x\sin\phif - y\left(\left\lvert \frac{q}{p}\right\rvert - 1\right) + y\Acpdec{f}.
\end{equation}
Finally, the dependence on the final state can be separated from the universal component by defining $\phif \equiv \phi_2 - \delta\phi_f$, see Ref.~\cite{Kagan:2020vri}, where $\phi_2$ is a final-state independent weak phase dubbed $\phi$ by the HFLAV collaboration~\cite{HFLAV18} and $\delta\phi_f$ is the same as above, obtaining
\begin{equation}
    \label{eq:DY_phenomenological}
    \DY{f}   \approx x\sin\phi_2 - y\left(\left\lvert \frac{q}{p}\right\rvert - 1\right) + y \Acpdec{f} \left(1 + \frac{x}{y}\cot\delta_f\right).
\end{equation}

\section{Upper bound on the size of \texorpdfstring{$\boldsymbol{\Delta Y_{\RS}}$}{DeltaY(K–pi+)}}
\label{app:delta_y_rs}
In this appendix the final states $\Km\pip$ and $\Kp\pim$ are denoted with $f$ and $\bar{f}$, respectively.
Furthermore, two weak phases $\phi_f^M$ and $\phi_f^\Gamma$, independent of those defined in Appendix~\ref{app:delta_y}, and the strong-phase difference between the doubly Cabibbo-suppressed (DCS) and the Cabibbo-favoured (CF) decay amplitudes, $\Delta_f$, are defined as~\cite{Kagan:2020vri}
\begin{equation}
    \begin{aligned}
        &\frac{M_{12}}{\lvert M_{12} \rvert} \frac{\af}{\abf}
                \equiv - \left\lvert\frac{\af}{\abf}\right\rvert e^{i(\phi_f^M - \Delta_f)},
        &\frac{\Gamma_{12}}{\lvert \Gamma_{12} \rvert} \frac{\af}{\abf}
                \equiv - \left\lvert\frac{\af}{\abf}\right\rvert e^{i(\phi_f^\Gamma - \Delta_f)},\\
        &\frac{M_{12}}{\lvert M_{12} \rvert} \frac{\afb}{\abfb}
                \equiv - \left\lvert\frac{\afb}{\abfb}\right\rvert e^{i(\phi_f^M + \Delta_f)},
        &\frac{\Gamma_{12}}{\lvert \Gamma_{12} \rvert} \frac{\afb}{\abfb}
                \equiv - \left\lvert\frac{\afb}{\abfb}\right\rvert e^{i(\phi_f^\Gamma + \Delta_f)}.
    \end{aligned}
\end{equation}
The time-dependent decay rates of \Dz and \Dzb right-sign decays are parametrised as
\begin{equation}
    \label{eq:decay_rates_rs}
    \begin{aligned}
        \Gamma(\decay{\Dz}{f,t}) &\equiv
            e^{-\tau}\lvert\af\rvert^{2}
            \left(1 + \sqrt{R_f}c^+_f \tau + c^{\prime +}_f \tau^2\right),\\
        \Gamma(\decay{\Dzb}{\bar{f},t}) &\equiv
            e^{-\tau}\lvert\abfb\rvert^{2}
            \left(1 + \sqrt{R_f}c^-_f \tau + c^{\prime -}_f \tau^2\right),
    \end{aligned}
\end{equation}
up to second order in the mixing parameters, where $R_f$ is the \CP-averaged ratio of DCS to CF decay rates, defined as
\begin{equation}
    R_f \equiv \frac{1}{2}\left(\left\lvert\frac{\afb}{\af}\right\rvert^{2} + \left\lvert\frac{\abf}{\abfb}\right\rvert^{2}\right),
\end{equation}
and the coefficients $c^\pm_f$ and $c^{\prime \pm}_{f}$ are equal to
\begin{align}
    \begin{split}c^\pm_f &\approx
                \big[1 \mp \tfrac{1}{2}(a^d_{\bar{f}} + a^d_f)\big]
                (- x_{12}\sin\Delta_f + y_{12}\cos\Delta_f)\\
                &\qquad\qquad\qquad\qquad\qquad\qquad\qquad\qquad
                \pm x_{12}\sin\phi_f^M\cos\Delta_f \pm y_{12} \sin\phi_f^\Gamma \sin\Delta_f,
    \end{split}\\
    c^{\prime \pm}_f &\approx
            \tfrac{1}{4} (y_{12}^2 - x_{12}^2)
            + \tfrac{1}{4} R_f\big[1 \mp (a^d_{\bar{f}} + a^d_f)\big](x_{12}^2 + y_{12}^2)
            \pm \tfrac{1}{2}R_f x_{12} y_{12}\sin\phi_{12},
\end{align}
up to second order in the \CP-violation parameters $\sin\phi^M_f$, $\sin\phi^\Gamma_f$, \Acpdec{f} and \Acpdec{\bar{f}}, where the last two parameters are the \CP asymmetries in the decay into the CF and DCS final states, defined as
\begin{equation}
    \Acpdec{f} \equiv \frac{\lvert \af \rvert^{2} - \lvert \abfb \rvert^{2}}
                          {\lvert \af \rvert^{2} + \lvert \abfb \rvert^{2}},
    \qquad
    \Acpdec{\bar{f}} \equiv \frac{\lvert \afb \rvert^{2} - \lvert \abf \rvert^{2}}
                          {\lvert \afb \rvert^{2} + \lvert \abf \rvert^{2}}.
\end{equation}

The analogue of Eqs.~(\ref{eq:acp},\ref{eq:acp_expansion}) for right-sign decays is
\begin{equation}
    A_{\CP}(f,t) \equiv \frac{\Gamma(\decay{\Dz}{f},t) - \Gamma(\decay{\Dzb}{\bar{f}},t)}
                             {\Gamma(\decay{\Dz}{f},t) + \Gamma(\decay{\Dzb}{\bar{f}},t)} \approx \Acpdec{f}
                              + \DY{f} \frac{t}{\tauDz},
\end{equation}
where the \DY{f} parameter is defined as
\begin{equation}
    \label{eq:deltaY_RS}
    \begin{aligned}
    \Delta Y_{f} &\equiv \sqrt{R_f} \times \frac{c^{+}_{f} - c^{-}_{f}}{2} \\
                &\approx \sqrt{R_f}\Big[
                x_{12}\sin\phi_f^M\cos\Delta_f + y_{12} \sin\phi_f^\Gamma \sin\Delta_f
                +\tfrac{1}{2}(a^d_{\bar{f}} + a^d_f) (x_{12}\sin\Delta_f - y_{12}\cos\Delta_f)
                \Big].
    \end{aligned}
\end{equation}
A global fit of the mixing and time-dependent \CP-violation parameters to all of the charm measurements except the present one, with the assumption that \Acpdec{f} is zero,\footnote{The asymmetries \Acpdec{f} and \Acpdec{\bar{f}} are expected to be negligible in the SM since CF and DCS decays are not sensitive to quantum chromodynamics penguin and chromomagnetic dipole operators.
    Experimentally, \mbox{$A_D \equiv (\lvert\afb/\af\rvert^2 - \lvert\abf/\abfb\rvert^2) / (\lvert\afb/\af\rvert^2 + \lvert\abf/\abfb\rvert^2) \approx \Acpdec{\bar{f}} - \Acpdec{f}$} is equal to $(-7 \pm 4) \times 10^{-3}$~\cite{HFLAV18}.}
provides $\lvert \DY{\RS}\rvert < 0.2 \times 10^{-4}$ at 95\% confidence level~\cite{pajero:2021}.
This number is around 40\% of the precision of the present measurement of \DY{\RS}.

\addcontentsline{toc}{section}{References}
\bibliographystyle{LHCb}
\bibliography{main,standard,LHCb-PAPER,LHCb-CONF,LHCb-DP,LHCb-TDR}

\newpage
\centerline
{\large\bf LHCb collaboration}
\begin
{flushleft}
\small
R.~Aaij$^{32}$,
C.~Abell{\'a}n~Beteta$^{50}$,
T.~Ackernley$^{60}$,
B.~Adeva$^{46}$,
M.~Adinolfi$^{54}$,
H.~Afsharnia$^{9}$,
C.A.~Aidala$^{85}$,
S.~Aiola$^{25}$,
Z.~Ajaltouni$^{9}$,
S.~Akar$^{65}$
% ...

{\footnotesize \it

% ...
$^{9}$Universit{\'e} Clermont Auvergne, CNRS/IN2P3, LPC, Clermont-Ferrand, France\\
% ...
$^{25}$INFN Sezione di Milano, Milano, Italy\\
% ...
$^{32}$Nikhef National Institute for Subatomic Physics, Amsterdam, Netherlands\\
% ...
$^{46}$Instituto Galego de F{\'\i}sica de Altas Enerx{\'\i}as (IGFAE), Universidade de Santiago de Compostela, Santiago de Compostela, Spain\\
% ...
$^{50}$Physik-Institut, Universit{\"a}t Z{\"u}rich, Z{\"u}rich, Switzerland\\
% ...
$^{54}$H.H. Wills Physics Laboratory, University of Bristol, Bristol, United Kingdom\\
% ...
$^{60}$Oliver Lodge Laboratory, University of Liverpool, Liverpool, United Kingdom\\
% ...
$^{65}$University of Cincinnati, Cincinnati, OH, United States\\
% ...
$^{68}$Syracuse University, Syracuse, NY, United States\\
% ...
$^{85}$University of Michigan, Ann Arbor, United States, associated to $^{68}$\\
\bigskip
% ...
\medskip
}
\end{flushleft}

\end{document}